\documentclass[twocolumn]{aastex63}

\usepackage{amsmath}
\usepackage{amssymb}

\usepackage{lineno}

\usepackage{comment}

\submitjournal{ApJS}
\shorttitle{The MOST Hosts Survey}
\shortauthors{Soumagnac et al.}

\begin{document}


\title{The MOST Hosts Survey: spectroscopic observation of\\ the  host galaxies of $\sim40{,}000$ transients using DESI}


\correspondingauthor{Maayane T. Soumagnac}
\email{mtsoumagnac@lbl.gov}
\author[0000-0001-6753-1488]{Maayane T. Soumagnac}
\affiliation{Department of Physics, Bar-Ilan University, Israel}
\affiliation{Lawrence Berkeley National Laboratory, 1 Cyclotron Road, Berkeley, CA 94720, USA}
\author[0000-0002-3389-0586]{Peter Nugent}
\affiliation{Lawrence Berkeley National Laboratory, 1 Cyclotron Road, Berkeley, CA 94720, USA}
\affiliation{Department of Astronomy, University of California, Berkeley, 501 Campbell Hall, Berkeley, CA 94720, USA}
\author{Robert A. Knop}
\affiliation{Lawrence Berkeley National Laboratory, 1 Cyclotron Road, Berkeley, CA 94720, USA}
\author{Anna Y. Q. Ho}
\affiliation{Department of Astronomy, Cornell University, Ithaca, NY, 14853, USA}
\author{William Hohensee}
\affiliation{Department of Astronomy, University of California, Berkeley, CA, 94720-3411, USA}
\author{Autumn Awbrey}
\affiliation{Department of Astronomy, University of California, Berkeley, CA, 94720-3411, USA}
\author{Alexis Andersen}
\affiliation{Department of Astronomy, University of California, Berkeley, CA, 94720-3411, USA}
\author{Greg Aldering}
\affiliation{Lawrence Berkeley National Laboratory, 1 Cyclotron Road, Berkeley, CA 94720, USA}
\author{Matan Ventura}
\affiliation{Department of Physics, Bar-Ilan University, Israel}

\author{Jessica N.~Aguilar}
\affiliation{Lawrence Berkeley National Laboratory, 1 Cyclotron Road, Berkeley, CA 94720, USA}
\author[0000-0001-6098-7247]{Steven Ahlen}
\affiliation{Physics Dept., Boston University, 590 Commonwealth Avenue, Boston, MA 02215, USA}

\author[0000-0001-5537-4710]{Segev Y. Benzvi}
 \affiliation{Department of Physics \& Astronomy, University of Rochester, Rochester,206 Bausch and Lomb Hall, P.O. Box 270171, Rochester, NY 14627-0171, USA}

\author{David Brooks}
\affiliation{Department of Physics \& Astronomy, University College London, Gower Street, London, WC1E 6BT, UK}

\author{Dillon Brout}
\affiliation{Department of Astronomy, Boston University, Boston, MA 02115, USA}
\affiliation{Department of Physics, Boston University, Boston, MA 02115, USA}

\author{Todd Claybaugh}
\affiliation{Lawrence Berkeley National Laboratory, 1 Cyclotron Road, Berkeley, CA 94720, USA}

\author{Tamara M.\ Davis}
\affiliation{School of Mathematics and Physics, The University of Queensland, St Lucia, QLD 4101, Australia}

\author{Kyle Dawson}
\affiliation{Department of Physics and Astronomy, The University of Utah, 115 South 1400 East, Salt Lake City, UT 84112, USA}

\author[0000-0002-1769-1640]{Axel de la Macorra}
\affiliation{Instituto de F\'{\i}sica, Universidad Nacional Aut\'{o}noma de M\'{e}xico,  Cd. de M\'{e}xico  C.P. 04510,  M\'{e}xico}

\author[0000-0002-4928-4003]{Arjun Dey}
\affiliation{NSF's NOIRLab, 950 N. Cherry Ave., Tucson, AZ 85719, USA}

\author[0000-0002-5665-7912]{Biprateep Dey}
\affiliation{Department of Physics \& Astronomy and Pittsburgh Particle Physics, Astrophysics, and Cosmology Center (PITT PACC), University of Pittsburgh, 3941 O'Hara Street, Pittsburgh, PA 15260, USA}

\author{Peter Doel}
\affiliation{Department of Physics \& Astronomy, University College London, Gower Street, London, WC1E 6BT, UK}

\author{Kelly A. Douglass}
 \affiliation{Department of Physics \& Astronomy, University of Rochester, Rochester, NY , USA}

\author[0000-0002-2890-3725]{Jaime E. Forero-Romero}
\affiliation{Departamento de F\'isica, Universidad de los Andes, Cra. 1 No. 18A-10, Edificio Ip, CP 111711, Bogot\'a, Colombia},
\affiliation{Observatorio Astron\'omico, Universidad de los Andes, Cra. 1 No. 18A-10, Edificio H, CP 111711 Bogot\'a, Colombia}

\author{Enrique Gazta{\~n}aga}
\affiliation{Institut d'Estudis Espacials de Catalunya (IEEC), 08034 Barcelona, Spain}
\affiliation{Institute of Cosmology \& Gravitation, University of Portsmouth, Dennis Sciama Building, Portsmouth, PO1 3FX, UK}
\affiliation{Institute of Space Sciences, ICE-CSIC, Campus UAB, Carrer de Can Magrans s/n, 08913 Bellaterra, Barcelona, Spain}

\author{Satya Gontcho A Gontcho}
 \affiliation{Lawrence Berkeley National Laboratory, 1 Cyclotron Road, Berkeley, CA 94720, USA}
 \affiliation{Department of Physics \& Astronomy, University of Rochester, Rochester, NY , USA}
 
\author{Or Graur}
 \affiliation{Institute of Cosmology and Gravitation, University of Portsmouth, Portsmouth, PO1 3FX, UK}
  \affiliation{Department of Astrophysics, American Museum of Natural History, Central Park West and 79th Street, New York, NY 10024, USA}

\author{Julien Guy}
\affiliation{Lawrence Berkeley National Laboratory, 1 Cyclotron Road, Berkeley, CA 94720, USA}

\author[0000-0003-1197-0902]{ChangHoon Hahn}
\affiliation{Department of Astrophysical Sciences, Princeton University, Princeton NJ 08544, USA}

\author{Klaus Honscheid}
\affiliation{Center for Cosmology and AstroParticle Physics, The Ohio State University, 191 West Woodruff Avenue, Columbus, OH 43210, USA}
\affiliation{Department of Physics, The Ohio State University, 191 West Woodruff Avenue, Columbus, OH 43210, USA}
\affiliation{The Ohio State University, Columbus, 43210 OH, USA}
  
\author{Cullan Howlett}
\affiliation{School of Mathematics and Physics, The University of Queensland, St Lucia, QLD 4101, Australia}

\author{Alex G. Kim}
\affiliation{Lawrence Berkeley National Laboratory, 1 Cyclotron Road, Berkeley, CA 94720, USA}

\author[0000-0003-3510-7134]{Theodore Kisner}
\affiliation{Lawrence Berkeley National Laboratory, 1 Cyclotron Road, Berkeley, CA 94720, USA}

\author[0000-0001-6356-7424]{Anthony Kremin}
\affiliation{Lawrence Berkeley National Laboratory, 1 Cyclotron Road, Berkeley, CA 94720, USA}

\author{Andrew Lambert}
\affiliation{Lawrence Berkeley National Laboratory, 1 Cyclotron Road, Berkeley, CA 94720, USA}

\author[0000-0003-1838-8528]{Martin Landriau}
\affiliation{Lawrence Berkeley National Laboratory, 1 Cyclotron Road, Berkeley, CA 94720, USA},

\author{Dustin Lang}
\affiliation{Dunlap Institute, University of Toronto, Toronto, ON M5S 3H4, Canada}
\affiliation{Department of Astronomy \& Astrophysics, University of Toronto, Toronto, ON M5S 3H4, Canada}
\affiliation{Perimeter Institute for Theoretical Physics, Waterloo, ON N2L 2Y5, Canada}

\author[0000-0001-7178-8868]{Laurent Le~Guillou}
\affiliation{Sorbonne Universit\'{e}, CNRS/IN2P3, Laboratoire de Physique Nucl\'{e}aire et de Hautes Energies (LPNHE), FR-75005 Paris, France}

\author[0000-0003-4962-8934]{Marc Manera}
\affiliation{Departament de F\'{i}sica, Serra H\'{u}nter, Universitat Aut\`{o}noma de Barcelona, 08193 Bellaterra (Barcelona), Spain}
\affiliation{Institut de F\'{i}sica dâ€™Altes Energies (IFAE), The Barcelona Institute of Science and Technology, Campus UAB, 08193 Bellaterra Barcelona, Spain}

\author[0000-0002-1125-7384]{Aaron Meisner}
\affiliation{NSF's NOIRLab, 950 N. Cherry Ave., Tucson, AZ 85719, USA}

\author{Ramon Miquel}
\affiliation{Instituci\'{o} Catalana de Recerca i Estudis Avan\c{c}ats, Passeig de Llu\'{\i}s Companys, 23, 08010 Barcelona, Spain}
\affiliation{Institut de F\'{i}sica dâ€™Altes Energies (IFAE), The Barcelona Institute of Science and Technology, Campus UAB, 08193 Bellaterra Barcelona, Spain}

\author{John Moustakas}
\affiliation{Department of Physics and Astronomy, Siena College, 515 Loudon Road, Loudonville, NY 12211, USA}

\author{Adam D. Myers}
\affiliation{Department of Physics \& Astronomy, University of Wyoming, 1000 E. University, Dept. 3905, Laramie, WY 82071, USA}
\author[0000-0001-6590-8122]{Jundan Nie}
\affiliation{National Astronomical Observatories, Chinese Academy of Sciences, A20 Datun Rd., Chaoyang District, Beijing, 100012, P.R. China}

\author[0000-0002-6011-0530]{Antonella Palmese}
\affiliation{McWilliams Center for Cosmology, Department of Physics, Carnegie Mellon University, Pittsburgh, PA 15213, USA}
\affiliation{Department of Physics, University of California, Berkeley, CA, 94720-3411, USA}
\affiliation{Lawrence Berkeley National Laboratory, 1 Cyclotron Road, Berkeley, CA 94720, USA}

\author[0000-0002-7464-2351]{David Parkinson}
\affiliation{Korea Astronomy and Space Science Institute, 776, Daedeokdae-ro, Yuseong-gu, Daejeon 34055, Republic of Korea}

\author{Claire Poppett}
\affiliation{Lawrence Berkeley National Laboratory, 1 Cyclotron Road, Berkeley, CA 94720, USA}
\affiliation{Space Sciences Laboratory, University of California, Berkeley, 7 Gauss Way, Berkeley, CA  94720, USA}
\affiliation{University of California, Berkeley, 110 Sproul Hall \#5800 Berkeley, CA 94720, USA}

\author[0000-0001-7145-8674]{Francisco Prada}
\affiliation{Instituto de Astrof\'{i}sica de Andaluc\'{i}a (CSIC), Glorieta de la Astronom\'{i}a, s/n, E-18008 Granada, Spain}

\author[0000-0001-7950-7864]{Fei Qin}
\affiliation{School of Physics, Korea Institute for Advanced Study, Hoegiro 85, Dongdaemun-gu,  Seoul 02455, Republic of Korea}

\author[0000-0001-5589-7116]{Mehdi Rezaie}
\affiliation{Department of Physics, Kansas State University, 116 Cardwell Hall, Manhattan, KS 66506, USA}

\author{Graziano Rossi}
\affiliation{Department of Physics and Astronomy, Sejong University, Seoul, 143-747, Korea}

\author[0000-0002-9646-8198]{Eusebio Sanchez}
\affiliation{CIEMAT, Avenida Complutense 40, E-28040 Madrid, Spain}

\author{David D.~Schlegel}
\affiliation{Lawrence Berkeley National Laboratory, 1 Cyclotron Road, Berkeley, CA 94720, USA}

\author{Michael Schubnell}
\affiliation{Department of Physics, University of Michigan, Ann Arbor, MI 48109, USA}
\affiliation{University of Michigan, Ann Arbor, MI 48109, USA}

\author[0000-0002-3461-0320]{Joseph H.~Silber}
\affiliation{Lawrence Berkeley National Laboratory, 1 Cyclotron Road, Berkeley, CA 94720, USA}

\author[0000-0003-1704-0781]{Gregory Tarl\'{e}}
\affiliation{University of Michigan, Ann Arbor, MI 48109, USA}


\author{Benjamin~A. Weaver}
\affiliation{NSF's NOIRLab, 950 N. Cherry Ave., Tucson, AZ 85719, USA}

\author[0000-0002-4135-0977]{Zhimin Zhou}
\affiliation{National Astronomical Observatories, Chinese Academy of Sciences, A20 Datun Rd., Chaoyang District, Beijing, 100012, P.R. China}

\begin{abstract}

We present the MOST Hosts survey (Multi-Object Spectroscopy of Transient Hosts). The survey is planned to run throughout the five years of operation of the Dark Energy Spectroscopic Instrument (DESI) and will generate a spectroscopic catalog of the hosts of most transients observed to date, in particular all the supernovae observed by most public, untargeted, wide-field, optical surveys (PTF/iPTF, SDSS II, ZTF, DECAT, DESIRT). Scientific questions for which the MOST Hosts survey will be useful include Type Ia supernova cosmology, fundamental plane and peculiar velocity measurements, and the understanding of the correlations between transients and their host galaxy properties. Here, we present the first release of the MOST Hosts survey: 
$21,931$ hosts of $20,235$ transients. These numbers represent $36\%$ of the final MOST Hosts sample, consisting of $60,212$ potential host galaxies of $38,603$ transients (a transient can be assigned multiple potential hosts). Of these galaxies, $40\%$ do not appear in the DESI primary target list and therefore require a specific program like MOST Hosts. Of all the transients in the MOST Hosts list, only $26.7\%$ have existing classifications, and so the survey will provide redshifts (and luminosities) for nearly $30,000$ transients. A preliminary Hubble diagram and a transient luminosity-duration diagram are shown as examples of future potential uses of the MOST Hosts survey.  The survey will also be timely, as a way to provide a training sample of spectroscopically observed transients for classifiers relying only on photometry, as we enter an era when most newly observed transients (e.g. those detected with the Rubin Observatory) will lack spectroscopic classification. The MOST Hosts DESI survey data will be released through the Wiserep platform on a rolling cadence and updated to match the DESI releases. Dates of future releases and updates are available through the {\href{https://mosthosts.desi.lbl.gov}{https://mosthosts.desi.lbl.gov}} website. 

\end{abstract}


\section{Introduction}

The Dark Energy Spectroscopic Instrument (DESI) project saw first light in May 2021. Using a robotic, $5020$-fiber-fed spectrograph \citep{instrument} mounted on the Mayall 4m telescope at the Kitt Peak National Observatory, DESI was designed to take spectra and study the distribution of $\sim 40$ million galaxies and quasars over its five years of operation \citep{DESI2016a,DESI2016b,SV}, in order to build a 3D map spanning from the nearby universe out to a distance of $11$ billion light years and to measure the effect of dark energy on the expansion of the universe with unprecedented accuracy. The project's main goals include deriving sub-percent constraints on the equation of state of dark energy and its time evolution \citep{Alam2017}. Redshift-space distortions measurements will also provide precise constraints on the growth of structure in the universe, aiming for percent-level precision on the cosmological parameters \citep{Guzzo2008, Blake2011,Pezzotta2017}.

A comprehensive description of the completed instrument can be found in \cite{instrument}; here, we quickly summarize the instrument main characteristics. The focal plane of the instrument is constructed of $5,020$ robotically controlled fiber positioners \citep{Silber2023}, each holding a unique fiber with a core diameter of $107\,\mu m$, which correspond to $\sim1.5\,\rm arcsec$ on the sky, most DESI galaxies being aboserve slightly larger than the fibers  size (Poppett et al. 2023 in prep.). Twenty fibers direct
light to a camera to monitor the sky brightness, while the remaining $5,000$ fibers direct the light of targeted objects to one of ten spectrographs. These spectrographs have three cameras, denoted
as B ($3600–5800\,\rm\r{A}$), R ($5760–7620\,\rm\r{A}$), and Z ($7520–
9824\,\rm\r{A}$), that provide a resolving power of roughly $2000$
at $3600$\,$\rm\r{A}$, increasing to roughly $5500$ at $9800$\,$\rm\r{A}$ (Jelinsky et al. 2023, in prep.). The full system has the sensitivity to measure and resolve the [OII] doublet down to fluxes of $8\times 10^{-17}\,\rm erg/s/cm^2$ in effective exposure times of 1000 seconds for galaxies $0.6 < z < 1.6$, with a median signal to noise ratio of $7$.

No specific lines are used to determine the redshift. Rather, a simultaneous template fit to the entire spectrum (i.e. including the continuum, emission, and absorption lines. $4000\,\rm\r{A}$ break etc) is performed. The redshift estimate is typically precise to four decimals. A detailed summary of the random/systematic redshift errors for each of the DESI individual samples is given in \cite{SV}.

Although the main goals of DESI are for cosmology using the large scale structure of galaxies, it will provide the largest spectroscopic map of both galaxies and stars to date, in a volume 10 times bigger than SDSS. As such, it opens opportunities for a variety of fields, including time-domain astronomy. In particular, a promising and substantial contribution by DESI to transient science is made possible by the overlap between the DESI footprint and the footprint of several untargeted, wide-field, optical transient surveys which have revolutionised the field in the last decade: ASAS-SN, ATLAS, PanSTARRS, PTF/iPTF \citep{Law2009}, SDSS-II \citep{Frieman2008}, ZTF \citep{Graham2019,Bellm2019}.

The Multi-Object Spectroscopy of Transient (MOST) Hosts survey seeks to leverage the emerging opportunities for new discoveries and breakthroughs at this very critical moment in the history of the field, when multi-object spectroscopy is reaching large scales and when new cutting-edge surveys are providing unprecedentedly copious and rich observations of transient phenomena. In this spirit, the MOST Hosts target list includes the host galaxies of most transients known to date.
MOST Hosts is not the first attempt to produce a large scale survey of the host galaxies of a large sample of transients. \cite{Gagliano2021} made a photometric version of such a survey with the Galaxies HOsting Supernova Transients (GHOST) database, which includes photometric properties of the host galaxies of 16,175 spectroscopically classified SNe from the Transient Name Server\footnote{ \href{https://wis-tns.weizmann.ac.il/}{https://wis-tns.weizmann.ac.il/}} and the Open Supernova Catalog\footnote{ \href{https://sne.space/}{https://sne.space/}} \citep{Guillochon2017}. 

The advantage of the MOST Hosts survey will be to provide spectroscopic and redshift information -- all provided by the same instrument -- about the hosts of a large sample of transient phenomena, which should allow one to quickly tackle several important science questions. 
Below, we detail a few example science cases of the MOST Hosts survey.

\subsection{Uncovering the relationship between supernovae and host galaxy types and physical properties}\label{sec:SNe}

Providing host spectroscopy and redshift information for most known supernovae (SNe) will allow one to study the relationship between the SN types, their physical properties and  their host galaxies. Recent studies have explored the correlation between a variety of SN type properties and their host galaxies. For example, \cite{2010ApJ...715..743K, 2010ApJ...722..566L, Sullivan2010, Childress2013a, Childress2013b, 2017ApJ...848...56U, 2020MNRAS.494.4426S} have explored the relationship between SN~Ia Hubble residuals and the integrated properties of their host galaxies; \cite{Pan2015} explored the correlation between SN spectral features and their host galaxies properties; \cite{Rigault2020} further showed a SN~Ia standardization bias dependent on the local specific star formation rate; \citet{Boone2021b} and \cite{2021ApJ...909...26B} have shown that these biases depend on how SNe~Ia are standardized; \cite{MassStepDust} found a correlation between the SNe brightness and the overall dust content of their host galaxy; \cite{Nugent2023} showed that the local environments of sub-Chandrasekhar mass explosions are more dust-affected than normal SNe~Ia. \cite{Perley2016} and \cite{Schulze2018} studied the host-galaxy properties of samples of Superluminous SNe (SLSNe) and showed that they are characterized by a low metallicity and that a short-lived stellar population is probably required to regulate the SLSN production. 

The overlap between the DESI footprint and the ZTF Bright Transient Survey (BTS; \citealt{Fremling2019}) offers a great opportunity to further explore the relationship between SN properties and the properties of their host galaxies. The BTS aimed to classify all extragalactic transients brighter than magnitude 19 (in either the g or r filter) and immediately announcing the classification to the public. Comprised of thousands of SNe, it is the largest flux-limited SN survey to date. However, only half of the hosts have known redshifts. For the BTS transients included in our sample (see section~\ref{sec:transients} for our transient selection strategy), the MOST Hosts survey will complete the BTS by providing the currently unknown host redshift information. 

With the imminent start of wide-field surveys such as the Rubin Observatory's Legacy Survey of Space and Time (LSST; \citealt{Tyson2002,lsst}), which will lead the vast majority of the newly observed SNe to lack a spectroscopic classification, different strategies are being explored to design transients classifiers which are based either on photometry only, or on a combination of photometry and host information (e.g. \citealt{Muthukrishna2019,Villar2019, Gomez2020a,Villar2020,Boone2021,Sanchez-Saez2021, Gagliano2023, Gomez2023a, Gomez2023b, Kisley2023} and the detailed review and classifier by \citealt{DeSoto2024}). The MOST Hosts survey will be key to provide a training sample of spectroscopically observed transients for such classifiers.

\subsection{Cosmology with SNe~Ia}\label{sec:sne1a}
Today's precision cosmology with SNe~Ia requires measurements with $\sim$1\% uncertainties such that the statistical and systematic uncertainties are on par with each other \citep{2024arXiv240102929D}. Thus uncertainties in the redshift need to be at the level of $\frac{\delta z}{z} \lesssim 0.0046$.  The ZTF SN~Ia sample, composed of several thousand spectroscopically classified SNe~Ia with well-sampled light curves, has the potential to significantly improve over existing low-redshift
SN~Ia samples.
The spectral resolution of the ``SED machine'' spectrograph (SEDm; \citealt{Blagorodnova2018}) at $R\sim100$ is too low to provide 
sufficiently good host redshifts, which prevents one from using all of this sample for cosmology. Most of these ZTF SNe have fairly bright host galaxies, which are already included in the DESI Bright Galaxy Survey (BGS, \citealt{BGS}), which will provide much of the missing redshift information needed to perform a cosmology analysis.

However, fainter SN~Ia hosts will not be in the BGS, and so are covered by MOST Hosts. These are critical for fully exploring residual dependencies on host galaxy stellar mass, star formation and dust that remain after light curve width and color standardization \citep{2006MNRAS.370..773M, 2010ApJ...715..743K, 2010MNRAS.406..782S, 2010ApJ...722..566L, Childress2013b, Rigault2013, 2015ApJ...802...20R, 2017ApJ...848...56U, 2018A&A...615A..68R, Rigault2020, 2020MNRAS.494.4426S,
2021ApJ...909...26B,
Boone2021b,
2021A&A...649A..74N,
2021MNRAS.501.4861K,
2022A&A...657A..22B}. Further, the assignment of a given SN to the image of a given host galaxy can be ambiguous \citep{Gupta2016, 2022AJ....164..195F} -- an ambiguity that usually can be broken using the redshift.

\subsection{Cosmology with SNe~IIps}
While Type~Ia supernovae have dominated the field of standardized candles in the last two decades, the use of SNe~IIp as cosmological probes has significantly increased in recent years \citep{Nugent2017}. Indeed, SNe~IIp explosions are  well-understood to be the result of a single red supergiant star undergoing core collapse, releasing $\sim10^{51}$\,ergs into its surroundings. Our good understanding of these explosions -- which is crucial in making them good standard candles -- is partially due to the fact that nearly two score of their progenitors were directly observed before the SNe occurred (e.g. \citealt{Smartt2009,Smartt2015}).  Methods to utilize SNe~IIp as cosmic candles range from detailed comparisons with explosion models to purely photometric approaches \citep{Nugent2017}. These methods offer a completely independent test of the cosmology inferred from SNe~Ia. 
SNe~IIp suffer some systematics that differ compared to SNe~Ia. Just as in the case of SNe~Ia, accurate DESI redshifts for a large sample of these objects will be key to effectively using them as cosmic probes.

\subsection{Peculiar Velocities Using SNe Tracers}

Both SNe~Ia and IIp are standardizable candles, meaning their relative distances can be precisely estimated. Because SNe~Ia provide relative distances accurate to 3 and 6\%, for spectroscopy and light curve standardization respectively \citep{Boone2021b}, they can be used to measure peculiar velocities \citep{Riess1997, Turnbull2012, Feindt2013, Boruah2020, Stahl2021}. 
Peculiar velocities trace the velocity field of matter, and so provide a direct probe of the gravitational field. Though there are far fewer peculiar velocity galaxies than the general population of galaxies, the cosmological measurements of the growth of structure from velocity data can be competitive with or exceed those from redshift-space distortions (at low redshift), since they are tracers of the velocity (and not density) field. In this way, peculiar velocities can provide a test of dark energy and modified gravity through the growth-of-structure parameter $\gamma$ that is different, but complementary, to those from the Hubble diagram \citep{Howlett2017,Kim2020}.

 ZTF-based distances combined with DESI redshifts provide a measure of per-object peculiar velocity. 
 The ZTF SNe Ia sample can be considered volume limited for $z < 0.06$ for normal Type Ia Supernovae (Rigault et al. in preparation, 2024) and realistic simulations of the ZTF survey show that this will provide a competitive and complementary measurement of the growth rate of structure $f\sigma_{8}$ at $z<0.1$ \citep{Carreres2023}.\footnote{$f \equiv {\rm d \ln}D(a)/{\rm d\ln }a$, which is related to the matter density $\Omega_m$ in 
standard Einstein gravity through the relation $f = \Omega_m^\gamma(z)$. $D$ is the 
linear growth factor and $a$ is the cosmic scale factor which is a function of time. The constant 
$\sigma_8$ is the root mean square of the mass fluctuation in spheres of 8 $h^{-1}$ Mpc.} Augmenting this with DESI redshifts could further enhance the size and quality of this sample. 

A subset of BGS galaxies will be used for Fundamental Plane (FP; \citealt{Dressler1987, Djorgovski1987}) and Tully-Fisher (TF; \citealt{Tully1977}) distance measurements. The large number of data points of these galaxy samples make the DESI Peculiar Velocity Survey \citep{Saulder2023} an important contributor to studies of peculiar motion and low redshift cosmology. Targeting spectra for DESI FP and TF galaxies that also happen to host a SN~Ia will help to provide a more precise calibration of the zero points of both these relations. The overlap between currently known SN~Ia distances and FP/TF galaxies of $\mathcal{O}(200)$ has limited number of data points \citep{Tully2023}. Comparing the overlap between the MOST Hosts target galaxies and the target selections of the DESI PV survey, we find that, after accounting for the probability for obtaining a successful PV from either the FP or TF method with DESI, there will be $\sim1500$ galaxies that have PV from either of these methods \textit{and} that host an SNe Ia. This is a factor of more than five greater than currently available, and studies using the DESI Early Data Release \citep{EDR} have demonstrated that the uncertainties in the zero-points of the TF and FP relations are the largest single contribution to the overall uncertainty in the individual galaxy distances and Hubble constant measurements (Said et. al., 2024; Douglass et. al., in prep.). These uncertainties are currently statistical, but will likely also be systematic in future, larger, data releases. The large number of accurate overlapping distances provided by MOST Hosts is hence expected to significantly improve the subsequent cosmological measurements available from the full DESI peculiar velocity sample\footnote{The MOST Hosts survey provides, through the procedure described in Figure~\ref{fig:decision}, the knowledge of which PV targets are likely to host SNe. In addition, being part of the MOST Hosts sample will upgrade the observation from BRIGHT to DARK, giving better data quality and redshift success rate for targets overlapping with the DESI PV survey.}.


%

\subsection{Nuclear transients}
ZTF has discovered thousands of nuclear transients in its first two years. These are comprised of Quasi-Stellar Object (QSOs), Active Galactic Nuclei (AGNs), and Tidal Disruption Events (TDEs), including the new class of ``changing look AGNs'' \citep{Frederick2019}. Spectroscopic classification of these classes of objects will be a key aid to the photometry-only classifications mentioned in section~\ref{sec:SNe} and play an important role to understand LSST's discoveries.

\section{Target Selection}\label{sec:moments}

\subsection{Transients}\label{sec:transients}

\begin{figure*}
\includegraphics[width=8cm]{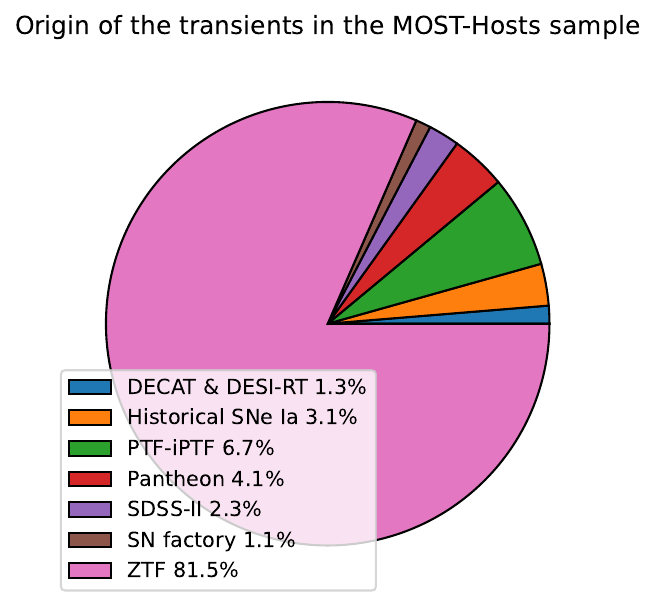}
\includegraphics[width=8cm]{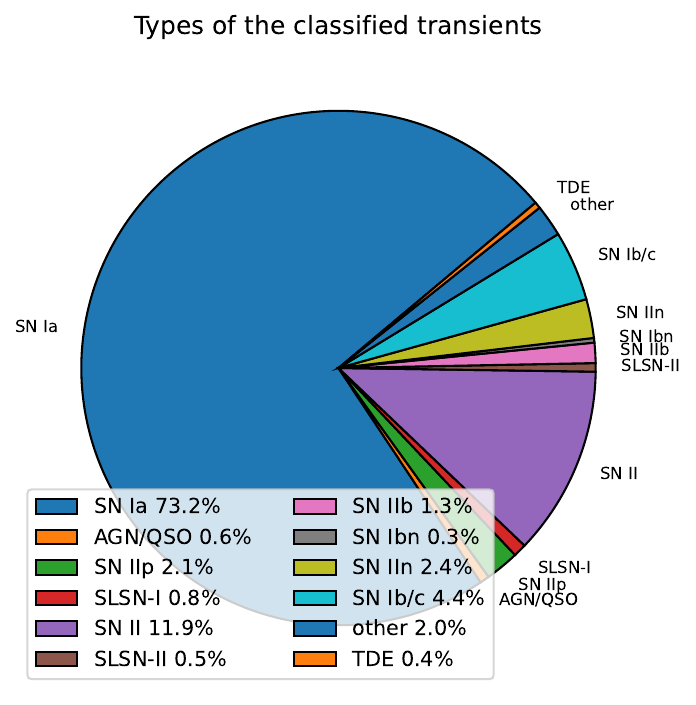}
\caption{Description of the list of transients. The left panel shows the fraction of transients originating from each of the surveys used. The right panel shows the types of the transients for which a classification is available ($26.7$\% of all transients in our list).}

\label{fig:pies}
\end{figure*}

We compiled a list of $38,603$ transients from several surveys: SDSS-II, PTF-iPTF, ZTF, DECAT-DDF, DESIRT. Below, we detail the methods we used to select those transients. 
Figure~\ref{fig:pies} shows the fraction of transients from each of these surveys, as well as the types of the transients that had a classification available (only $26.7$\%). The vast majority of our transients were detected by ZTF (section~\ref{sec:ztf}), followed by PTF-iPTF (section~\ref{sec:ptf}) and SDSS-II (section~\ref{sec:sdssii}). The majority of classified SNe are SNe~Ia. Table~\ref{tab:origin_transients} summarizes the number of transients from each of these surveys.

\begin{deluxetable*}{lccc}
\tablecaption{{\bf Origin of the transients in the MOST Hosts survey}}
\label{tab:origin_transients}
\tablecolumns{4}
\tablewidth{0pt}
\tablehead{\colhead{Survey}&\colhead{Number of transients in the final survey}&\colhead{Classified transient}&\colhead{Unclassified transients}}
\startdata
ZTF               & $31,478$ ($81.5\%$) & $5,343$ ($13.8\%$) & $26,135$ ($67.7\%$) \\
PTF-iPTF            & $2,577$ ($6.7\%$) & $2,577$ ($6.7\%$)& $0$ ($0.0\%$)\\
Pantheon             & $1,567$ ($4.1\%$) & $0$ ($0.0\%$)& $1,567$ ($4.1\%$)\\
Historical SNe Ia   & $1,179$ ($3.1\%$) & $1,109$ ($2.9\%$)& $70$ ($0.2\%$)\\
SDSS-II              & $874$ ($2.3\%$)&$874$ ($2.3\%$)& $0$ ($0.0\%$)\\
DECAT \& DESI-RT     & $503$ ($1.3\%$)&$0$ ($0.0\%)$& $503$ ($1.3\%$)\\
SNfactory           & $425$ ($1.1\%$)&$408$($1.0\%$)& $17$ ($0.1\%)$\\
\hline
Total & $38,603$ ($100.0\%$) & $10,311$ ($26.7\%$)& $28,292$ ($73.3\%$)
\enddata
\end{deluxetable*}

\subsubsection{ZTF targets}\label{sec:ztf}
We selected candidates using data from ZTF observations spanning 2018--2021.
ZTF images are processed and reference-subtracted
by the IPAC ZTF pipeline \citep{Masci2019}.
Every 5$\sigma$ point-source detection is saved as an ``alert.''
Alerts are distributed in Avro format \citep{Patterson2019}.

We used \texttt{ztfquery} \citep{Rigault2018} to identify fields in the primary grid with $E(B-V)<0.3\,$mag at the central field coordinate.
For each field-night, we searched the alert database \texttt{Kowalski} to identify candidates using the following steps:

\begin{enumerate}
    \item We applied basic cuts to remove artefacts and stellar phenomena. We kept sources with a real-bogus score $\texttt{rb}>0.5$ \citep{Mahabal2019}
    and a deep learning score $\texttt{braai}>0.8$ \citep{Duev2019}.
    The \texttt{braai} score
    corresponds to a false positive rate of 0.7\% and a false negative rate of 3\% \citep{Duev2019}.
    We removed sources within 2$^{\prime\prime}$ of a counterpart with a star-galaxy score greater than 0.76 \citep{Tachibana2018},
    and sources within 15$^{\prime\prime}$ of a bright ($r<15\,$mag) star.
    We removed sources that arose from negative subtractions. This left $\sim4.5\,$M unique sources (ZTF generates half a million alerts per night).
    \item Next, we made cuts on the light curves using public-survey data. We required that each source be detected in at least two alerts. The time from the first to last alert had to be greater than 0.5d (so, spanning at least two nights). The time between one pair of alerts had to be less than 12 nights (12.5\,d). In addition, there had to be alerts in more than one filter. This left $51,711$ unique sources. 
\end{enumerate}

In addition to these selected candidates, to avoid missing any obvious known transient, we added to this list all the transients from ZTF up to October 2020, that have been assigned a classification different than ``star'', ``varstar'', ``unknown'', ``duplicate'', ``rock'', ``Bogus'', ``nova''. A fraction of this compiled list of ZTF transient falls outside of the DESI footprint: Table~\ref{tab:origin_transients} and Figure~\ref{fig:pies} shows the relative amount of the ZTF objects in the overall transient list, as well as the types of the transients for which a classifications is available.
\subsubsection{PTF-iPTF targets}\label{sec:ptf}
We include in our list all the SNe observed by the PTF and iPTF project between 2009 and 2018. \citep{PTF2010ApJ...721..777A,PTF2014MNRAS.438.1391P,PTF2018ApJ...855....2Q,PTF2019A&A...621A..71T,PTF2019MNRAS.486.2308F,PTF2021ApJS..255...29S, PTF2021MNRAS.500.5142F}.
Figure~\ref{fig:pies} shows the relative amount of these objects in the overall transient list.

\subsubsection{SDSS-II targets}\label{sec:sdssii}
We included in our list all the transients from the SDSS-II data realease presented in \cite{Sako2018}.
\subsubsection{DECAT-DDF targets}

The ``DECam Alliance for Transients'' (DECAT) is a loosely-defined collaboration of a number of time-domain DECam programs.  Targets here were selected from the DECAT-DDF (``Deep Drilling Field'') project \citep{Graham2023}.  As part of this project, five extragalactic equatorial fields are targeted and observed approximately every three days.  From the observations through the Spring semester 2022, supernovae were identified first by selecting objects that had multiple detections that passed the survey's real/bogus cut, but that did not have more than a few detections outside a several-month window (so as to reject recurrent sources such as AGN).  This automatic criterion still leaves hundreds of candidates, most of which are not relevant for our program, so the remaining light curves were visually scanned to identify supernovae.  The ones whose light curves clearly showed a transient object were included in the MOST Hosts list. 
Figure~\ref{fig:pies} shows the relative amount of these objects in the overall transient list.

\subsubsection{DESIRT targets}

The DECam Survey of Intermediate Redshift Transients (DESIRT; \citealt{desirt}; PI: Palmese \& Wang), another member of DECAT, is a multi-band time domain program being carried out with DECam since 2021. DESIRT repeatedly observes regions of the sky that DESI is also likely to cover around the same time frame of weeks to months, so that both serendipitous DESI spectroscopic observations and targets of opportunity (in the form of DESI spare fiber allocation) are possible (see section 3.2.2 of \citealt{Myers2023} and section 6.4 of \citealt{Schlafly2023}). The DESIRT observations have a cadence of $\sim 3-4$ days on average, and cover $grz$ observations of $\sim 100$ sq.~deg., reaching a $5\sigma$ magnitude depth of $23-23.2$ on most nights. This depth allows the detection of a range of transients that are significantly fainter than those from ZTF and PTF,  but brighter than most objects from the DDF, matching well the typical redshift range of the BGS targets. As a result $\sim 40\%$ of the DESIRT transients' host galaxies have a redshift from the BGS sample without the need for a secondary target program, such as the MOST Hosts survey. We have added the remaining DESIRT hosts from semesters 2021A and 2022A as targets for the MOST Hosts survey, for a total of $\sim 1,600$ targets (including those observed serendipitously by BGS). Figure~\ref{fig:pies} shows the relative amount of these objects in the overall transient list.

\subsection{Nearby Supernova Factory and ``historical'' SNe}
An additional 1572 transients not already in the BGS or selected above, either from the Nearby Supernova Factory search and spectroscopic follow-up program \citep[SNfactory;][]{Aldering2002}, or major SN~Ia compilations used to measure dark energy (e.g., Union3: \citealt{Rubin2023}, Pantheon+: \citealt{Brout2022,Carr2022})
, and lacking a redshift, were added. The spectral types of some probable SNe or AGN from the SNfactory are unknown because at the time of discovery they had been deemed unlikely to be an early SN~Ia or as having a redshift beyond which discovery selection effects would make them useful for cosmological constraints. Redshifts for these, especially if the redshift places the transient beyond the limits of a volume-limited survey, can be used probabilistically in calculations of search efficiency and volumetric transient rates. Figure~\ref{fig:pies} shows the relative amount of these objects in the overall transient list.

\subsection{Transient hosts}
Once a list of transients was completed, we queried the Legacy Survey data (DR9; \citealt{Dey2019}) to identify the most likely host galaxies of these transients. Without spectroscopic data, matching transients with their host is a delicate task and the identity of the host galaxy of a given transient often remains ambiguous. Previous attempts to perform such a task include the algorithm constructed for the GHOST database, which employs deep postage stamps of the field surrounding each transient \citep{Gagliano2021} and the Directional Light Radius technique \citep{Sullivan2006rates,Gupta2016}. Below, we present the algorithm we used to identify the most likely hosts of our sample of transients.

\subsubsection{Basic Algorithm}

\begin{figure*}
\begin{center}
  \includegraphics[angle=90,height=0.94\textheight]{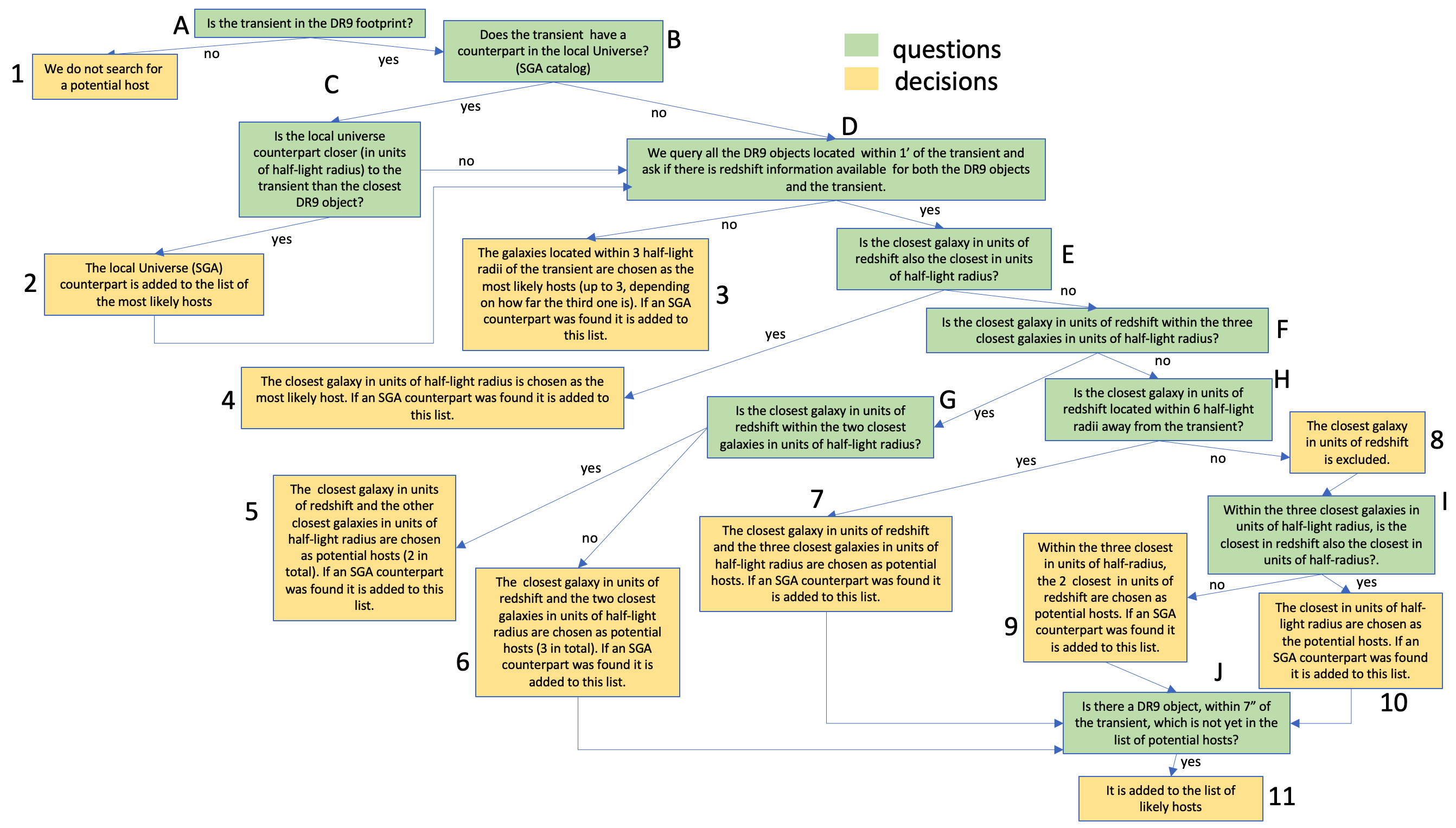}
    \caption{Description of the algorithm used to select one or several potential hosts.}
    \label{fig:decision}
\end{center}
\end{figure*}

Figure~\ref{fig:decision} shows the steps we used in our host-identification algorithm. We start by querying all galaxies in the DR9 data release located within $1'$ from each transient, which turned out to be an effective way to exclude objects outside of the DESI footprint 
(blocks ``A'' and ``1'' of Figure~\ref{fig:decision}). If the query does include one or more objects, we follow the question-decision tree shown in Figure~\ref{fig:decision}. 

The {\tt Tractor} modeling code \citep{Lang2016b,Lang2016} tends to model large galaxies as multiple smaller ones, referred to as ``shreds''. To overcome this issue, an important component of our algorithm consisted of querying the database of the Siena Galaxy Atlas (SGA, \citealt{Moustakas2023}), consisting of $383,620$ nearby galaxies (e.g. block ``B'' of the diagram). However, we always query the DR9 in addition to the SGA database, even when a good SGA candidate has been identified, as this has proven important to avoid missing other likely hosts. Although this step allows us to get rid of many {\tt Tractor} shreds, it does not help overcoming the shredding of galaxies smaller than $\sim 20$\,arcsec, which are not included in the SGA catalog. For those cases, we added a final filtering layer to our algorithm which is detailed in Section\;\ref{sec:shreds}.

The metric used by our algorithm in order to evaluate the proximity of a transient to a potential host on the celestial sphere is the distance in units of light-radii of the host (i.e., the radius at which a given fraction of the total light of a galaxy is emitted). In addition, our algorithm incorporates redshift information by comparing the redshift of the transient -- when available -- and the photometric redshift of the host provided by DR9. However, ``distance'' in units of redshift is never a reason to exclude a potential host. It is only used to add an additional possible likely host to a host that has been identified to be ``close'' in units of light radii. 

Within the list of $57,960$ transients we compiled, 
about one third are outside of the DESI footprint. Among the $38,603$ remaining ones, 
$16\%$ had one of their identified potential hosts in the SGA, and 
$99.99\%$ had an identified potential host in the Legacy survey DR9 release. For 
$55\%$ of these transients, our algorithm selected multiple potential hosts, and the total number of galaxies identified as potential hosts amounts to $73,665$. After removing shreds with the procedure described in Section\,\ref{sec:shreds}, the total number of hosts is $60,212$. Sixty percent of these galaxies already appear in the DESI target list, and $57\%$ are part of the DESI BGS survey.

\subsubsection{Host identification quality assessment}\label{sec:assesment}

In order to assess the quality of our host identification, we compared the most likely hosts provided by our algorithm with the host galaxies identified by overlapping projects.
In particular, we performed a comparison of the hosts identified for a subsample of our transient targets overlapping with the GHOST survey \citep{Gagliano2021} and the Pantheon+ survey \citep{Carr2022}, which both had to perform the same host identification task for their own purposes and already published their results.

The Pantheon+ sample is composed of supernovae taken from a diverse array of samples, listed in \cite{Carr2022}, ranging from $z$ of 0.01 to 2.4. Hosts were identified using a combination of methods, and the sample of hosts, redshifts, and peculiar velocities is presented in \cite{Carr2022}. At high redshift, host assignments were determined from the directional light radius computations of the individual surveys and associated galaxy catalogs, while at low redshift visual inspection was used. 
For 1944 SNe from the Pantheon+ sample, we were able to compare the hosts identified to the one identified by \cite{Carr2022}. In $98 \%$ of the cases, we find good agreement between the host identifications, i.e. the host identified by our algorithm -- and in the case of multiple hosts being identified as likely candidates, at least one of the multiple potential hosts is within $2$\,arcsec of the host identified by \cite{Carr2022}. The remaining $2\%$ of cases are well understood, and are due to one of the following reasons: (1) the {\tt Tractor} code models a large galaxy with several smaller ones, leading to a wrong host localization, a well-documented behavior called ``shredding'' which is addressed in section~\ref{sec:shreds}; (2) the Pantheon+ host was identified as a star in DR9 ($0.4\%$); (3) the transient is slightly outside the DR9 footprint, leading to a wrong host localization.



\subsubsection{Additions}
 An additional $237$ targets were added to the host galaxy sample compiled through the above host identification steps, either because no host could be found in Legacy Survey or other deep imaging, or occasionally, because different host galaxy candidates had already been identified through visual inspection. In the former case, fibers were placed at the positions of the transients themselves, in the hope that the host galaxy might be an undetected emission line galaxy. This approach was justified by the work of \cite{Childress2011, Childress2013a}, where emission-line redshifts were obtained with the Keck 10-m telescope for faint dwarf galaxies co-spatial with spectroscopically identified SNe~Ia. 


\subsubsection{Ambiguous host identifications}
\begin{figure*}[htp]
    \centering \includegraphics[width=1\textwidth]{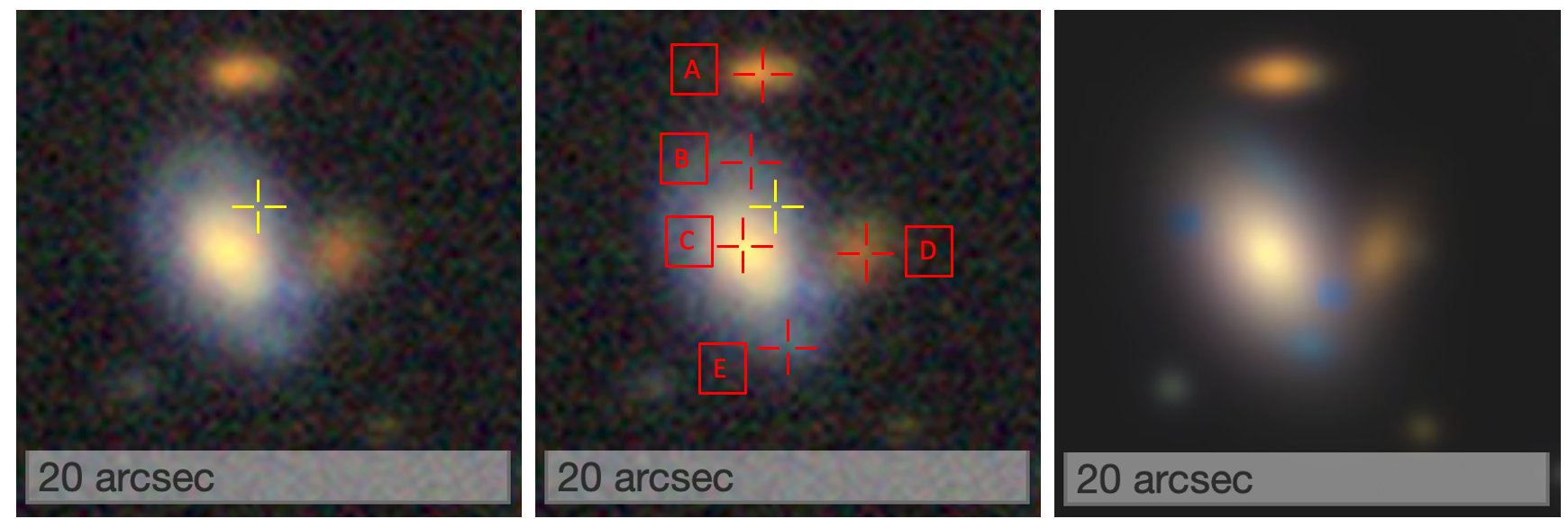}
    \caption{{\bf Example of multiple potential host identifications}. The left panel shows the Legacy survey image, with the location of the transient (yellow) at $\alpha=67.217^\circ$ and $\delta=-9.379^\circ$. The middle panel shows the hosts proposed by our algorithm as likely potential hosts, before filtering out {\tt Tractor} shreds. The hosts annotated as `E' and `B' are {\tt Tractor} shreds: {\tt Tractor} modeled the arm of the galaxy annotated as `C' with multiple smaller galaxies, a well known and documented behavior that is visible in the right panel, where we show the {\tt Tractor} model for this image. `E' and `B' were filtered out by the procedure described in section~\ref{sec:shreds}, leaving us with the hosts `A', `C'  and `D' as the potential hosts.
    The host annotated `A' is probably not the correct host, but the algorithm was tweaked conservatively, resulting in a higher likelihood of proposing wrong hosts rather than missing correct ones. Although `C' seems like the most plausible host, without additional information, such as the redshift, it is hard to exclude `D' as the correct host.}
    \label{fig:ambiguous}
\end{figure*}

Matching a transient with its host, while all we have is imaging information about surrounding galaxies (and maybe some spectroscopic information in the case of the transient) is a delicate question which often has no straightforward answer. To avoid missing the real host, we programmed the algorithm presented in Figure~\ref{fig:decision} to suggest multiple potential hosts when the answer is ambiguous. 
In Figure~\ref{fig:ambiguous}, we show one of the cases where several hosts are suggested by our algorithm as the potential host. The annotated host `A' is probably not the correct host, but the algorithm was tweaked conservatively, resulting in a higher likelihood of proposing wrong hosts rather than missing correct ones. Although `C' seems like the most plausible host, it is difficult to distinguish between galaxies `C' and `D' without additional information, such as the redshift. This tweaking was fine-tuned using comparison with other samples such as the Pantheon+ sample, as detailed in Section~\ref{sec:assesment}. The figure also shows that sometimes, the reason why our algorithm proposes multiple hosts is due to large galaxies being ``shredded'' by {\tt Tractor}; we describe how we mitigate this in the next section.

\subsection{Removing shredding of small galaxies}\label{sec:shreds}
The {\tt Tractor} code tends to model large galaxies with multiple smaller ones. This well-documented behavior is called ``shredding'' and is expected to improve in future releases of the Legacy Survey data. 
Our algorythm partially overcomes shredding by querying the SGA catalog of nearby galaxies \citep{Moustakas2023}, however, galaxies smaller than $\sim 20$\,arcsec, which are not included in the SGA catalog, are still subject to shredding. To filter out shreds of small galaxies, we removed galaxies fulfilling the following criteria: 
\begin{itemize} 
\item The galaxy was selected as the potential host of a transient for which three or more galaxies were suggested. This is motivated by the fact that shredding usually results in multiple fake hosts. 
\item The {\tt fracflux} parameters of the galaxy (the profile-weighted fraction of the flux from other sources divided by the total flux in a given band) in both the $g$ and $r$ band is higher than $1$. This is a somewhat heuristic criterion which was confirmed (through visual inspections) to be very effective to remove shreds. Its effectiveness is due to the fact that shreds are usually modeled as small, faint galaxies near larger, brighter ones.
\end{itemize}
In Figure~\ref{fig:ambiguous}, the galaxy annotated as `C' is an example of a galaxy too small to be included in the SGA catalog, yet subject to shredding. The galaxies `B' and `E' are {\tt Tractor} shreds which were removed by this procedure.

\section{Results}
\begin{figure*}[htp]
    \centering \includegraphics[width=0.75\textwidth]{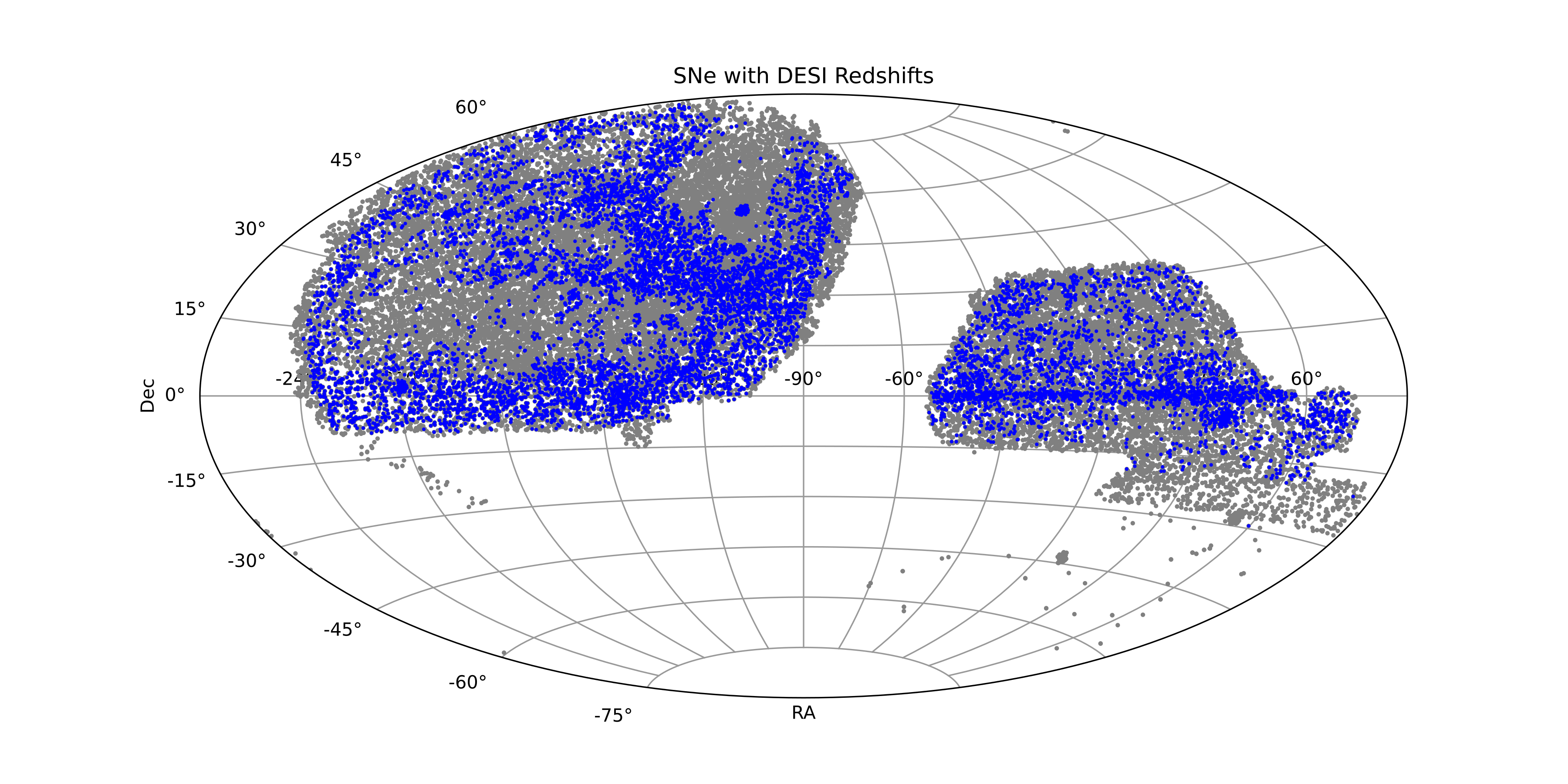}
    \caption{Distribution of MOST Hosts candidates on the sky.  Grey dots represent everything in the MOST Hosts target list.  Blue dots represent candidates for which at least one host has been observed as of MOST Hosts DR1.  (At this scale, it is impossible to separate multiple hosts for the same candidate).}
    \label{fig:skydist}
\end{figure*}

\begin{figure*}[htp]
    \includegraphics[width=0.5\textwidth]{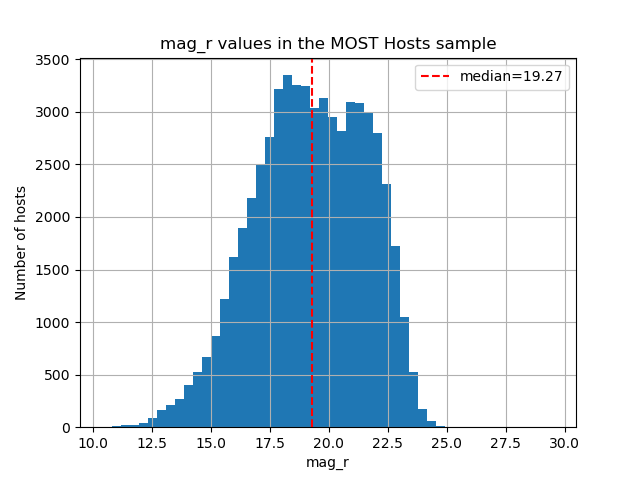}
    \includegraphics[width=0.5\textwidth]{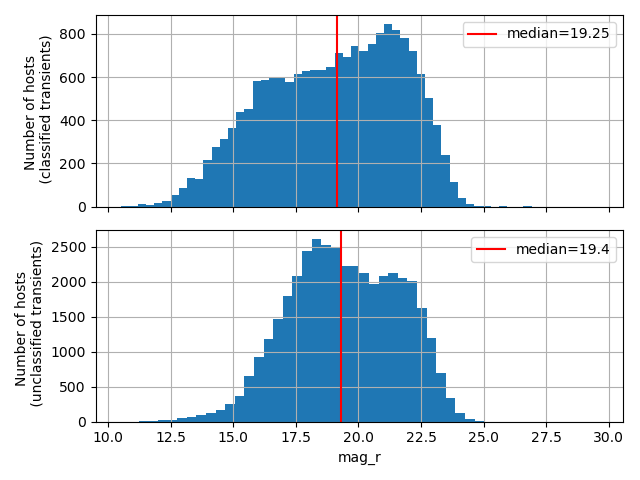}
    \caption{Distribution of magnitudes (dereddened) of the MOST Hosts candidates. The left panel shows the distribution for the entire sample. On the right pannel, we show the distribution for hosts of transients for which a classification exists (top) and for hosts for which no classification is known (bottom).}
    \label{fig:mag_r}
\end{figure*}

\begin{figure*}[htp]
    \centering \includegraphics[width=0.75\textwidth]{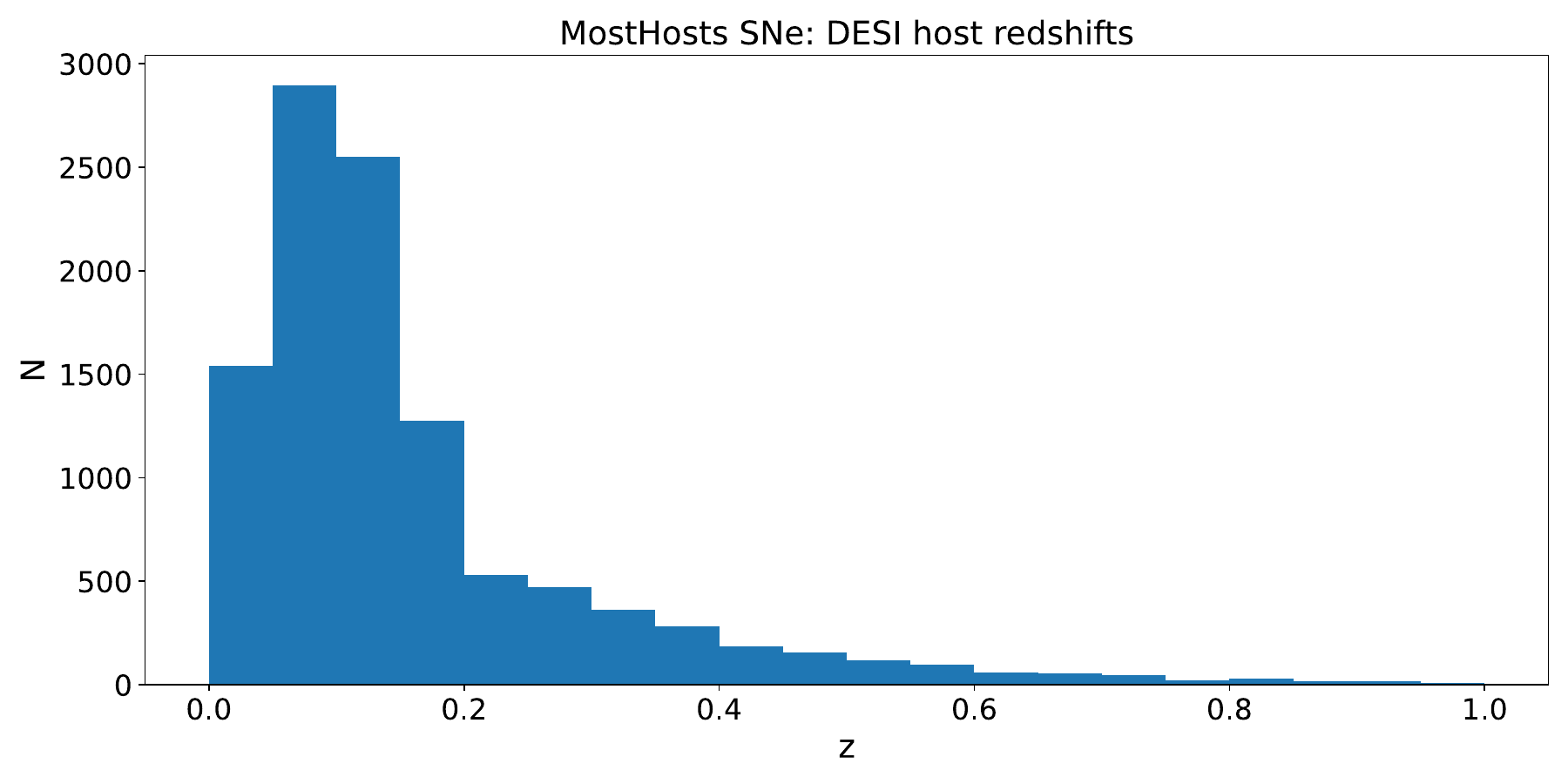}
    \caption{Histogram of host redshifts observed by DESI that will be included in DR1.  Each transient is included once in the histogram; in cases where more than one potential host for a transient has been observed by DESI, one of the observations is chosen randomly for the histogram.}
    \label{fig:zhist}
\end{figure*}
\subsection{Content of the first release}
The first release of the MOST Hosts survey consists of all the hosts observed as of February 14, 2024: a total of $21,931$ hosts which make $36\%$ of the final sample. The figures shown in the paper have been computed with the DESI DR1 release (iron, DESI Collaboration et al. 2024 I in preparation), which contains $11,451$ of these galaxies ($19\%$ of the final sample). The sucess rate in our sample (i.e. the fraction of galaxies which were both observed and have a redshift measurement, tagged with ${\tt zwarn}=0$), is $96\%$. Figure~\ref{fig:skydist} illustrates the distribution of the entire MOST Hosts sample on the sky and the fraction of hosts which have already been observed. Figure~\ref{fig:mag_r} shows the magnitude distribution of the galaxies targeted in the MOST-Host sample. Figure~\ref{fig:zhist} shows the distribution of redshifts of the hosts that have been observed so far.

\subsection{Preliminary Results}

\subsubsection{Hubble Diagram}

\begin{figure*}[htp]
    \centering

    \includegraphics[width=12cm]{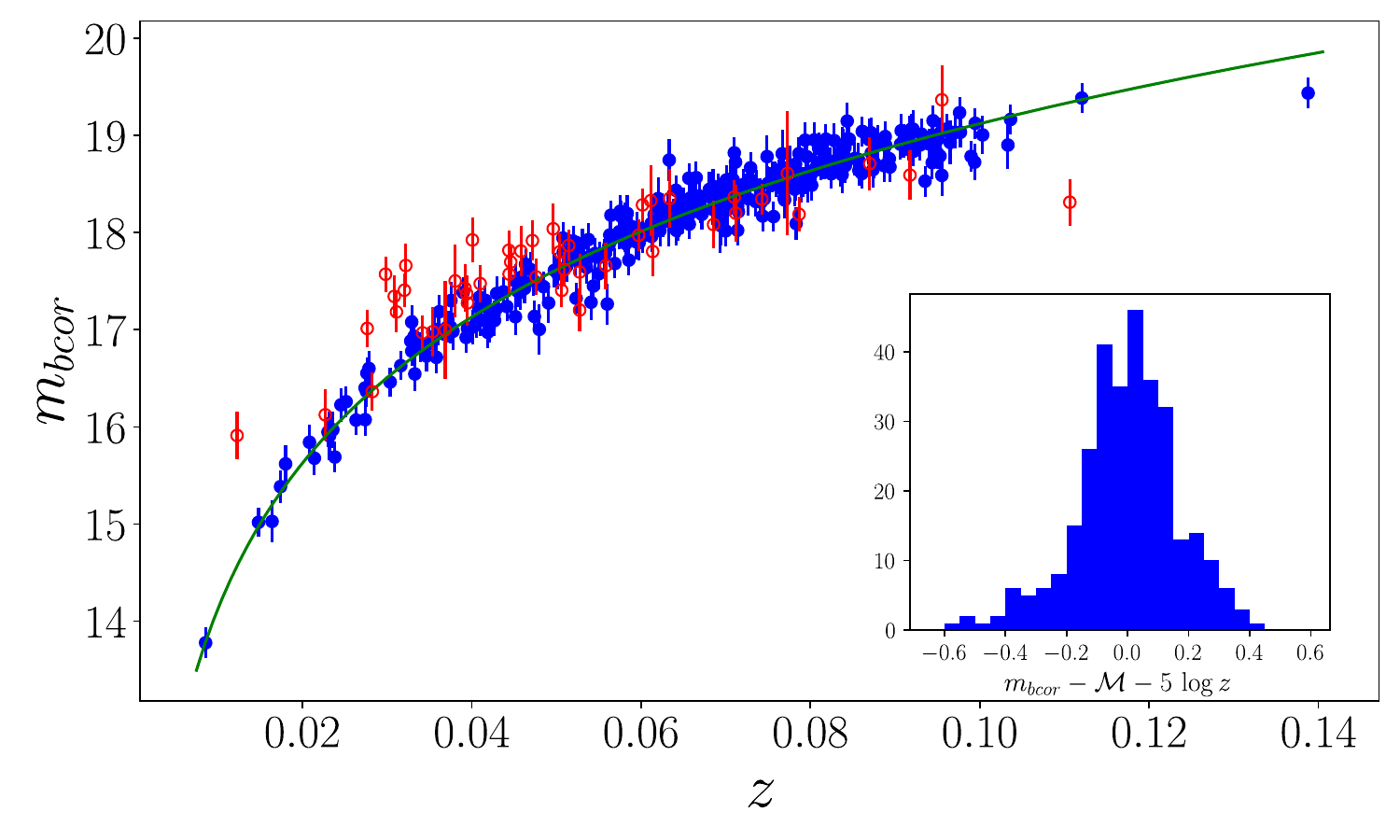}
    \caption{Preliminary Hubble diagram of a subset of the Type~Ia supernova hosts observed so far with the MOST Hosts project.  The plot shows supernovae from the MOST Hosts sample that have hosts observed by DESI, and that are part of the ZTF Bright Transient Survey \citep{Perley2020}.  The redshifts are the DESI redshifts of the host galaxy of the supernova.  Red points are ones that do not pass the cuts $|c|<0.3$, $|x_1|<3$, $\sigma_c<0.2$, $\sigma_{x_1}<1$ \citep{pbb+2021}; these objects were not included in the fit to Equation~\ref{eq:hubfit}.  Error bars on $m_{bcor}$ include an 0.15-magnitude intrinsic dispersion.  Inset: histogram of residuals of the data points that passed the cuts relative to the fit Hubble law.}
    \label{fig:Hubble}
\end{figure*}

\begin{figure*}[htp]
    \centering

    \includegraphics[width=8cm]{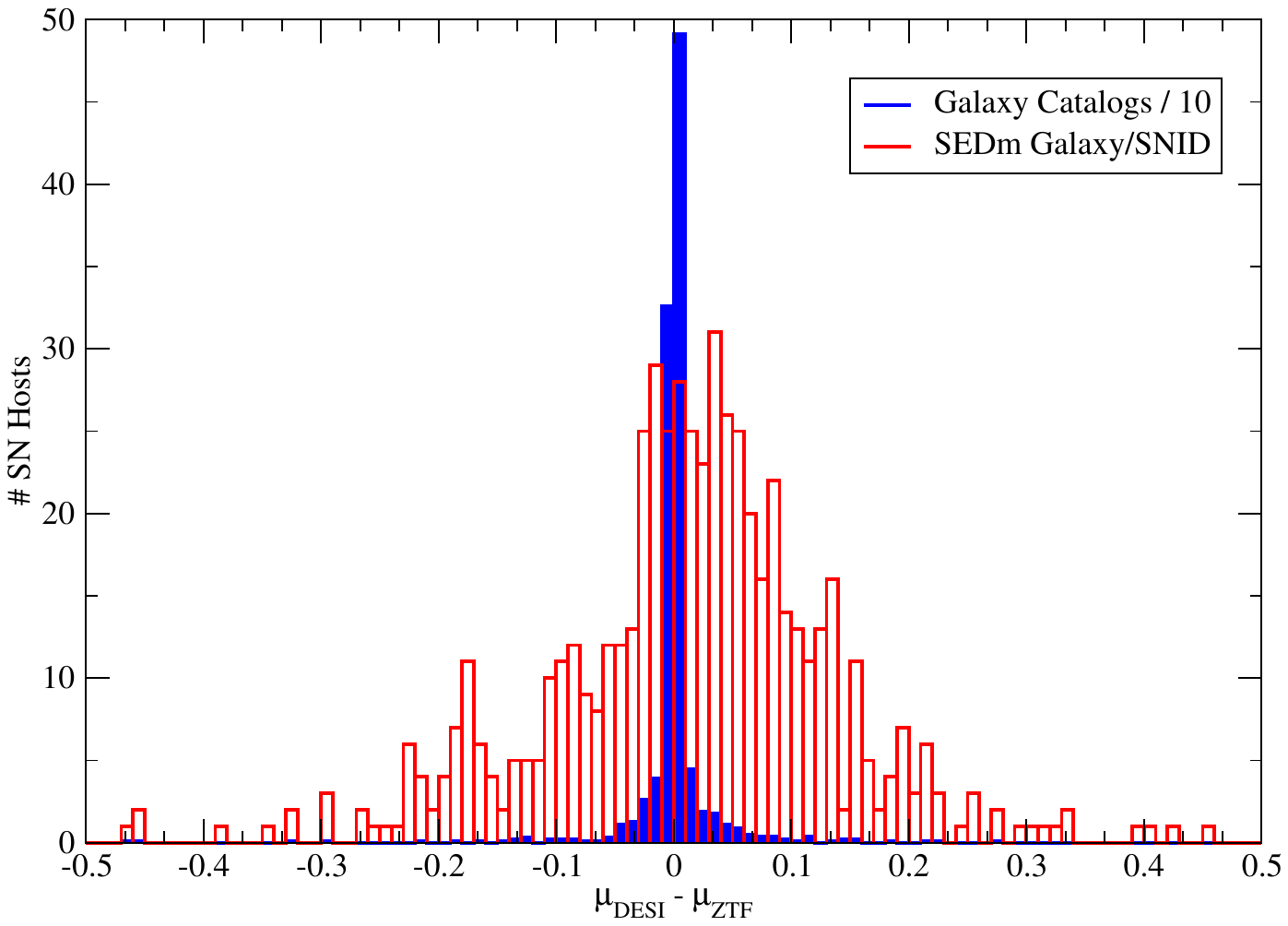}
    \includegraphics[width=8cm]{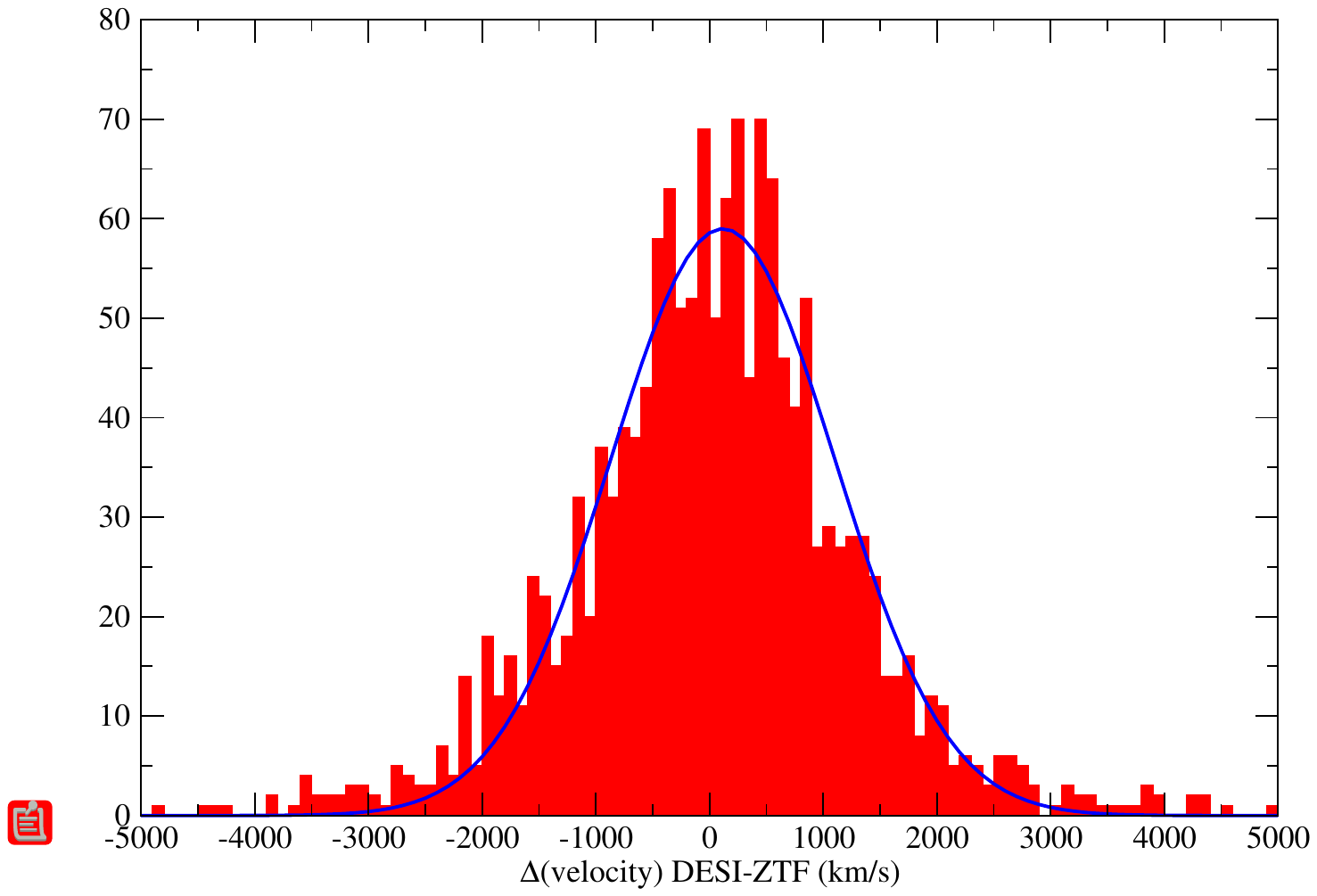}
    \caption{{\bf Left panel:} a histogram of the difference in distance modulus between those obtained by the DESI MOST Hosts project and from ZTF, either from pre-existing galaxay catalogs (blue, totaling 1064 SNe~Ia) or from those obtained by the lower-resolution SEDm (red, totaling 593 SNe~Ia). The increase in the scatter is $\sim0.08$ mag. for those obtained by SEDm. {\bf Right panel:} a histogram of the difference in velocities obtained with the DESI MOST Hosts project and with ZTF using the SEDm, including 1548 with a SNID \citep{Blondin2007} measurement from \cite{Rigault2024}. The gaussian that best fits this histogram corresponds to $1\sigma=986\,\rm km/s$.}
    \label{fig:ztf-desi}
\end{figure*}

Figure~\ref{fig:Hubble} shows a preliminary Hubble diagram of a subset of the Type~Ia supernova hosts observed so far for this project.  This Hubble diagram is preliminary, and intended as a demonstration of what may be done with host reshifts from the MOST Hosts project; we have not performed a careful analysis of systematics, so we do not quote any of the fit parameters.  The plot shows supernovae from the MOST Hosts sample that have hosts observed by DESI and that are part of the ZTF Bright Transient Survey \citep{Perley2020}.  The redshifts are the DESI redshifts of the host galaxy of the supernova. The magnitudes come from 
light curves produced by performing aperture photometry on difference images.  Original images are in the ZTF $g$, $r$, and (where available) $i$ bands, and were downloaded from the public ZTF data server.\footnote{\texttt{https://sites.astro.caltech.edu/ztf/bts/explorer.php}, accessed 2023-11-28} For each band and supernova, we built a reference from 6--11 of the best-seeing public ZTF images available either before the supernova exploded or after it had faded, and performed an Alard/Lupton subtraction \citep{AlardLup1998,BeckerHotpants} of that reference from each of the images containing the supernova.  To each of these light curves we performed a SALT2 fit \citep{Guy2010} using {\tt sncosmo} \citep{SNCosmo}, producing an effective $B$-band magnitude.  (Plotted magnitudes are corrected for light curve width and light curve color using the $\alpha$ and $\beta$ parameters from the fit described below.)  There were 422 BTS SNe~Ia that had at least one candidate host observed in the DESI data release and for which we were able to construct light curve reference images in at least two colors. For 59 of these objects, the host redshift was different enough - by more than 0.01 - from the supernova redshift that it is likely that the observed candidate host is not the right host for the supernova. Indeed, Figure~\ref{fig:ztf-desi} shows that empirically, SEDm and DESI redshifts are consistent within 0.01 in 99\% of the cases. Of those 59 objects,  52 have additional host candidates that have not yet been observed by DESI.  Another 3 objects were omitted because of bad subtractions (due to edge proximity or a bright host that always subtracted poorly), and another 3 were omitted with visually poor SALT2 fits (which upon review of their light curves and spectra were not in fact SNe~Ia), leaving the 357 supernovae with DESI-measured host redshifts shown in Figure~\ref{fig:Hubble}.

The fit in the plot comes from a MCMC model fit to the \citet{Tripp1998} relation:
\begin{equation}
    m_{b*}\ =\ \mathcal{M}\,+\,5\log{z}\,-\,\alpha\,x_1\,+\,\beta\,c
    \label{eq:hubfit}
\end{equation}
where $z$ is the host redshift, $m_{b*}$, $x_1$, and $c$ are from the SALT2 fits. The three fit parameters are $\mathcal{M}$ and the nuisance parameters $\alpha$ (for light curve width) and $\beta$ (for color).  $\mathcal{M}$ is related to the absolute $B$-magnitude $M$ of a SN~Ia by $\mathcal{M}=M+5\log(c\,/\,H_0\,d_0)$, $d_0$ being the standard 10~pc used for absolute magnitudes.  We added an 0.15-magnitude intrinsic dispersion to $m_{b*}$ so that the reduced $\chi^2$ would be equal to 1. 

 As another example of the impact of the MOST Hosts survey on supernova astrophysics and cosmology, we turn our attention to the upcoming release of the Zwicky Transient Facility DR2 data set of over 3600 confirmed SNe~Ia having a spectrum and ZTF lightcurve \citep{Rigault2024}. Of the $2,200$ SNe~Ia in their sample which have a galaxy catalog redshift, over 70\% come from the MOST Hosts project. An example of the impact of MOST Hosts on future ZTF SN~Ia cosmology measurements can be seen in Figure~\ref{fig:ztf-desi}. Here we have made a histogram of the difference in distance modulus between DESI and ZTF where the latter have been taken from other galaxy catalogs (e.g. NED,  SDSS, etc.) or from the lower resolution (R$\sim$100) measurements from the SEDm \citep{2018PASP..130c5003B}. While the effects are minimal when compared to previous high-resolution observations, the scatter in a potential Hubble diagram from those SNe~Ia and their hosts observed only by SEDm  worsens by 0.08 magnitudes for the nearly 600 SNe~Ia which fell into this category.

\subsubsection{Transient Luminosity-Duration Diagram}

\begin{figure}[htp]
    \centering
    \includegraphics[width=8cm]{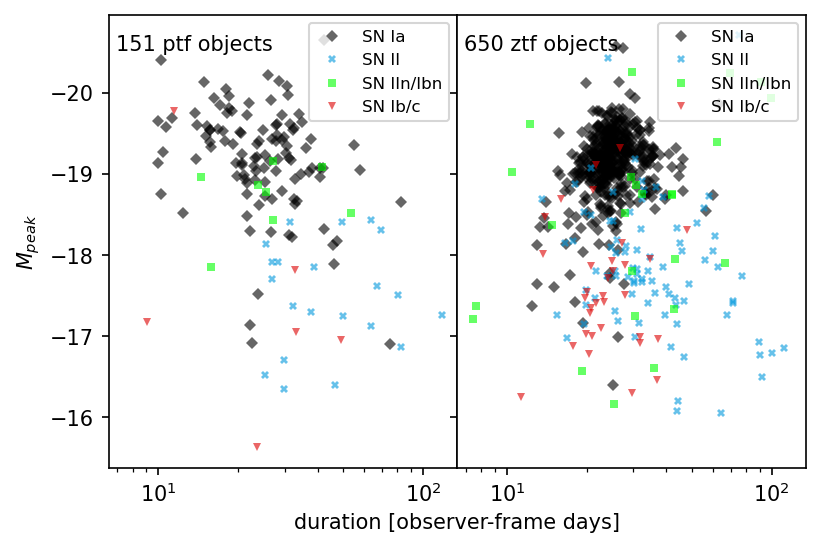}
    \caption{Peak absolute magnitude, calculated from DESI redshift measurements, plotted against duration for two transient datasets, from PTF/iPTF and ZTF. The duration calculation differed slightly between the two sets, as described in the text.}
    \label{fig:LuminosityDurationPlot}
\end{figure}

The vast majority of transients explode in galaxies with unknown redshift, and luminosity is important for photometric classification of these transients.
Figure~\ref{fig:LuminosityDurationPlot} shows the luminosity and duration of the PTF/iPTF and ZTF transients, based on the DESI redshift of the host galaxy and photometry of the transient emission. First we obtained a list of all MOST Hosts redshifts as of May 29, 2023. For the ZTF transients, we used light curve measurements from \citet{Perley2020} of duration and peak apparent magnitude. Using the DESI redshift data we converted from apparent to absolute magnitude following a Flat $\Lambda$CDM cosmological model\footnote{See Table 4 of \cite{Planck2016} (TT, TE, EE + lowP + lensing + ext)} \citep{Planck2016}. 

For the PTF/iPTF transients, we obtained the PTFIDE \citep{Cao2016} $r$-band light curve for the objects and applied a series of quality cuts to the curves based on \citet{Perley2020}. The cuts we applied to the PTF/iPTF transients are:

\begin{enumerate}
    \item The transient must have been observed at least once significantly prior to peak light, (here ``peak'' is  defined as the brightest measured point, i.e. the point with the highest measured flux). We required at least one observation between 7.5 and 16.5 days prior to the brightest recorded detection.
    The transient need not have been detected in this pre-peak observation, but we required that the observation be deep enough to be constraining (limiting magnitude of the observation $m_{\text{lim}} > 19$ mag).
    \item In addition to the measurement at time of peak light, we required at least one more observation near this time. Specifically, the transient must have an additional observation either 2.5 to 7.5 days before \emph{or} 2.5 to 7.5 days after this measurement, also with $m_{\text{lim}} > 19$ mag.
    \item The transient must have been observed at least once significantly after peak light. We required an observation between 7.5 and 16.5 days after the maximum. This observation must also have a limiting magnitude $m_{\text{lim}} > 19$ mag.
\end{enumerate}
 
The remaining light curves were then roughly analyzed by fitting a Gaussian to the data in flux space, allowing us to determine an approximate duration and peak apparent magnitude from the FWHM and peak of the fit. We then used the DESI redshifts to convert to absolute magnitude. Although the two data sets used are not new, DESI redshift data allows for SNe luminosity comparisons at a much higher accuracy than was previously accessible.

The process of analyzing the PTF/iPTF light curves was as follows. After removing the transients which did not pass the quality cuts, a Gaussian was fit in flux space using \texttt{scipy.optimize.curve\_fit} with unconstrained bounds and using the ``lm'' method. An issue we encountered was that the PTF/iPTF light curve data had a high occurrence of non-detections which coincided with periods of positive detections, which caused issues with the Gaussian fitting model and resulted in fits that, upon visual inspection, poorly represented the transient light curve. As an approximate solution, we visually identified the transients with this issue and dropped all non-detections between the first and last positive detection, which improved the fit substantially. A more accurate analysis of the PTF/iPTF data set is certainly possible, using a better light curve approximation than a simple Gaussian and a more complex treatment of the non-detection issue, but this approach yielded a sufficiently good approximation to begin investigating the research potential of this method.

In order to increase the accuracy of our data, we conducted a manual review of the DESI-generated redshift values. To confirm the redshift of the object, we visually examined the DESI spectrum of each host using Skyportal \citep{Skyportal}. 
Galaxy emission and absorption lines were plotted in accordance with the proposed redshift. If the plotted lines matched with 2 or more absorption or emission peaks, such as the emission lines like the [O\,II]\,3727\AA\ doublet, [O\,III]\,4959/5007\AA, and Balmer lines, or absorption lines like Ca~II (H, K, NIR triplet), Magnesium (Mg) and Sodium (Na), the redshift was deemed reliable. If there was too much noise, no spectra existed for said object, or none of the expected spectral lines matched up with any peaks, the redshift was not reliable and we estimated an updated redshift when possible. 
Of the 183 PTF objects which survived the cuts, 169 were determined to have reliable redshifts, and 151 were determined to additionally have reasonable gaussian fit approximations. Of the 682 ZTF objects, 650 were determined to have reliable redshifts. These numbers are consistent with the success rate reported for the BGS sample in \cite{EDR}, within $1\sigma$ error bars computed using the method by \cite{Cameron2011}. 


\section{Conclusion}

The MOST Hosts DESI survey will provide over $60{,}000$ spectra of potential hosts for approximately $40{,}000$ transients detected with most of the public, untargeted, wide-field, optical transient surveys to date.  In doing so, the survey will provide redshifts (and luminosities) for nearly $30,000$ unclassified transients.
The first release of the MOST Hosts survey consists of all spectra that have been observed as of February 14 2024: a total of $21,931$ hosts of $20,235$ transients have been observed, which is $36\%$ of the final hosts sample. 

The figures shown in this paper are based on DESI DR1 (iron), containing about $20\%$ of the final MOST Hosts spectra ($11,451$) and representing at least one of the potential hosts of around $30\%$ of the transients in the sample ($10,834$). 

In an upcoming paper \citep{Rigault2024}, ZTF will present their first Hubble diagram analysis with over $3,600$ SNe~Ia. For these SNe~Ia, the MOST Hosts project has targeted $1,980$ host galaxies within the DESI footprint, of which $1,034$ have already been observed and are part of this release.

 The MOST Hosts survey is highly versatile, supporting a wide range of scientific investigations, including: (1) cosmological measurements through diverse methods; (2) exploring the complex connections between transient properties and their environments; (3) identifying rare outliers in the luminosity-duration space, such as superluminous supernovae and gap transients; and (4) enhancing the performance of photometry-only classifiers as we approach the LSST era, a time when most supernovae will not have spectroscopic classifications. Several proofs-of-concept have been conducted to assess the future potential of this survey. 

\section{Data releases and data availability}

The first release of the MOST Hosts survey will be made accessible through the Wiserep platform\footnote{\href{https://www.wiserep.org}{https://www.wiserep.org}} near the date of the publication of this paper. Updates and dates of future releases, which will be made on a rolling cadence, will be provided on a regular basis through the ``Status \& Schedule'' section of the {\href{https://mosthosts.desi.lbl.gov}{https://mosthosts.desi.lbl.gov}} website. The data releases will include: the transient name, host identifier (with a mention of whether it is one of several potential hosts to be observed for this transient), all the {\tt Tractor} DR9/SGA parameters (which include photometry, uncertainties, best fit galaxy types etc), the measured DESI redshift and a median spectrum. The released spectra are a combination of spectra from DR1 (iron) and spectra from the DESI ``daily'' reduction: spectra that are reduced quickly with a pipeline ran every night and using the latest code and calibrations, thus resulting in a dataset which is not a completely self-consistent over time.
Updates to these spectra will be performed with public releases of the DESI data, to ensure that the spectra in the MOST Hosts releases are completely consistent with those in the data releases. 

All data points used in the figures in this work are available in the following Zenodo repository: {\href{https://doi.org/10.5281/zenodo.10716918}{https://doi.org/10.5281/zenodo.10716918}}

\acknowledgments
We thank Joshua Bloom and Guy Nir for useful discussions.

This material is based upon work supported by the U.S. Department of Energy (DOE), Office of Science, Office of High-Energy Physics, under Contract No. DE–AC02–05CH11231, and by the National Energy Research Scientific Computing Center, a DOE Office of Science User Facility under the same contract. Additional support for DESI was provided by the U.S. National Science Foundation (NSF), Division of Astronomical Sciences under Contract No. AST-0950945 to the NSF’s National Optical-Infrared Astronomy Research Laboratory; the Science and Technology Facilities Council of the United Kingdom; the Gordon and Betty Moore Foundation; the Heising-Simons Foundation; the French Alternative Energies and Atomic Energy Commission (CEA); the National Council of Science and Technology of Mexico (CONACYT); the Ministry of Science and Innovation of Spain (MICINN), and by the DESI Member Institutions: \url{https://www.desi.lbl.gov/collaborating-institutions}. Any opinions, findings, and conclusions or recommendations expressed in this material are those of the author(s) and do not necessarily reflect the views of the U. S. National Science Foundation, the U. S. Department of Energy, or any of the listed funding agencies.

The authors are honored to be permitted to conduct scientific research on Iolkam Du’ag (Kitt Peak), a mountain with particular significance to the Tohono O’odham Nation.

The DESI Legacy Imaging Surveys consist of three individual and complementary projects: the Dark Energy Camera Legacy Survey (DECaLS), the Beijing-Arizona Sky Survey (BASS), and the Mayall z-band Legacy Survey (MzLS). DECaLS, BASS and MzLS together include data obtained, respectively, at the Blanco telescope, Cerro Tololo Inter-American Observatory, NSF’s NOIRLab; the Bok telescope, Steward Observatory, University of Arizona; and the Mayall telescope, Kitt Peak National Observatory, NOIRLab. NOIRLab is operated by the Association of Universities for Research in Astronomy (AURA) under a cooperative agreement with the National Science Foundation. Pipeline processing and analyses of the data were supported by NOIRLab and the Lawrence Berkeley National Laboratory. Legacy Surveys also uses data products from the Near-Earth Object Wide-field Infrared Survey Explorer (NEOWISE), a project of the Jet Propulsion Laboratory/California Institute of Technology, funded by the National Aeronautics and Space Administration. Legacy Surveys was supported by: the Director, Office of Science, Office of High Energy Physics of the U.S. Department of Energy; the National Energy Research Scientific Computing Center, a DOE Office of Science User Facility; the U.S. National Science Foundation, Division of Astronomical Sciences; the National Astronomical Observatories of China, the Chinese Academy of Sciences and the Chinese National Natural Science Foundation. LBNL is managed by the Regents of the University of California under contract to the U.S. Department of Energy. The complete acknowledgments can be found at {\href{https://www.legacysurvey.org/}{https://www.legacysurvey.org/}}.


NOIRLab is operated by the Association of Universities for Research in Astronomy (AURA) under a cooperative agreement with the National Science Foundation. LBNL is managed by the Regents of the University of California under contract to the U.S. Department of Energy.

This project used data obtained with the Dark Energy Camera (DECam), which was constructed by the Dark Energy Survey (DES) collaboration. Funding for the DES Projects has been provided by the U.S. Department of Energy, the U.S. National Science Foundation, the Ministry of Science and Education of Spain, the Science and Technology Facilities Council of the United Kingdom, the Higher Education Funding Council for England, the National Center for Supercomputing Applications at the University of Illinois at Urbana-Champaign, the Kavli Institute of Cosmological Physics at the University of Chicago, Center for Cosmology and Astro-Particle Physics at the Ohio State University, the Mitchell Institute for Fundamental Physics and Astronomy at Texas A\&M University, Financiadora de Estudos e Projetos, Fundacao Carlos Chagas Filho de Amparo, Financiadora de Estudos e Projetos, Fundacao Carlos Chagas Filho de Amparo a Pesquisa do Estado do Rio de Janeiro, Conselho Nacional de Desenvolvimento Cientifico e Tecnologico and the Ministerio da Ciencia, Tecnologia e Inovacao, the Deutsche Forschungsgemeinschaft and the Collaborating Institutions in the Dark Energy Survey. The Collaborating Institutions are Argonne National Laboratory, the University of California at Santa Cruz, the University of Cambridge, Centro de Investigaciones Energeticas, Medioambientales y Tecnologicas-Madrid, the University of Chicago, University College London, the DES-Brazil Consortium, the University of Edinburgh, the Eidgenossische Technische Hochschule (ETH) Zurich, Fermi National Accelerator Laboratory, the University of Illinois at Urbana-Champaign, the Institut de Ciencies de l’Espai (IEEC/CSIC), the Institut de Fisica d’Altes Energies, Lawrence Berkeley National Laboratory, the Ludwig Maximilians Universitat Munchen and the associated Excellence Cluster Universe, the University of Michigan, NSF’s NOIRLab, the University of Nottingham, the Ohio State University, the University of Pennsylvania, the University of Portsmouth, SLAC National Accelerator Laboratory, Stanford University, the University of Sussex, and Texas A\&M University.

Based on observations at Cerro Tololo Inter-American Observatory, NSF’s NOIRLab (NOIRLab Prop. ID 2021A-0148, 2022A-388025,
2022B-297190; PI: A. Palmese \& L. Wang), which is managed by the Association of Universities for Research in Astronomy (AURA) under a cooperative agreement with the National Science Foundation. 

BASS is a key project of the Telescope Access Program (TAP), which has been funded by the National Astronomical Observatories of China, the Chinese Academy of Sciences (the Strategic Priority Research Program “The Emergence of Cosmological Structures” Grant \# XDB09000000), and the Special Fund for Astronomy from the Ministry of Finance. The BASS is also supported by the External Cooperation Program of Chinese Academy of Sciences (Grant \# 114A11KYSB20160057), and Chinese National Natural Science Foundation (Grant \# 12120101003, \# 11433005).

The Legacy Survey team makes use of data products from the Near-Earth Object Wide-field Infrared Survey Explorer (NEOWISE), which is a project of the Jet Propulsion Laboratory/California Institute of Technology. NEOWISE is funded by the National Aeronautics and Space Administration.

The Legacy Surveys imaging of the DESI footprint is supported by the Director, Office of Science, Office of High Energy Physics of the U.S. Department of Energy under Contract No. DE-AC02-05CH1123, by the National Energy Research Scientific Computing Center, a DOE Office of Science User Facility under the same contract; and by the U.S. National Science Foundation, Division of Astronomical Sciences under Contract No. AST-0950945 to NOAO.

The Siena Galaxy Atlas was made possible by funding support from the U.S. Department of Energy, Office of Science, Office of High Energy Physics under Award Number DE-SC0020086 and from the National Science Foundation under grant AST-1616414.

\bibliographystyle{apj} 
\bibliography{bibliograph.bib}

\begin{thebibliography}{}
\expandafter\ifx\csname natexlab\endcsname\relax\def\natexlab#1{#1}\fi

\bibitem[{{Alam} {et~al.}(2017){Alam}, {Ata}, {Bailey}, {Beutler}, {Bizyaev}, {Blazek}, {Bolton}, {Brownstein}, {Burden}, {Chuang}, {Comparat}, {Cuesta}, {Dawson}, {Eisenstein}, {Escoffier}, {Gil-Mar{\'\i}n}, {Grieb}, {Hand}, {Ho}, {Kinemuchi}, {Kirkby}, {Kitaura}, {Malanushenko}, {Malanushenko}, {Maraston}, {McBride}, {Nichol}, {Olmstead}, {Oravetz}, {Padmanabhan}, {Palanque-Delabrouille}, {Pan}, {Pellejero-Ibanez}, {Percival}, {Petitjean}, {Prada}, {Price-Whelan}, {Reid}, {Rodr{\'\i}guez-Torres}, {Roe}, {Ross}, {Ross}, {Rossi}, {Rubi{\~n}o-Mart{\'\i}n}, {Saito}, {Salazar-Albornoz}, {Samushia}, {S{\'a}nchez}, {Satpathy}, {Schlegel}, {Schneider}, {Sc{\'o}ccola}, {Seo}, {Sheldon}, {Simmons}, {Slosar}, {Strauss}, {Swanson}, {Thomas}, {Tinker}, {Tojeiro}, {Maga{\~n}a}, {Vazquez}, {Verde}, {Wake}, {Wang}, {Weinberg}, {White}, {Wood-Vasey}, {Y{\`e}che}, {Zehavi}, {Zhai}, \& {Zhao}}]{Alam2017}
{Alam}, S., {Ata}, M., {Bailey}, S., {et~al.} 2017, \mnras, 470, 2617

\bibitem[{{Alard} \& {Lupton}(1998)}]{AlardLup1998}
{Alard}, C., \& {Lupton}, R.~H. 1998, \apj, 503, 325

\bibitem[{{Aldering} {et~al.}(2002){Aldering}, {Adam}, {Antilogus}, {Astier}, {Bacon}, {Bongard}, {Bonnaud}, {Copin}, {Hardin}, {Henault}, {Howell}, {Lemonnier}, {Levy}, {Loken}, {Nugent}, {Pain}, {Pecontal}, {Pecontal}, {Perlmutter}, {Quimby}, {Schahmaneche}, {Smadja}, \& {Wood-Vasey}}]{Aldering2002}
{Aldering}, G., {Adam}, G., {Antilogus}, P., {et~al.} 2002, in Society of Photo-Optical Instrumentation Engineers (SPIE) Conference Series, Vol. 4836, Survey and Other Telescope Technologies and Discoveries, ed. J.~A. {Tyson} \& S.~{Wolff}, 61--72

\bibitem[{{Arcavi} {et~al.}(2010){Arcavi}, {Gal-Yam}, {Kasliwal}, {Quimby}, {Ofek}, {Kulkarni}, {Nugent}, {Cenko}, {Bloom}, {Sullivan}, {Howell}, {Poznanski}, {Filippenko}, {Law}, {Hook}, {J{\"o}nsson}, {Blake}, {Cooke}, {Dekany}, {Rahmer}, {Hale}, {Smith}, {Zolkower}, {Velur}, {Walters}, {Henning}, {Bui}, {McKenna}, \& {Jacobsen}}]{PTF2010ApJ...721..777A}
{Arcavi}, I., {Gal-Yam}, A., {Kasliwal}, M.~M., {et~al.} 2010, \apj, 721, 777

\bibitem[{Barbary {et~al.}(2023)Barbary, Bailey, Barentsen, Barclay, Biswas, Boone, Craig, Feindt, Friesen, Goldstein, Jha, Jones, Mondon, Papadogiannakis, Perrefort, Pierel, Rodney, Rose, Saunders, Sipőcz, Sofiatti, Thomas, van Santen, Vincenzi, Wang, \& Wood-Vasey}]{SNCosmo}
Barbary, K., Bailey, S., Barentsen, G., {et~al.} 2023, SNCosmo

\bibitem[{{Becker}(2015)}]{BeckerHotpants}
{Becker}, A. 2015, {HOTPANTS: High Order Transform of PSF ANd Template Subtraction}, Astrophysics Source Code Library, record ascl:1504.004, ascl:1504.004

\bibitem[{{Bellm} {et~al.}(2019){Bellm}, {Kulkarni}, {Graham}, {Dekany}, {Smith}, {Riddle}, {Masci}, {Helou}, {Prince}, {Adams}, {Barbarino}, {Barlow}, {Bauer}, {Beck}, {Belicki}, {Biswas}, {Blagorodnova}, {Bodewits}, {Bolin}, {Brinnel}, {Brooke}, {Bue}, {Bulla}, {Burruss}, {Cenko}, {Chang}, {Connolly}, {Coughlin}, {Cromer}, {Cunningham}, {De}, {Delacroix}, {Desai}, {Duev}, {Eadie}, {Farnham}, {Feeney}, {Feindt}, {Flynn}, {Franckowiak}, {Frederick}, {Fremling}, {Gal-Yam}, {Gezari}, {Giomi}, {Goldstein}, {Golkhou}, {Goobar}, {Groom}, {Hacopians}, {Hale}, {Henning}, {Ho}, {Hover}, {Howell}, {Hung}, {Huppenkothen}, {Imel}, {Ip}, {Ivezi{\'c}}, {Jackson}, {Jones}, {Juric}, {Kasliwal}, {Kaspi}, {Kaye}, {Kelley}, {Kowalski}, {Kramer}, {Kupfer}, {Landry}, {Laher}, {Lee}, {Lin}, {Lin}, {Lunnan}, {Giomi}, {Mahabal}, {Mao}, {Miller}, {Monkewitz}, {Murphy}, {Ngeow}, {Nordin}, {Nugent}, {Ofek}, {Patterson}, {Penprase}, {Porter}, {Rauch}, {Rebbapragada}, {Reiley}, {Rigault}, {Rodriguez}, {van Roestel}, {Rusholme}, {van
  Santen}, {Schulze}, {Shupe}, {Singer}, {Soumagnac}, {Stein}, {Surace}, {Sollerman}, {Szkody}, {Taddia}, {Terek}, {Van Sistine}, {van Velzen}, {Vestrand}, {Walters}, {Ward}, {Ye}, {Yu}, {Yan}, \& {Zolkower}}]{Bellm2019}
{Bellm}, E.~C., {Kulkarni}, S.~R., {Graham}, M.~J., {et~al.} 2019, Publications of the Astronomical Society of the Pacific, 131, 018002

\bibitem[{{Blagorodnova} {et~al.}(2018{\natexlab{a}}){Blagorodnova}, {Neill}, {Walters}, {Kulkarni}, {Fremling}, {Ben-Ami}, {Dekany}, {Fucik}, {Konidaris}, {Nash}, {Ngeow}, {Ofek}, {O' Sullivan}, {Quimby}, {Ritter}, \& {Vyhmeister}}]{Blagorodnova2018}
{Blagorodnova}, N., {Neill}, J.~D., {Walters}, R., {et~al.} 2018{\natexlab{a}}, \pasp, 130, 035003

\bibitem[{{Blagorodnova} {et~al.}(2018{\natexlab{b}}){Blagorodnova}, {Neill}, {Walters}, {Kulkarni}, {Fremling}, {Ben-Ami}, {Dekany}, {Fucik}, {Konidaris}, {Nash}, {Ngeow}, {Ofek}, {O' Sullivan}, {Quimby}, {Ritter}, \& {Vyhmeister}}]{2018PASP..130c5003B}
---. 2018{\natexlab{b}}, \pasp, 130, 035003

\bibitem[{{Blake} {et~al.}(2011){Blake}, {Brough}, {Colless}, {Contreras}, {Couch}, {Croom}, {Davis}, {Drinkwater}, {Forster}, {Gilbank}, {Gladders}, {Glazebrook}, {Jelliffe}, {Jurek}, {Li}, {Madore}, {Martin}, {Pimbblet}, {Poole}, {Pracy}, {Sharp}, {Wisnioski}, {Woods}, {Wyder}, \& {Yee}}]{Blake2011}
{Blake}, C., {Brough}, S., {Colless}, M., {et~al.} 2011, \mnras, 415, 2876

\bibitem[{{Blondin} \& {Tonry}(2007)}]{Blondin2007}
{Blondin}, S., \& {Tonry}, J.~L. 2007, \apj, 666, 1024

\bibitem[{{Boone}(2021)}]{Boone2021}
{Boone}, K. 2021, \aj, 162, 275

\bibitem[{{Boone} {et~al.}(2021){Boone}, {Aldering}, {Antilogus}, {Aragon}, {Bailey}, {Baltay}, {Bongard}, {Buton}, {Copin}, {Dixon}, {Fouchez}, {Gangler}, {Gupta}, {Hayden}, {Hillebrandt}, {Kim}, {Kowalski}, {K{\"u}sters}, {L{\'e}get}, {Mondon}, {Nordin}, {Pain}, {Pecontal}, {Pereira}, {Perlmutter}, {Ponder}, {Rabinowitz}, {Rigault}, {Rubin}, {Runge}, {Saunders}, {Smadja}, {Suzuki}, {Tao}, {Taubenberger}, {Thomas}, \& {Vincenzi}}]{Boone2021b}
{Boone}, K., {Aldering}, G., {Antilogus}, P., {et~al.} 2021, \apj, 912, 71

\bibitem[{{Boruah} {et~al.}(2020){Boruah}, {Hudson}, \& {Lavaux}}]{Boruah2020}
{Boruah}, S.~S., {Hudson}, M.~J., \& {Lavaux}, G. 2020, \mnras, 498, 2703

\bibitem[{{Briday} {et~al.}(2022){Briday}, {Rigault}, {Graziani}, {Copin}, {Aldering}, {Amenouche}, {Brinnel}, {Kim}, {Kim}, {Lezmy}, {Nicolas}, {Nordin}, {Perlmutter}, {Rosnet}, \& {Smith}}]{2022A&A...657A..22B}
{Briday}, M., {Rigault}, M., {Graziani}, R., {et~al.} 2022, \aap, 657, A22

\bibitem[{{Brout} \& {Scolnic}(2021)}]{2021ApJ...909...26B}
{Brout}, D., \& {Scolnic}, D. 2021, \apj, 909, 26

\bibitem[{{Brout} {et~al.}(2022){Brout}, {Scolnic}, {Popovic}, {Riess}, {Carr}, {Zuntz}, {Kessler}, {Davis}, {Hinton}, {Jones}, {Kenworthy}, {Peterson}, {Said}, {Taylor}, {Ali}, {Armstrong}, {Charvu}, {Dwomoh}, {Meldorf}, {Palmese}, {Qu}, {Rose}, {Sanchez}, {Stubbs}, {Vincenzi}, {Wood}, {Brown}, {Chen}, {Chambers}, {Coulter}, {Dai}, {Dimitriadis}, {Filippenko}, {Foley}, {Jha}, {Kelsey}, {Kirshner}, {M{\"o}ller}, {Muir}, {Nadathur}, {Pan}, {Rest}, {Rojas-Bravo}, {Sako}, {Siebert}, {Smith}, {Stahl}, \& {Wiseman}}]{Brout2022}
{Brout}, D., {Scolnic}, D., {Popovic}, B., {et~al.} 2022, \apj, 938, 110

\bibitem[{{Cameron}(2011)}]{Cameron2011}
{Cameron}, E. 2011, \pasa, 28, 128

\bibitem[{{Cao} {et~al.}(2016){Cao}, {Nugent}, \& {Kasliwal}}]{Cao2016}
{Cao}, Y., {Nugent}, P.~E., \& {Kasliwal}, M.~M. 2016, \pasp, 128, 114502

\bibitem[{{Carr} {et~al.}(2022){Carr}, {Davis}, {Scolnic}, {Said}, {Brout}, {Peterson}, \& {Kessler}}]{Carr2022}
{Carr}, A., {Davis}, T.~M., {Scolnic}, D., {et~al.} 2022, \pasa, 39, e046

\bibitem[{{Carreres} {et~al.}(2023){Carreres}, {Bautista}, {Feinstein}, {Fouchez}, {Racine}, {Smith}, {Amenouche}, {Aubert}, {Dhawan}, {Ginolin}, {Goobar}, {Gris}, {Lacroix}, {Nuss}, {Regnault}, {Rigault}, {Robert}, {Rosnet}, {Sommer}, {Dekany}, {Groom}, {Sravan}, {Masci}, \& {Purdum}}]{Carreres2023}
{Carreres}, B., {Bautista}, J.~E., {Feinstein}, F., {et~al.} 2023, \aap, 674, A197

\bibitem[{{Childress} {et~al.}(2011){Childress}, {Aldering}, {Aragon}, {Antilogus}, {Bailey}, {Baltay}, {Bongard}, {Buton}, {Canto}, {Chotard}, {Copin}, {Fakhouri}, {Gangler}, {Kerschhaggl}, {Kowalski}, {Hsiao}, {Loken}, {Nugent}, {Paech}, {Pain}, {Pecontal}, {Pereira}, {Perlmutter}, {Rabinowitz}, {Runge}, {Scalzo}, {Thomas}, {Smadja}, {Tao}, {Weaver}, \& {Wu}}]{Childress2011}
{Childress}, M., {Aldering}, G., {Aragon}, C., {et~al.} 2011, \apj, 733, 3

\bibitem[{{Childress} {et~al.}(2013{\natexlab{a}}){Childress}, {Aldering}, {Antilogus}, {Aragon}, {Bailey}, {Baltay}, {Bongard}, {Buton}, {Canto}, {Cellier-Holzem}, {Chotard}, {Copin}, {Fakhouri}, {Gangler}, {Guy}, {Hsiao}, {Kerschhaggl}, {Kim}, {Kowalski}, {Loken}, {Nugent}, {Paech}, {Pain}, {Pecontal}, {Pereira}, {Perlmutter}, {Rabinowitz}, {Rigault}, {Runge}, {Scalzo}, {Smadja}, {Tao}, {Thomas}, {Weaver}, \& {Wu}}]{Childress2013a}
{Childress}, M., {Aldering}, G., {Antilogus}, P., {et~al.} 2013{\natexlab{a}}, \apj, 770, 107

\bibitem[{{Childress} {et~al.}(2013{\natexlab{b}}){Childress}, {Aldering}, {Antilogus}, {Aragon}, {Bailey}, {Baltay}, {Bongard}, {Buton}, {Canto}, {Cellier-Holzem}, {Chotard}, {Copin}, {Fakhouri}, {Gangler}, {Guy}, {Hsiao}, {Kerschhaggl}, {Kim}, {Kowalski}, {Loken}, {Nugent}, {Paech}, {Pain}, {Pecontal}, {Pereira}, {Perlmutter}, {Rabinowitz}, {Rigault}, {Runge}, {Scalzo}, {Smadja}, {Tao}, {Thomas}, {Weaver}, \& {Wu}}]{Childress2013b}
---. 2013{\natexlab{b}}, \apj, 770, 108

\bibitem[{{de Soto} {et~al.}(2024){de Soto}, {Villar}, {Berger}, {Gomez}, {Hosseinzadeh}, {Branton}, {Campos}, {DeLucchi}, {Kubica}, {Lynn}, {Malanchev}, \& {Malz}}]{DeSoto2024}
{de Soto}, K.~M., {Villar}, A., {Berger}, E., {et~al.} 2024, arXiv e-prints, arXiv:2403.07975

\bibitem[{{DES Collaboration} {et~al.}(2024){DES Collaboration}, {Abbott}, {Acevedo}, {Aguena}, {Alarcon}, {Allam}, {Alves}, {Amon}, {Andrade-Oliveira}, {Annis}, {Armstrong}, {Asorey}, {Avila}, {Bacon}, {Bassett}, {Bechtol}, {Bernardinelli}, {Bernstein}, {Bertin}, {Blazek}, {Bocquet}, {Brooks}, {Brout}, {Buckley-Geer}, {Burke}, {Camacho}, {Camilleri}, {Campos}, {Carnero Rosell}, {Carollo}, {Carr}, {Carretero}, {Castander}, {Cawthon}, {Chang}, {Chen}, {Choi}, {Conselice}, {Costanzi}, {da Costa}, {Crocce}, {Davis}, {DePoy}, {Desai}, {Diehl}, {Dixon}, {Dodelson}, {Doel}, {Doux}, {Drlica-Wagner}, {Elvin-Poole}, {Everett}, {Ferrero}, {Fert{\'e}}, {Flaugher}, {Foley}, {Fosalba}, {Friedel}, {Frieman}, {Frohmaier}, {Galbany}, {Garc{\'\i}a-Bellido}, {Gatti}, {Gaztanaga}, {Giannini}, {Glazebrook}, {Graur}, {Gruen}, {Gruendl}, {Gutierrez}, {Hartley}, {Herner}, {Hinton}, {Hollowood}, {Honscheid}, {Huterer}, {Jain}, {James}, {Jeffrey}, {Kelsey}, {Kent}, {Kessler}, {Kim}, {Kirshner}, {Kovacs}, {Kuehn}, {Lahav}, {Lee},
  {Lee}, {Lewis}, {Li}, {Lidman}, {Lin}, {Marshall}, {Martini}, {Mena-Fern{\'a}ndez}, {Menanteau}, {Miquel}, {Mohr}, {Mould}, {Muir}, {M{\"o}ller}, {Neilsen}, {Nichol}, {Nugent}, {Ogando}, {Palmese}, {Pan}, {Paterno}, {Percival}, {Pereira}, {Pieres}, {Plazas Malag{\'o}n}, {Popovic}, {Porredon}, {Prat}, {Qu}, {Raveri}, {Rodr{\'\i}guez-Monroy}, {Romer}, {Roodman}, {Rose}, {Sako}, {Sanchez}, {Sanchez Cid}, {Schubnell}, {Scolnic}, {Sevilla-Noarbe}, {Shah}, {Allyn. Smith}, {Smith}, {Soares-Santos}, {Suchyta}, {Sullivan}, {Suntzeff}, {Swanson}, {S{\'a}nchez}, {Tarle}, {Taylor}, {Thomas}, {To}, {Toy}, {Troxel}, {Tucker}, {Tucker}, {Uddin}, {Vincenzi}, {Walker}, {Weaverdyck}, {Wechsler}, {Weller}, {Wester}, {Wiseman}, {Yamamoto}, {Yuan}, {Zhang}, \& {Zhang}}]{2024arXiv240102929D}
{DES Collaboration}, {Abbott}, T.~M.~C., {Acevedo}, M., {et~al.} 2024, arXiv e-prints, arXiv:2401.02929

\bibitem[{{DESI Collaboration} {et~al.}(2016{\natexlab{a}}){DESI Collaboration}, {Aghamousa}, {Aguilar}, {Ahlen}, {Alam}, {Allen}, {Allende Prieto}, {Annis}, {Bailey}, {Balland}, {Ballester}, {Baltay}, {Beaufore}, {Bebek}, {Beers}, {Bell}, {Bernal}, {Besuner}, {Beutler}, {Blake}, {Bleuler}, {Blomqvist}, {Blum}, {Bolton}, {Briceno}, {Brooks}, {Brownstein}, {Buckley-Geer}, {Burden}, {Burtin}, {Busca}, {Cahn}, {Cai}, {Cardiel-Sas}, {Carlberg}, {Carton}, {Casas}, {Castander}, {Cervantes-Cota}, {Claybaugh}, {Close}, {Coker}, {Cole}, {Comparat}, {Cooper}, {Cousinou}, {Crocce}, {Cuby}, {Cunningham}, {Davis}, {Dawson}, {de la Macorra}, {De Vicente}, {Delubac}, {Derwent}, {Dey}, {Dhungana}, {Ding}, {Doel}, {Duan}, {Ealet}, {Edelstein}, {Eftekharzadeh}, {Eisenstein}, {Elliott}, {Escoffier}, {Evatt}, {Fagrelius}, {Fan}, {Fanning}, {Farahi}, {Farihi}, {Favole}, {Feng}, {Fernandez}, {Findlay}, {Finkbeiner}, {Fitzpatrick}, {Flaugher}, {Flender}, {Font-Ribera}, {Forero-Romero}, {Fosalba}, {Frenk}, {Fumagalli}, {Gaensicke},
  {Gallo}, {Garcia-Bellido}, {Gaztanaga}, {Pietro Gentile Fusillo}, {Gerard}, {Gershkovich}, {Giannantonio}, {Gillet}, {Gonzalez-de-Rivera}, {Gonzalez-Perez}, {Gott}, {Graur}, {Gutierrez}, {Guy}, {Habib}, {Heetderks}, {Heetderks}, {Heitmann}, {Hellwing}, {Herrera}, {Ho}, {Holland}, {Honscheid}, {Huff}, {Hutchinson}, {Huterer}, {Hwang}, {Illa Laguna}, {Ishikawa}, {Jacobs}, {Jeffrey}, {Jelinsky}, {Jennings}, {Jiang}, {Jimenez}, {Johnson}, {Joyce}, {Jullo}, {Juneau}, {Kama}, {Karcher}, {Karkar}, {Kehoe}, {Kennamer}, {Kent}, {Kilbinger}, {Kim}, {Kirkby}, {Kisner}, {Kitanidis}, {Kneib}, {Koposov}, {Kovacs}, {Koyama}, {Kremin}, {Kron}, {Kronig}, {Kueter-Young}, {Lacey}, {Lafever}, {Lahav}, {Lambert}, {Lampton}, {Landriau}, {Lang}, {Lauer}, {Le Goff}, {Le Guillou}, {Le Van Suu}, {Lee}, {Lee}, {Leitner}, {Lesser}, {Levi}, {L'Huillier}, {Li}, {Liang}, {Lin}, {Linder}, {Loebman}, {Luki{\'c}}, {Ma}, {MacCrann}, {Magneville}, {Makarem}, {Manera}, {Manser}, {Marshall}, {Martini}, {Massey}, {Matheson}, {McCauley},
  {McDonald}, {McGreer}, {Meisner}, {Metcalfe}, {Miller}, {Miquel}, {Moustakas}, {Myers}, {Naik}, {Newman}, {Nichol}, {Nicola}, {Nicolati da Costa}, {Nie}, {Niz}, {Norberg}, {Nord}, {Norman}, {Nugent}, {O'Brien}, {Oh}, {Olsen}, {Padilla}, {Padmanabhan}, {Padmanabhan}, {Palanque-Delabrouille}, {Palmese}, {Pappalardo}, {P{\^a}ris}, {Park}, {Patej}, {Peacock}, {Peiris}, {Peng}, {Percival}, {Perruchot}, {Pieri}, {Pogge}, {Pollack}, {Poppett}, {Prada}, {Prakash}, {Probst}, {Rabinowitz}, {Raichoor}, {Ree}, {Refregier}, {Regal}, {Reid}, {Reil}, {Rezaie}, {Rockosi}, {Roe}, {Ronayette}, {Roodman}, {Ross}, {Ross}, {Rossi}, {Rozo}, {Ruhlmann-Kleider}, {Rykoff}, {Sabiu}, {Samushia}, {Sanchez}, {Sanchez}, {Schlegel}, {Schneider}, {Schubnell}, {Secroun}, {Seljak}, {Seo}, {Serrano}, {Shafieloo}, {Shan}, {Sharples}, {Sholl}, {Shourt}, {Silber}, {Silva}, {Sirk}, {Slosar}, {Smith}, {Smoot}, {Som}, {Song}, {Sprayberry}, {Staten}, {Stefanik}, {Tarle}, {Sien Tie}, {Tinker}, {Tojeiro}, {Valdes}, {Valenzuela}, {Valluri},
  {Vargas-Magana}, {Verde}, {Walker}, {Wang}, {Wang}, {Weaver}, {Weaverdyck}, {Wechsler}, {Weinberg}, {White}, {Yang}, {Yeche}, {Zhang}, {Zhao}, {Zheng}, {Zhou}, {Zhou}, {Zhu}, {Zou}, \& {Zu}}]{DESI2016a}
{DESI Collaboration}, {Aghamousa}, A., {Aguilar}, J., {et~al.} 2016{\natexlab{a}}, arXiv e-prints, arXiv:1611.00036

\bibitem[{{DESI Collaboration} {et~al.}(2016{\natexlab{b}}){DESI Collaboration}, {Aghamousa}, {Aguilar}, {Ahlen}, {Alam}, {Allen}, {Allende Prieto}, {Annis}, {Bailey}, {Balland}, {Ballester}, {Baltay}, {Beaufore}, {Bebek}, {Beers}, {Bell}, {Bernal}, {Besuner}, {Beutler}, {Blake}, {Bleuler}, {Blomqvist}, {Blum}, {Bolton}, {Briceno}, {Brooks}, {Brownstein}, {Buckley-Geer}, {Burden}, {Burtin}, {Busca}, {Cahn}, {Cai}, {Cardiel-Sas}, {Carlberg}, {Carton}, {Casas}, {Castander}, {Cervantes-Cota}, {Claybaugh}, {Close}, {Coker}, {Cole}, {Comparat}, {Cooper}, {Cousinou}, {Crocce}, {Cuby}, {Cunningham}, {Davis}, {Dawson}, {de la Macorra}, {De Vicente}, {Delubac}, {Derwent}, {Dey}, {Dhungana}, {Ding}, {Doel}, {Duan}, {Ealet}, {Edelstein}, {Eftekharzadeh}, {Eisenstein}, {Elliott}, {Escoffier}, {Evatt}, {Fagrelius}, {Fan}, {Fanning}, {Farahi}, {Farihi}, {Favole}, {Feng}, {Fernandez}, {Findlay}, {Finkbeiner}, {Fitzpatrick}, {Flaugher}, {Flender}, {Font-Ribera}, {Forero-Romero}, {Fosalba}, {Frenk}, {Fumagalli}, {Gaensicke},
  {Gallo}, {Garcia-Bellido}, {Gaztanaga}, {Pietro Gentile Fusillo}, {Gerard}, {Gershkovich}, {Giannantonio}, {Gillet}, {Gonzalez-de-Rivera}, {Gonzalez-Perez}, {Gott}, {Graur}, {Gutierrez}, {Guy}, {Habib}, {Heetderks}, {Heetderks}, {Heitmann}, {Hellwing}, {Herrera}, {Ho}, {Holland}, {Honscheid}, {Huff}, {Hutchinson}, {Huterer}, {Hwang}, {Illa Laguna}, {Ishikawa}, {Jacobs}, {Jeffrey}, {Jelinsky}, {Jennings}, {Jiang}, {Jimenez}, {Johnson}, {Joyce}, {Jullo}, {Juneau}, {Kama}, {Karcher}, {Karkar}, {Kehoe}, {Kennamer}, {Kent}, {Kilbinger}, {Kim}, {Kirkby}, {Kisner}, {Kitanidis}, {Kneib}, {Koposov}, {Kovacs}, {Koyama}, {Kremin}, {Kron}, {Kronig}, {Kueter-Young}, {Lacey}, {Lafever}, {Lahav}, {Lambert}, {Lampton}, {Landriau}, {Lang}, {Lauer}, {Le Goff}, {Le Guillou}, {Le Van Suu}, {Lee}, {Lee}, {Leitner}, {Lesser}, {Levi}, {L'Huillier}, {Li}, {Liang}, {Lin}, {Linder}, {Loebman}, {Luki{\'c}}, {Ma}, {MacCrann}, {Magneville}, {Makarem}, {Manera}, {Manser}, {Marshall}, {Martini}, {Massey}, {Matheson}, {McCauley},
  {McDonald}, {McGreer}, {Meisner}, {Metcalfe}, {Miller}, {Miquel}, {Moustakas}, {Myers}, {Naik}, {Newman}, {Nichol}, {Nicola}, {Nicolati da Costa}, {Nie}, {Niz}, {Norberg}, {Nord}, {Norman}, {Nugent}, {O'Brien}, {Oh}, {Olsen}, {Padilla}, {Padmanabhan}, {Padmanabhan}, {Palanque-Delabrouille}, {Palmese}, {Pappalardo}, {P{\^a}ris}, {Park}, {Patej}, {Peacock}, {Peiris}, {Peng}, {Percival}, {Perruchot}, {Pieri}, {Pogge}, {Pollack}, {Poppett}, {Prada}, {Prakash}, {Probst}, {Rabinowitz}, {Raichoor}, {Ree}, {Refregier}, {Regal}, {Reid}, {Reil}, {Rezaie}, {Rockosi}, {Roe}, {Ronayette}, {Roodman}, {Ross}, {Ross}, {Rossi}, {Rozo}, {Ruhlmann-Kleider}, {Rykoff}, {Sabiu}, {Samushia}, {Sanchez}, {Sanchez}, {Schlegel}, {Schneider}, {Schubnell}, {Secroun}, {Seljak}, {Seo}, {Serrano}, {Shafieloo}, {Shan}, {Sharples}, {Sholl}, {Shourt}, {Silber}, {Silva}, {Sirk}, {Slosar}, {Smith}, {Smoot}, {Som}, {Song}, {Sprayberry}, {Staten}, {Stefanik}, {Tarle}, {Sien Tie}, {Tinker}, {Tojeiro}, {Valdes}, {Valenzuela}, {Valluri},
  {Vargas-Magana}, {Verde}, {Walker}, {Wang}, {Wang}, {Weaver}, {Weaverdyck}, {Wechsler}, {Weinberg}, {White}, {Yang}, {Yeche}, {Zhang}, {Zhao}, {Zheng}, {Zhou}, {Zhou}, {Zhu}, {Zou}, \& {Zu}}]{DESI2016b}
---. 2016{\natexlab{b}}, arXiv e-prints, arXiv:1611.00037

\bibitem[{{DESI Collaboration} {et~al.}(2022){DESI Collaboration}, {Abareshi}, {Aguilar}, {Ahlen}, {Alam}, {Alexander}, {Alfarsy}, {Allen}, {Allende Prieto}, {Alves}, {Ameel}, {Armengaud}, {Asorey}, {Aviles}, {Bailey}, {Balaguera-Antol{\'\i}nez}, {Ballester}, {Baltay}, {Bault}, {Beltran}, {Benavides}, {BenZvi}, {Berti}, {Besuner}, {Beutler}, {Bianchi}, {Blake}, {Blanc}, {Blum}, {Bolton}, {Bose}, {Bramall}, {Brieden}, {Brodzeller}, {Brooks}, {Brownewell}, {Buckley-Geer}, {Cahn}, {Cai}, {Canning}, {Capasso}, {Carnero Rosell}, {Carton}, {Casas}, {Castander}, {Cervantes-Cota}, {Chabanier}, {Chaussidon}, {Chuang}, {Circosta}, {Cole}, {Cooper}, {da Costa}, {Cousinou}, {Cuceu}, {Davis}, {Dawson}, {de la Cruz-Noriega}, {de la Macorra}, {de Mattia}, {Della Costa}, {Demmer}, {Derwent}, {Dey}, {Dey}, {Dhungana}, {Ding}, {Dobson}, {Doel}, {Donald-McCann}, {Donaldson}, {Douglass}, {Duan}, {Dunlop}, {Edelstein}, {Eftekharzadeh}, {Eisenstein}, {Enriquez-Vargas}, {Escoffier}, {Evatt}, {Fagrelius}, {Fan}, {Fanning},
  {Fawcett}, {Ferraro}, {Ereza}, {Flaugher}, {Font-Ribera}, {Forero-Romero}, {Frenk}, {Fromenteau}, {G{\"a}nsicke}, {Garcia-Quintero}, {Garrison}, {Gazta{\~n}aga}, {Gerardi}, {Gil-Mar{\'\i}n}, {Gontcho a Gontcho}, {Gonzalez-Morales}, {Gonzalez-de-Rivera}, {Gonzalez-Perez}, {Gordon}, {Graur}, {Green}, {Grove}, {Gruen}, {Gutierrez}, {Guy}, {Hahn}, {Harris}, {Herrera}, {Herrera-Alcantar}, {Honscheid}, {Howlett}, {Huterer}, {Ir{\v{s}}i{\v{c}}}, {Ishak}, {Jelinsky}, {Jiang}, {Jimenez}, {Jing}, {Joyce}, {Jullo}, {Juneau}, {Kara{\c{c}}ayl{\i}}, {Karamanis}, {Karcher}, {Karim}, {Kehoe}, {Kent}, {Kirkby}, {Kisner}, {Kitaura}, {Koposov}, {Kov{\'a}cs}, {Kremin}, {Krolewski}, {L'Huillier}, {Lahav}, {Lambert}, {Lamman}, {Lan}, {Landriau}, {Lane}, {Lang}, {Lange}, {Lasker}, {Le Guillou}, {Leauthaud}, {Le Van Suu}, {Levi}, {Li}, {Magneville}, {Manera}, {Manser}, {Marshall}, {Martini}, {McCollam}, {McDonald}, {Meisner}, {Mena-Fern{\'a}ndez}, {Meneses-Rizo}, {Mezcua}, {Miller}, {Miquel}, {Montero-Camacho}, {Moon},
  {Moustakas}, {Mueller}, {Mu{\~n}oz-Guti{\'e}rrez}, {Myers}, {Nadathur}, {Najita}, {Napolitano}, {Neilsen}, {Newman}, {Nie}, {Ning}, {Niz}, {Norberg}, {Noriega}, {O'Brien}, {Obuljen}, {Palanque-Delabrouille}, {Palmese}, {Zhiwei}, {Pappalardo}, {PENG}, {Percival}, {Perruchot}, {Pogge}, {Poppett}, {Porredon}, {Prada}, {Prochaska}, {Pucha}, {P{\'e}rez-Fern{\'a}ndez}, {P{\'e}rez-R{\`a}fols}, {Rabinowitz}, {Raichoor}, {Ramirez-Solano}, {Ram{\'\i}rez-P{\'e}rez}, {Ravoux}, {Reil}, {Rezaie}, {Rocher}, {Rockosi}, {Roe}, {Roodman}, {Ross}, {Rossi}, {Ruggeri}, {Ruhlmann-Kleider}, {Sabiu}, {Safonova}, {Said}, {Saintonge}, {Salas Catonga}, {Samushia}, {Sanchez}, {Saulder}, {Schaan}, {Schlafly}, {Schlegel}, {Schmoll}, {Scholte}, {Schubnell}, {Secroun}, {Seo}, {Serrano}, {Sharples}, {Sholl}, {Silber}, {Silva}, {Sirk}, {Siudek}, {Smith}, {Sprayberry}, {Staten}, {Stupak}, {Tan}, {Tarl{\'e}}, {Tie}, {Tojeiro}, {Ure{\~n}a-L{\'o}pez}, {Valdes}, {Valenzuela}, {Valluri}, {Vargas-Maga{\~n}a}, {Verde}, {Walther}, {Wang}, {Wang},
  {Weaver}, {Weaverdyck}, {Wechsler}, {Wilson}, {Yang}, {Yu}, {Yuan}, {Y{\`e}che}, {Zhang}, {Zhang}, {Zhao}, {Zhou}, {Zhou}, {Zou}, {Zou}, {Zou}, {Zu}, \& {DESI Collaboration}}]{instrument}
{DESI Collaboration}, {Abareshi}, B., {Aguilar}, J., {et~al.} 2022, \aj, 164, 207

\bibitem[{{DESI Collaboration} {et~al.}(2023{\natexlab{a}}){DESI Collaboration}, {Adame}, {Aguilar}, {Ahlen}, {Alam}, {Aldering}, {Alexander}, {Alfarsy}, {Allende Prieto}, {Alvarez}, {Alves}, {Anand}, {Andrade-Oliveira}, {Armengaud}, {Asorey}, {Avila}, {Aviles}, {Bailey}, {Balaguera-Antol{\'\i}nez}, {Ballester}, {Baltay}, {Bault}, {Bautista}, {Behera}, {Beltran}, {BenZvi}, {Beraldo e Silva}, {Bermejo-Climent}, {Berti}, {Besuner}, {Beutler}, {Bianchi}, {Blake}, {Blum}, {Bolton}, {Brieden}, {Brodzeller}, {Brooks}, {Brown}, {Buckley-Geer}, {Burtin}, {Cabayol-Garcia}, {Cai}, {Canning}, {Cardiel-Sas}, {Carnero Rosell}, {Castander}, {Cervantes-Cota}, {Chabanier}, {Chaussidon}, {Chaves-Montero}, {Chen}, {Chuang}, {Claybaugh}, {Cole}, {Cooper}, {Cuceu}, {Davis}, {Dawson}, {de Belsunce}, {de la Cruz}, {de la Macorra}, {de Mattia}, {Demina}, {Demirbozan}, {DeRose}, {Dey}, {Dey}, {Dhungana}, {Ding}, {Ding}, {Doel}, {Doshi}, {Douglass}, {Edge}, {Eftekharzadeh}, {Eisenstein}, {Elliott}, {Escoffier}, {Fagrelius}, {Fan},
  {Fanning}, {Fawcett}, {Ferraro}, {Ereza}, {Flaugher}, {Font-Ribera}, {Forero-S{\'a}nchez}, {Forero-Romero}, {Frenk}, {G{\"a}nsicke}, {Garc{\'\i}a}, {Garc{\'\i}a-Bellido}, {Garcia-Quintero}, {Garrison}, {Gil-Mar{\'\i}n}, {Golden-Marx}, {Gontcho}, {Gonzalez-Morales}, {Gonzalez-Perez}, {Gordon}, {Graur}, {Green}, {Gruen}, {Guy}, {Hadzhiyska}, {Hahn}, {Han}, {Hanif}, {Herrera-Alcantar}, {Honscheid}, {Hou}, {Howlett}, {Huterer}, {Ir{\v{s}}i{\v{c}}}, {Ishak}, {Jacques}, {Jana}, {Jiang}, {Jimenez}, {Jing}, {Joudaki}, {Jullo}, {Juneau}, {Kizhuprakkat}, {Kara{\c{c}}ayl{\i}}, {Karim}, {Kehoe}, {Kent}, {Khederlarian}, {Kim}, {Kirkby}, {Kisner}, {Kitaura}, {Kneib}, {Koposov}, {Kov{\'a}cs}, {Kremin}, {Krolewski}, {L'Huillier}, {Lambert}, {Lamman}, {Lan}, {Landriau}, {Lang}, {Lange}, {Lasker}, {Le Guillou}, {Leauthaud}, {Levi}, {Li}, {Linder}, {Lyons}, {Magneville}, {Manera}, {Manser}, {Margala}, {Martini}, {McDonald}, {Medina}, {Medina-Varela}, {Meisner}, {Mena-Fern{\'a}ndez}, {Meneses-Rizo}, {Mezcua}, {Miquel},
  {Montero-Camacho}, {Moon}, {Moore}, {Moustakas}, {Mueller}, {Mundet}, {Mu{\~n}oz-Guti{\'e}rrez}, {Myers}, {Nadathur}, {Napolitano}, {Neveux}, {Newman}, {Nie}, {Nikutta}, {Niz}, {Norberg}, {Noriega}, {Paillas}, {Palanque-Delabrouille}, {Palmese}, {Zhiwei}, {Parkinson}, {Penmetsa}, {Percival}, {P{\'e}rez-Fern{\'a}ndez}, {P{\'e}rez-R{\`a}fols}, {Pieri}, {Poppett}, {Porredon}, {Pothier}, {Prada}, {Pucha}, {Raichoor}, {Ram{\'\i}rez-P{\'e}rez}, {Ramirez-Solano}, {Rashkovetskyi}, {Ravoux}, {Rocher}, {Rockosi}, {Ross}, {Rossi}, {Ruggeri}, {Ruhlmann-Kleider}, {Sabiu}, {Said}, {Saintonge}, {Samushia}, {Sanchez}, {Saulder}, {Schaan}, {Schlafly}, {Schlegel}, {Scholte}, {Schubnell}, {Seo}, {Shafieloo}, {Sharples}, {Sheu}, {Silber}, {Sinigaglia}, {Siudek}, {Slepian}, {Smith}, {Sprayberry}, {Stephey}, {Su{\'a}rez-P{\'e}rez}, {Sun}, {Tan}, {Tarl{\'e}}, {Tojeiro}, {Ure{\~n}a-L{\'o}pez}, {Vaisakh}, {Valcin}, {Valdes}, {Valluri}, {Vargas-Maga{\~n}a}, {Variu}, {Verde}, {Walther}, {Wang}, {Wang}, {Weaver}, {Weaverdyck},
  {Wechsler}, {White}, {Xie}, {Yang}, {Y{\`e}che}, {Yu}, {Yuan}, {Zhang}, {Zhang}, {Zhao}, {Zheng}, {Zhou}, {Zhou}, {Zou}, {Zou}, \& {Zu}}]{EDR}
{DESI Collaboration}, {Adame}, A.~G., {Aguilar}, J., {et~al.} 2023{\natexlab{a}}, arXiv e-prints, arXiv:2306.06308

\bibitem[{{DESI Collaboration} {et~al.}(2023{\natexlab{b}}){DESI Collaboration}, {Adame}, {Aguilar}, {Ahlen}, {Alam}, {Aldering}, {Alexander}, {Alfarsy}, {Allende Prieto}, {Alvarez}, {Alves}, {Anand}, {Andrade-Oliveira}, {Armengaud}, {Asorey}, {Avila}, {Aviles}, {Bailey}, {Balaguera-Antol{\'\i}nez}, {Ballester}, {Baltay}, {Bault}, {Bautista}, {Behera}, {Beltran}, {BenZvi}, {Beraldo e Silva}, {Bermejo-Climent}, {Berti}, {Besuner}, {Beutler}, {Bianchi}, {Blake}, {Blum}, {Bolton}, {Brieden}, {Brodzeller}, {Brooks}, {Brown}, {Buckley-Geer}, {Burtin}, {Cabayol-Garcia}, {Cai}, {Canning}, {Cardiel-Sas}, {Carnero Rosell}, {Castander}, {Cervantes-Cota}, {Chabanier}, {Chaussidon}, {Chaves-Montero}, {Chen}, {Chuang}, {Claybaugh}, {Cole}, {Cooper}, {Cuceu}, {Davis}, {Dawson}, {de Belsunce}, {de la Cruz}, {de la Macorra}, {de Mattia}, {Demina}, {Demirbozan}, {DeRose}, {Dey}, {Dey}, {Dhungana}, {Ding}, {Ding}, {Doel}, {Doshi}, {Douglass}, {Edge}, {Eftekharzadeh}, {Eisenstein}, {Elliott}, {Escoffier}, {Fagrelius}, {Fan},
  {Fanning}, {Fawcett}, {Ferraro}, {Ereza}, {Flaugher}, {Font-Ribera}, {Forero-S{\'a}nchez}, {Forero-Romero}, {Frenk}, {G{\"a}nsicke}, {Garc{\'\i}a}, {Garc{\'\i}a-Bellido}, {Garcia-Quintero}, {Garrison}, {Gil-Mar{\'\i}n}, {Golden-Marx}, {Gontcho}, {Gonzalez-Morales}, {Gonzalez-Perez}, {Gordon}, {Graur}, {Green}, {Gruen}, {Guy}, {Hadzhiyska}, {Hahn}, {Han}, {Hanif}, {Herrera-Alcantar}, {Honscheid}, {Hou}, {Howlett}, {Huterer}, {Ir{\v{s}}i{\v{c}}}, {Ishak}, {Jana}, {Jiang}, {Jimenez}, {Jing}, {Joudaki}, {Jullo}, {Juneau}, {Kizhuprakkat}, {Kara{\c{c}}ayl{\i}}, {Karim}, {Kehoe}, {Kent}, {Khederlarian}, {Kim}, {Kirkby}, {Kisner}, {Kitaura}, {Kneib}, {Koposov}, {Kov{\'a}cs}, {Kremin}, {Krolewski}, {L'Huillier}, {Lambert}, {Lamman}, {Lan}, {Landriau}, {Lang}, {Lange}, {Lasker}, {Le Guillou}, {Leauthaud}, {Levi}, {Li}, {Linder}, {Lyons}, {Magneville}, {Manera}, {Manser}, {Margala}, {Martini}, {McDonald}, {Medina}, {Medina-Varela}, {Meisner}, {Mena-Fern{\'a}ndez}, {Meneses-Rizo}, {Mezcua}, {Miquel}, {Montero-Camacho},
  {Moon}, {Moore}, {Moustakas}, {Mueller}, {Mundet}, {Mu{\~n}oz-Guti{\'e}rrez}, {Myers}, {Nadathur}, {Napolitano}, {Neveux}, {Newman}, {Nie}, {Niz}, {Norberg}, {Noriega}, {Paillas}, {Palanque-Delabrouille}, {Palmese}, {Zhiwei}, {Parkinson}, {Penmetsa}, {Percival}, {P{\'e}rez-Fern{\'a}ndez}, {P{\'e}rez-R{\`a}fols}, {Pieri}, {Poppett}, {Porredon}, {Prada}, {Pucha}, {Raichoor}, {Ram{\'\i}rez-P{\'e}rez}, {Ramirez-Solano}, {Rashkovetskyi}, {Ravoux}, {Rocher}, {Rockosi}, {Ross}, {Rossi}, {Ruggeri}, {Ruhlmann-Kleider}, {Sabiu}, {Said}, {Saintonge}, {Samushia}, {Sanchez}, {Saulder}, {Schaan}, {Schlafly}, {Schlegel}, {Scholte}, {Schubnell}, {Seo}, {Shafieloo}, {Sharples}, {Sheu}, {Silber}, {Sinigaglia}, {Siudek}, {Slepian}, {Smith}, {Sprayberry}, {Stephey}, {Su{\'a}rez-P{\'e}rez}, {Sun}, {Tan}, {Tarl{\'e}}, {Tojeiro}, {Ure{\~n}a-L{\'o}pez}, {Vaisakh}, {Valcin}, {Valdes}, {Valluri}, {Vargas-Maga{\~n}a}, {Variu}, {Verde}, {Walther}, {Wang}, {Wang}, {Weaver}, {Weaverdyck}, {Wechsler}, {White}, {Xie}, {Yang}, {Y{\`e}che},
  {Yu}, {Yuan}, {Zhang}, {Zhang}, {Zhao}, {Zheng}, {Zhou}, {Zhou}, {Zou}, {Zou}, \& {Zu}}]{SV}
---. 2023{\natexlab{b}}, arXiv e-prints, arXiv:2306.06307

\bibitem[{{Dey} {et~al.}(2019){Dey}, {Schlegel}, {Lang}, {Blum}, {Burleigh}, {Fan}, {Findlay}, {Finkbeiner}, {Herrera}, {Juneau}, {Landriau}, {Levi}, {McGreer}, {Meisner}, {Myers}, {Moustakas}, {Nugent}, {Patej}, {Schlafly}, {Walker}, {Valdes}, {Weaver}, {Y{\`e}che}, {Zou}, {Zhou}, {Abareshi}, {Abbott}, {Abolfathi}, {Aguilera}, {Alam}, {Allen}, {Alvarez}, {Annis}, {Ansarinejad}, {Aubert}, {Beechert}, {Bell}, {BenZvi}, {Beutler}, {Bielby}, {Bolton}, {Brice{\~n}o}, {Buckley-Geer}, {Butler}, {Calamida}, {Carlberg}, {Carter}, {Casas}, {Castander}, {Choi}, {Comparat}, {Cukanovaite}, {Delubac}, {DeVries}, {Dey}, {Dhungana}, {Dickinson}, {Ding}, {Donaldson}, {Duan}, {Duckworth}, {Eftekharzadeh}, {Eisenstein}, {Etourneau}, {Fagrelius}, {Farihi}, {Fitzpatrick}, {Font-Ribera}, {Fulmer}, {G{\"a}nsicke}, {Gaztanaga}, {George}, {Gerdes}, {Gontcho}, {Gorgoni}, {Green}, {Guy}, {Harmer}, {Hernandez}, {Honscheid}, {Huang}, {James}, {Jannuzi}, {Jiang}, {Joyce}, {Karcher}, {Karkar}, {Kehoe}, {Kneib}, {Kueter-Young}, {Lan},
  {Lauer}, {Le Guillou}, {Le Van Suu}, {Lee}, {Lesser}, {Perreault Levasseur}, {Li}, {Mann}, {Marshall}, {Mart{\'\i}nez-V{\'a}zquez}, {Martini}, {du Mas des Bourboux}, {McManus}, {Meier}, {M{\'e}nard}, {Metcalfe}, {Mu{\~n}oz-Guti{\'e}rrez}, {Najita}, {Napier}, {Narayan}, {Newman}, {Nie}, {Nord}, {Norman}, {Olsen}, {Paat}, {Palanque-Delabrouille}, {Peng}, {Poppett}, {Poremba}, {Prakash}, {Rabinowitz}, {Raichoor}, {Rezaie}, {Robertson}, {Roe}, {Ross}, {Ross}, {Rudnick}, {Safonova}, {Saha}, {S{\'a}nchez}, {Savary}, {Schweiker}, {Scott}, {Seo}, {Shan}, {Silva}, {Slepian}, {Soto}, {Sprayberry}, {Staten}, {Stillman}, {Stupak}, {Summers}, {Sien Tie}, {Tirado}, {Vargas-Maga{\~n}a}, {Vivas}, {Wechsler}, {Williams}, {Yang}, {Yang}, {Yapici}, {Zaritsky}, {Zenteno}, {Zhang}, {Zhang}, {Zhou}, \& {Zhou}}]{Dey2019}
{Dey}, A., {Schlegel}, D.~J., {Lang}, D., {et~al.} 2019, \aj, 157, 168

\bibitem[{{Djorgovski} \& {Davis}(1987)}]{Djorgovski1987}
{Djorgovski}, S., \& {Davis}, M. 1987, \apj, 313, 59

\bibitem[{{Dressler} {et~al.}(1987){Dressler}, {Faber}, {Burstein}, {Davies}, {Lynden-Bell}, {Terlevich}, \& {Wegner}}]{Dressler1987}
{Dressler}, A., {Faber}, S.~M., {Burstein}, D., {et~al.} 1987, \apjl, 313, L37

\bibitem[{{Duev} {et~al.}(2019){Duev}, {Mahabal}, {Masci}, {Graham}, {Rusholme}, {Walters}, {Karmarkar}, {Frederick}, {Kasliwal}, {Rebbapragada}, \& {Ward}}]{Duev2019}
{Duev}, D.~A., {Mahabal}, A., {Masci}, F.~J., {et~al.} 2019, \mnras, 489, 3582

\bibitem[{{Feindt} {et~al.}(2013){Feindt}, {Kerschhaggl}, {Kowalski}, {Aldering}, {Antilogus}, {Aragon}, {Bailey}, {Baltay}, {Bongard}, {Buton}, {Canto}, {Cellier-Holzem}, {Childress}, {Chotard}, {Copin}, {Fakhouri}, {Gangler}, {Guy}, {Kim}, {Nugent}, {Nordin}, {Paech}, {Pain}, {Pecontal}, {Pereira}, {Perlmutter}, {Rabinowitz}, {Rigault}, {Runge}, {Saunders}, {Scalzo}, {Smadja}, {Tao}, {Thomas}, {Weaver}, \& {Wu}}]{Feindt2013}
{Feindt}, U., {Kerschhaggl}, M., {Kowalski}, M., {et~al.} 2013, \aap, 560, A90

\bibitem[{{F{\"o}rster} {et~al.}(2022){F{\"o}rster}, {Mu{\~n}oz Arancibia}, {Reyes-Jainaga}, {Gagliano}, {Britt}, {Cuellar-Carrillo}, {Figueroa-Tapia}, {Polzin}, {Yousef}, {Arredondo}, {Rodr{\'\i}guez-Mancini}, {Correa-Orellana}, {Bayo}, {Bauer}, {Catelan}, {Cabrera-Vives}, {Dastidar}, {Est{\'e}vez}, {Pignata}, {Hern{\'a}ndez-Garc{\'\i}a}, {Huijse}, {Reyes}, {S{\'a}nchez-S{\'a}ez}, {Ram{\'\i}rez}, {Grand{\'o}n}, {Pineda-Garc{\'\i}a}, {Chabour-Barra}, \& {Silva-Farf{\'a}n}}]{2022AJ....164..195F}
{F{\"o}rster}, F., {Mu{\~n}oz Arancibia}, A.~M., {Reyes-Jainaga}, I., {et~al.} 2022, \aj, 164, 195

\bibitem[{{Frederick} {et~al.}(2019){Frederick}, {Gezari}, {Graham}, {Cenko}, {van Velzen}, {Stern}, {Blagorodnova}, {Kulkarni}, {Yan}, {De}, {Fremling}, {Hung}, {Kara}, {Shupe}, {Ward}, {Bellm}, {Dekany}, {Duev}, {Feindt}, {Giomi}, {Kupfer}, {Laher}, {Masci}, {Miller}, {Neill}, {Ngeow}, {Patterson}, {Porter}, {Rusholme}, {Sollerman}, \& {Walters}}]{Frederick2019}
{Frederick}, S., {Gezari}, S., {Graham}, M.~J., {et~al.} 2019, \apj, 883, 31

\bibitem[{{Fremling} {et~al.}(2019){Fremling}, {Miller}, {Sharma}, {Dugas}, {Perley}, {Taggart}, {Sollerman}, {Goobar}, {Graham}, {Neill}, {Nordin}, {Rigault}, {Walters}, {Andreoni}, {Bagdasaryan}, {Belicki}, {Cannella}, {Bellm}, {Cenko}, {De}, {Dekany}, {Frederick}, {Golkhou}, {Graham}, {Helou}, {Ho}, {Kasliwal}, {Kupfer}, {Laher}, {Mahabal}, {Masci}, {Riddle}, {Rusholme}, {Schulze}, {Shupe}, {Smith}, {Yan}, {Yao}, {Zhuang}, \& {Kulkarni}}]{Fremling2019}
{Fremling}, U.~C., {Miller}, A.~A., {Sharma}, Y., {et~al.} 2019, arXiv e-prints, arXiv:1910.12973

\bibitem[{{Frieman} {et~al.}(2008){Frieman}, {Bassett}, {Becker}, {Choi}, {Cinabro}, {DeJongh}, {Depoy}, {Dilday}, {Doi}, {Garnavich}, {Hogan}, {Holtzman}, {Im}, {Jha}, {Kessler}, {Konishi}, {Lampeitl}, {Marriner}, {Marshall}, {McGinnis}, {Miknaitis}, {Nichol}, {Prieto}, {Riess}, {Richmond}, {Romani}, {Sako}, {Schneider}, {Smith}, {Takanashi}, {Tokita}, {van der Heyden}, {Yasuda}, {Zheng}, {Adelman-McCarthy}, {Annis}, {Assef}, {Barentine}, {Bender}, {Blandford}, {Boroski}, {Bremer}, {Brewington}, {Collins}, {Crotts}, {Dembicky}, {Eastman}, {Edge}, {Edmondson}, {Elson}, {Eyler}, {Filippenko}, {Foley}, {Frank}, {Goobar}, {Gueth}, {Gunn}, {Harvanek}, {Hopp}, {Ihara}, {Ivezi{\'c}}, {Kahn}, {Kaplan}, {Kent}, {Ketzeback}, {Kleinman}, {Kollatschny}, {Kron}, {Krzesi{\'n}ski}, {Lamenti}, {Leloudas}, {Lin}, {Long}, {Lucey}, {Lupton}, {Malanushenko}, {Malanushenko}, {McMillan}, {Mendez}, {Morgan}, {Morokuma}, {Nitta}, {Ostman}, {Pan}, {Rockosi}, {Romer}, {Ruiz-Lapuente}, {Saurage}, {Schlesinger}, {Snedden}, {Sollerman},
  {Stoughton}, {Stritzinger}, {Subba Rao}, {Tucker}, {Vaisanen}, {Watson}, {Watters}, {Wheeler}, {Yanny}, \& {York}}]{Frieman2008}
{Frieman}, J.~A., {Bassett}, B., {Becker}, A., {et~al.} 2008, \aj, 135, 338

\bibitem[{{Frohmaier} {et~al.}(2019){Frohmaier}, {Sullivan}, {Nugent}, {Smith}, {Dimitriadis}, {Bloom}, {Cenko}, {Kasliwal}, {Kulkarni}, {Maguire}, {Ofek}, {Poznanski}, \& {Quimby}}]{PTF2019MNRAS.486.2308F}
{Frohmaier}, C., {Sullivan}, M., {Nugent}, P.~E., {et~al.} 2019, \mnras, 486, 2308

\bibitem[{{Frohmaier} {et~al.}(2021){Frohmaier}, {Angus}, {Vincenzi}, {Sullivan}, {Smith}, {Nugent}, {Cenko}, {Gal-Yam}, {Kulkarni}, {Law}, \& {Quimby}}]{PTF2021MNRAS.500.5142F}
{Frohmaier}, C., {Angus}, C.~R., {Vincenzi}, M., {et~al.} 2021, \mnras, 500, 5142

\bibitem[{{Gagliano} {et~al.}(2023){Gagliano}, {Contardo}, {Foreman-Mackey}, {Malz}, \& {Aleo}}]{Gagliano2023}
{Gagliano}, A., {Contardo}, G., {Foreman-Mackey}, D., {Malz}, A.~I., \& {Aleo}, P.~D. 2023, \apj, 954, 6

\bibitem[{{Gagliano} {et~al.}(2021){Gagliano}, {Narayan}, {Engel}, {Carrasco Kind}, \& {LSST Dark Energy Science Collaboration}}]{Gagliano2021}
{Gagliano}, A., {Narayan}, G., {Engel}, A., {Carrasco Kind}, M., \& {LSST Dark Energy Science Collaboration}. 2021, \apj, 908, 170

\bibitem[{{Gomez} {et~al.}(2023{\natexlab{a}}){Gomez}, {Berger}, {Blanchard}, {Hosseinzadeh}, {Nicholl}, {Hiramatsu}, {Villar}, \& {Yin}}]{Gomez2023a}
{Gomez}, S., {Berger}, E., {Blanchard}, P.~K., {et~al.} 2023{\natexlab{a}}, \apj, 949, 114

\bibitem[{{Gomez} {et~al.}(2020){Gomez}, {Berger}, {Blanchard}, {Hosseinzadeh}, {Nicholl}, {Villar}, \& {Yin}}]{Gomez2020a}
---. 2020, \apj, 904, 74

\bibitem[{{Gomez} {et~al.}(2023{\natexlab{b}}){Gomez}, {Villar}, {Berger}, {Gezari}, {van Velzen}, {Nicholl}, {Blanchard}, \& {Alexander}}]{Gomez2023b}
{Gomez}, S., {Villar}, V.~A., {Berger}, E., {et~al.} 2023{\natexlab{b}}, \apj, 949, 113

\bibitem[{{Graham} {et~al.}(2019){Graham}, {Kulkarni}, {Bellm}, {Adams}, {Barbarino}, {Blagorodnova}, {Bodewits}, {Bolin}, {Brady}, {Cenko}, {Chang}, {Coughlin}, {De}, {Eadie}, {Farnham}, {Feindt}, {Franckowiak}, {Fremling}, {Gal-yam}, {Gezari}, {Ghosh}, {Goldstein}, {Golkhou}, {Goobar}, {Ho}, {Huppenkothen}, {Ivezic}, {Jones}, {Juric}, {Kaplan}, {Kasliwal}, {Kelley}, {Kupfer}, {Lee}, {Lin}, {Lunnan}, {Mahabal}, {Miller}, {Ngeow}, {Nugent}, {Ofek}, {Prince}, {Rauch}, {van Roestel}, {Schulze}, {Singer}, {Sollerman}, {Taddia}, {Yan}, {Ye}, {Yu}, {Andreoni}, {Barlow}, {Bauer}, {Beck}, {Belicki}, {Biswas}, {Brinnel}, {Brooke}, {Bue}, {Bulla}, {Burdge}, {Burruss}, {Connolly}, {Cromer}, {Cunningham}, {Dekany}, {Delacroix}, {Desai}, {Duev}, {Hacopians}, {Hale}, {Helou}, {Henning}, {Hover}, {Hillenbrand}, {Howell}, {Hung}, {Imel}, {Ip}, {Jackson}, {Kaspi}, {Kaye}, {Kowalski}, {Kramer}, {Kuhn}, {Land ry}, {Laher}, {Mao}, {Masci}, {Monkewitz}, {Murphy}, {Nordin}, {Patterson}, {Penprase}, {Porter}, {Rebbapragada},
  {Reiley}, {Riddle}, {Rigault}, {Rodriguez}, {Rusholme}, {van Santen}, {Shupe}, {Smith}, {Soumagnac}, {Stein}, {Surace}, {Szkody}, {Terek}, {van Sistine}, {van Velzen}, {Vestrand}, {Walters}, {Ward}, {Zhang}, \& {Zolkower}}]{Graham2019}
{Graham}, M.~J., {Kulkarni}, S.~R., {Bellm}, E.~C., {et~al.} 2019, arXiv e-prints, arXiv:1902.01945

\bibitem[{{Graham} {et~al.}(2023){Graham}, {Knop}, {Kennedy}, {Nugent}, {Bellm}, {Catelan}, {Patel}, {Smotherman}, {Soraisam}, {Stetzler}, {Aldoroty}, {Awbrey}, {Baeza-Villagra}, {Bernardinelli}, {Bianco}, {Brout}, {Clarke}, {Clarkson}, {Collett}, {Davenport}, {Fu}, {Gizis}, {Heinze}, {Hu}, {Jha}, {Juri{\'c}}, {Kalmbach}, {Kim}, {Lee}, {Lidman}, {Magee}, {Mart{\'\i}nez-V{\'a}zquez}, {Matheson}, {Narayan}, {Palmese}, {Phillips}, {Rabus}, {Rest}, {Rodr{\'\i}guez-Segovia}, {Street}, {Vivas}, {Wang}, {Wolf}, \& {Yang}}]{Graham2023}
{Graham}, M.~L., {Knop}, R.~A., {Kennedy}, T.~D., {et~al.} 2023, \mnras, 519, 3881

\bibitem[{{Guillochon} {et~al.}(2017){Guillochon}, {Parrent}, {Kelley}, \& {Margutti}}]{Guillochon2017}
{Guillochon}, J., {Parrent}, J., {Kelley}, L.~Z., \& {Margutti}, R. 2017, \apj, 835, 64

\bibitem[{{Gupta} {et~al.}(2016){Gupta}, {Kuhlmann}, {Kovacs}, {Spinka}, {Kessler}, {Goldstein}, {Liotine}, {Pomian}, {D'Andrea}, {Sullivan}, {Carretero}, {Castander}, {Nichol}, {Finley}, {Fischer}, {Foley}, {Kim}, {Papadopoulos}, {Sako}, {Scolnic}, {Smith}, {Tucker}, {Uddin}, {Wolf}, {Yuan}, {Abbott}, {Abdalla}, {Benoit-L{\'e}vy}, {Bertin}, {Brooks}, {Carnero Rosell}, {Carrasco Kind}, {Cunha}, {da Costa}, {Desai}, {Doel}, {Eifler}, {Evrard}, {Flaugher}, {Fosalba}, {Gazta{\~n}aga}, {Gruen}, {Gruendl}, {James}, {Kuehn}, {Kuropatkin}, {Maia}, {Marshall}, {Miquel}, {Plazas}, {Romer}, {S{\'a}nchez}, {Schubnell}, {Sevilla-Noarbe}, {Sobreira}, {Suchyta}, {Swanson}, {Tarle}, {Walker}, \& {Wester}}]{Gupta2016}
{Gupta}, R.~R., {Kuhlmann}, S., {Kovacs}, E., {et~al.} 2016, \aj, 152, 154

\bibitem[{{Guy} {et~al.}(2010){Guy}, {Sullivan}, {Conley}, {Regnault}, {Astier}, {Balland}, {Basa}, {Carlberg}, {Fouchez}, {Hardin}, {Hook}, {Howell}, {Pain}, {Palanque-Delabrouille}, {Perrett}, {Pritchet}, {Rich}, {Ruhlmann-Kleider}, {Balam}, {Baumont}, {Ellis}, {Fabbro}, {Fakhouri}, {Fourmanoit}, {Gonz{\'a}lez-Gait{\'a}n}, {Graham}, {Hsiao}, {Kronborg}, {Lidman}, {Mourao}, {Perlmutter}, {Ripoche}, {Suzuki}, \& {Walker}}]{Guy2010}
{Guy}, J., {Sullivan}, M., {Conley}, A., {et~al.} 2010, \aap, 523, A7

\bibitem[{{Guzzo} {et~al.}(2008){Guzzo}, {Pierleoni}, {Meneux}, {Branchini}, {Le F{\`e}vre}, {Marinoni}, {Garilli}, {Blaizot}, {De Lucia}, {Pollo}, {McCracken}, {Bottini}, {Le Brun}, {Maccagni}, {Picat}, {Scaramella}, {Scodeggio}, {Tresse}, {Vettolani}, {Zanichelli}, {Adami}, {Arnouts}, {Bardelli}, {Bolzonella}, {Bongiorno}, {Cappi}, {Charlot}, {Ciliegi}, {Contini}, {Cucciati}, {de la Torre}, {Dolag}, {Foucaud}, {Franzetti}, {Gavignaud}, {Ilbert}, {Iovino}, {Lamareille}, {Marano}, {Mazure}, {Memeo}, {Merighi}, {Moscardini}, {Paltani}, {Pell{\`o}}, {Perez-Montero}, {Pozzetti}, {Radovich}, {Vergani}, {Zamorani}, \& {Zucca}}]{Guzzo2008}
{Guzzo}, L., {Pierleoni}, M., {Meneux}, B., {et~al.} 2008, \nat, 451, 541

\bibitem[{{Hahn} {et~al.}(2023){Hahn}, {Wilson}, {Ruiz-Macias}, {Cole}, {Weinberg}, {Moustakas}, {Kremin}, {Tinker}, {Smith}, {Wechsler}, {Ahlen}, {Alam}, {Bailey}, {Brooks}, {Cooper}, {Davis}, {Dawson}, {Dey}, {Dey}, {Eftekharzadeh}, {Eisenstein}, {Fanning}, {Forero-Romero}, {Frenk}, {Gazta{\~n}aga}, {Gontcho A Gontcho}, {Guy}, {Honscheid}, {Ishak}, {Juneau}, {Kehoe}, {Kisner}, {Lan}, {Landriau}, {Le Guillou}, {Levi}, {Magneville}, {Martini}, {Meisner}, {Myers}, {Nie}, {Norberg}, {Palanque-Delabrouille}, {Percival}, {Poppett}, {Prada}, {Raichoor}, {Ross}, {Safonova}, {Saulder}, {Schlafly}, {Schlegel}, {Sierra-Porta}, {Tarle}, {Weaver}, {Y{\`e}che}, {Zarrouk}, {Zhou}, {Zhou}, \& {Zou}}]{BGS}
{Hahn}, C., {Wilson}, M.~J., {Ruiz-Macias}, O., {et~al.} 2023, \aj, 165, 253

\bibitem[{{Howlett} {et~al.}(2017){Howlett}, {Robotham}, {Lagos}, \& {Kim}}]{Howlett2017}
{Howlett}, C., {Robotham}, A. S.~G., {Lagos}, C. D.~P., \& {Kim}, A.~G. 2017, \apj, 847, 128

\bibitem[{{Ivezi{\'c}} {et~al.}(2019){Ivezi{\'c}}, {Kahn}, {Tyson}, {Abel}, {Acosta}, {Allsman}, {Alonso}, {AlSayyad}, {Anderson}, {Andrew}, {Angel}, {Angeli}, {Ansari}, {Antilogus}, {Araujo}, {Armstrong}, {Arndt}, {Astier}, {Aubourg}, {Auza}, {Axelrod}, {Bard}, {Barr}, {Barrau}, {Bartlett}, {Bauer}, {Bauman}, {Baumont}, {Bechtol}, {Bechtol}, {Becker}, {Becla}, {Beldica}, {Bellavia}, {Bianco}, {Biswas}, {Blanc}, {Blazek}, {Blandford}, {Bloom}, {Bogart}, {Bond}, {Booth}, {Borgland}, {Borne}, {Bosch}, {Boutigny}, {Brackett}, {Bradshaw}, {Brandt}, {Brown}, {Bullock}, {Burchat}, {Burke}, {Cagnoli}, {Calabrese}, {Callahan}, {Callen}, {Carlin}, {Carlson}, {Chandrasekharan}, {Charles-Emerson}, {Chesley}, {Cheu}, {Chiang}, {Chiang}, {Chirino}, {Chow}, {Ciardi}, {Claver}, {Cohen-Tanugi}, {Cockrum}, {Coles}, {Connolly}, {Cook}, {Cooray}, {Covey}, {Cribbs}, {Cui}, {Cutri}, {Daly}, {Daniel}, {Daruich}, {Daubard}, {Daues}, {Dawson}, {Delgado}, {Dellapenna}, {de Peyster}, {de Val-Borro}, {Digel}, {Doherty}, {Dubois},
  {Dubois-Felsmann}, {Durech}, {Economou}, {Eifler}, {Eracleous}, {Emmons}, {Fausti Neto}, {Ferguson}, {Figueroa}, {Fisher-Levine}, {Focke}, {Foss}, {Frank}, {Freemon}, {Gangler}, {Gawiser}, {Geary}, {Gee}, {Geha}, {Gessner}, {Gibson}, {Gilmore}, {Glanzman}, {Glick}, {Goldina}, {Goldstein}, {Goodenow}, {Graham}, {Gressler}, {Gris}, {Guy}, {Guyonnet}, {Haller}, {Harris}, {Hascall}, {Haupt}, {Hernandez}, {Herrmann}, {Hileman}, {Hoblitt}, {Hodgson}, {Hogan}, {Howard}, {Huang}, {Huffer}, {Ingraham}, {Innes}, {Jacoby}, {Jain}, {Jammes}, {Jee}, {Jenness}, {Jernigan}, {Jevremovi{\'c}}, {Johns}, {Johnson}, {Johnson}, {Jones}, {Juramy-Gilles}, {Juri{\'c}}, {Kalirai}, {Kallivayalil}, {Kalmbach}, {Kantor}, {Karst}, {Kasliwal}, {Kelly}, {Kessler}, {Kinnison}, {Kirkby}, {Knox}, {Kotov}, {Krabbendam}, {Krughoff}, {Kub{\'a}nek}, {Kuczewski}, {Kulkarni}, {Ku}, {Kurita}, {Lage}, {Lambert}, {Lange}, {Langton}, {Le Guillou}, {Levine}, {Liang}, {Lim}, {Lintott}, {Long}, {Lopez}, {Lotz}, {Lupton}, {Lust}, {MacArthur}, {Mahabal},
  {Mandelbaum}, {Markiewicz}, {Marsh}, {Marshall}, {Marshall}, {May}, {McKercher}, {McQueen}, {Meyers}, {Migliore}, {Miller}, {Mills}, {Miraval}, {Moeyens}, {Moolekamp}, {Monet}, {Moniez}, {Monkewitz}, {Montgomery}, {Morrison}, {Mueller}, {Muller}, {Mu{\~n}oz Arancibia}, {Neill}, {Newbry}, {Nief}, {Nomerotski}, {Nordby}, {O'Connor}, {Oliver}, {Olivier}, {Olsen}, {O'Mullane}, {Ortiz}, {Osier}, {Owen}, {Pain}, {Palecek}, {Parejko}, {Parsons}, {Pease}, {Peterson}, {Peterson}, {Petravick}, {Libby Petrick}, {Petry}, {Pierfederici}, {Pietrowicz}, {Pike}, {Pinto}, {Plante}, {Plate}, {Plutchak}, {Price}, {Prouza}, {Radeka}, {Rajagopal}, {Rasmussen}, {Regnault}, {Reil}, {Reiss}, {Reuter}, {Ridgway}, {Riot}, {Ritz}, {Robinson}, {Roby}, {Roodman}, {Rosing}, {Roucelle}, {Rumore}, {Russo}, {Saha}, {Sassolas}, {Schalk}, {Schellart}, {Schindler}, {Schmidt}, {Schneider}, {Schneider}, {Schoening}, {Schumacher}, {Schwamb}, {Sebag}, {Selvy}, {Sembroski}, {Seppala}, {Serio}, {Serrano}, {Shaw}, {Shipsey}, {Sick}, {Silvestri},
  {Slater}, {Smith}, {Smith}, {Sobhani}, {Soldahl}, {Storrie-Lombardi}, {Stover}, {Strauss}, {Street}, {Stubbs}, {Sullivan}, {Sweeney}, {Swinbank}, {Szalay}, {Takacs}, {Tether}, {Thaler}, {Thayer}, {Thomas}, {Thornton}, {Thukral}, {Tice}, {Trilling}, {Turri}, {Van Berg}, {Vanden Berk}, {Vetter}, {Virieux}, {Vucina}, {Wahl}, {Walkowicz}, {Walsh}, {Walter}, {Wang}, {Wang}, {Warner}, {Wiecha}, {Willman}, {Winters}, {Wittman}, {Wolff}, {Wood-Vasey}, {Wu}, {Xin}, {Yoachim}, \& {Zhan}}]{lsst}
{Ivezi{\'c}}, {\v{Z}}., {Kahn}, S.~M., {Tyson}, J.~A., {et~al.} 2019, \apj, 873, 111

\bibitem[{{Kelly} {et~al.}(2010){Kelly}, {Hicken}, {Burke}, {Mandel}, \& {Kirshner}}]{2010ApJ...715..743K}
{Kelly}, P.~L., {Hicken}, M., {Burke}, D.~L., {Mandel}, K.~S., \& {Kirshner}, R.~P. 2010, \apj, 715, 743

\bibitem[{{Kelsey} {et~al.}(2021){Kelsey}, {Sullivan}, {Smith}, {Wiseman}, {Brout}, {Davis}, {Frohmaier}, {Galbany}, {Grayling}, {Guti{\'e}rrez}, {Hinton}, {Kessler}, {Lidman}, {M{\"o}ller}, {Sako}, {Scolnic}, {Uddin}, {Vincenzi}, {Abbott}, {Aguena}, {Allam}, {Annis}, {Avila}, {Bacon}, {Bertin}, {Brooks}, {Burke}, {Carnero Rosell}, {Carrasco Kind}, {Carretero}, {Castander}, {Costanzi}, {da Costa}, {Desai}, {Diehl}, {Doel}, {Everett}, {Ferrero}, {Fert{\'e}}, {Flaugher}, {Fosalba}, {Garc{\'\i}a-Bellido}, {Gerdes}, {Gruen}, {Gruendl}, {Gschwend}, {Gutierrez}, {Hollowood}, {Honscheid}, {James}, {Kim}, {Kuehn}, {Kuropatkin}, {Lahav}, {Lima}, {Marshall}, {Martini}, {Menanteau}, {Miquel}, {Morgan}, {Ogando}, {Palmese}, {Paz-Chinch{\'o}n}, {Plazas}, {Romer}, {S{\'a}nchez}, {Sanchez}, {Serrano}, {Sevilla-Noarbe}, {Suchyta}, {Tarle}, {Thomas}, {To}, {Varga}, {Walker}, {Wilkinson}, \& {DES Collaboration}}]{2021MNRAS.501.4861K}
{Kelsey}, L., {Sullivan}, M., {Smith}, M., {et~al.} 2021, \mnras, 501, 4861

\bibitem[{{Kim} \& {Linder}(2020)}]{Kim2020}
{Kim}, A.~G., \& {Linder}, E.~V. 2020, \prd, 101, 023516

\bibitem[{{Kisley} {et~al.}(2023){Kisley}, {Qin}, {Zabludoff}, {Barnard}, \& {Ko}}]{Kisley2023}
{Kisley}, M., {Qin}, Y.-J., {Zabludoff}, A., {Barnard}, K., \& {Ko}, C.-L. 2023, \apj, 942, 29

\bibitem[{{Lampeitl} {et~al.}(2010){Lampeitl}, {Smith}, {Nichol}, {Bassett}, {Cinabro}, {Dilday}, {Foley}, {Frieman}, {Garnavich}, {Goobar}, {Im}, {Jha}, {Marriner}, {Miquel}, {Nordin}, {{\"O}stman}, {Riess}, {Sako}, {Schneider}, {Sollerman}, \& {Stritzinger}}]{2010ApJ...722..566L}
{Lampeitl}, H., {Smith}, M., {Nichol}, R.~C., {et~al.} 2010, \apj, 722, 566

\bibitem[{{Lang} {et~al.}(2016{\natexlab{a}}){Lang}, {Hogg}, \& {Mykytyn}}]{Lang2016b}
{Lang}, D., {Hogg}, D.~W., \& {Mykytyn}, D. 2016{\natexlab{a}}, {The Tractor: Probabilistic astronomical source detection and measurement}, Astrophysics Source Code Library, record ascl:1604.008, ascl:1604.008

\bibitem[{{Lang} {et~al.}(2016{\natexlab{b}}){Lang}, {Hogg}, \& {Schlegel}}]{Lang2016}
{Lang}, D., {Hogg}, D.~W., \& {Schlegel}, D.~J. 2016{\natexlab{b}}, \aj, 151, 36

\bibitem[{{Law} {et~al.}(2009){Law}, {Kulkarni}, {Dekany}, {Ofek}, {Quimby}, {Nugent}, {Surace}, {Grillmair}, {Bloom}, {Kasliwal}, {Bildsten}, {Brown}, {Cenko}, {Ciardi}, {Croner}, {Djorgovski}, {van Eyken}, {Filippenko}, {Fox}, {Gal-Yam}, {Hale}, {Hamam}, {Helou}, {Henning}, {Howell}, {Jacobsen}, {Laher}, {Mattingly}, {McKenna}, {Pickles}, {Poznanski}, {Rahmer}, {Rau}, {Rosing}, {Shara}, {Smith}, {Starr}, {Sullivan}, {Velur}, {Walters}, \& {Zolkower}}]{Law2009}
{Law}, N.~M., {Kulkarni}, S.~R., {Dekany}, R.~G., {et~al.} 2009, \pasp, 121, 1395

\bibitem[{{Mahabal} {et~al.}(2019){Mahabal}, {Rebbapragada}, {Walters}, {Masci}, {Blagorodnova}, {van Roestel}, {Ye}, {Biswas}, {Burdge}, {Chang}, {Duev}, {Golkhou}, {Miller}, {Nordin}, {Ward}, {Adams}, {Bellm}, {Branton}, {Bue}, {Cannella}, {Connolly}, {Dekany}, {Feindt}, {Hung}, {Fortson}, {Frederick}, {Fremling}, {Gezari}, {Graham}, {Groom}, {Kasliwal}, {Kulkarni}, {Kupfer}, {Lin}, {Lintott}, {Lunnan}, {Parejko}, {Prince}, {Riddle}, {Rusholme}, {Saunders}, {Sedaghat}, {Shupe}, {Singer}, {Soumagnac}, {Szkody}, {Tachibana}, {Tirumala}, {van Velzen}, \& {Wright}}]{Mahabal2019}
{Mahabal}, A., {Rebbapragada}, U., {Walters}, R., {et~al.} 2019, \pasp, 131, 038002

\bibitem[{{Mannucci} {et~al.}(2006){Mannucci}, {Della Valle}, \& {Panagia}}]{2006MNRAS.370..773M}
{Mannucci}, F., {Della Valle}, M., \& {Panagia}, N. 2006, \mnras, 370, 773

\bibitem[{{Masci} {et~al.}(2019){Masci}, {Laher}, {Rusholme}, {Shupe}, {Groom}, {Surace}, {Jackson}, {Monkewitz}, {Beck}, {Flynn}, {Terek}, {Landry}, {Hacopians}, {Desai}, {Howell}, {Brooke}, {Imel}, {Wachter}, {Ye}, {Lin}, {Cenko}, {Cunningham}, {Rebbapragada}, {Bue}, {Miller}, {Mahabal}, {Bellm}, {Patterson}, {Juri{\'c}}, {Golkhou}, {Ofek}, {Walters}, {Graham}, {Kasliwal}, {Dekany}, {Kupfer}, {Burdge}, {Cannella}, {Barlow}, {Van Sistine}, {Giomi}, {Fremling}, {Blagorodnova}, {Levitan}, {Riddle}, {Smith}, {Helou}, {Prince}, \& {Kulkarni}}]{Masci2019}
{Masci}, F.~J., {Laher}, R.~R., {Rusholme}, B., {et~al.} 2019, \pasp, 131, 018003

\bibitem[{Meldorf {et~al.}(2022)Meldorf, Palmese, Brout, Chen, Scolnic, Kelsey, Galbany, Hartley, Davis, Drlica-Wagner, Vincenzi, Annis, Dixon, Graur, Lidman, Möller, Nugent, Rose, Smith, Allam, Tucker, Asorey, Calcino, Carollo, Glazebrook, Lewis, Taylor, Tucker, Kim, Diehl, Aguena, Andrade-Oliveira, Bacon, Bertin, Bocquet, Brooks, Burke, Carretero, Kind, Castander, Costanzi, da~Costa, Desai, Doel, Everett, Ferrero, Friedel, Frieman, García-Bellido, Gatti, Gruen, Gschwend, Gutierrez, Hinton, Hollowood, Honscheid, James, Kuehn, March, Marshall, Menanteau, Miquel, Morgan, Paz-Chinchón, Pereira, Malagón, Sanchez, Scarpine, Sevilla-Noarbe, Suchyta, Tarle, Varga, \& DES}]{MassStepDust}
Meldorf, C., Palmese, A., Brout, D., {et~al.} 2022, Monthly Notices of the Royal Astronomical Society, https://academic.oup.com/mnras/advance-article-pdf/doi/10.1093/mnras/stac3056/46648726/stac3056.pdf, stac3056

\bibitem[{{Moustakas} {et~al.}(2023){Moustakas}, {Lang}, {Dey}, {Juneau}, {Meisner}, {Myers}, {Schlafly}, {Schlegel}, {Valdes}, {Weaver}, \& {Zhou}}]{Moustakas2023}
{Moustakas}, J., {Lang}, D., {Dey}, A., {et~al.} 2023, arXiv e-prints, arXiv:2307.04888

\bibitem[{{Muthukrishna} {et~al.}(2019){Muthukrishna}, {Narayan}, {Mandel}, {Biswas}, \& {Hlo{\v{z}}ek}}]{Muthukrishna2019}
{Muthukrishna}, D., {Narayan}, G., {Mandel}, K.~S., {Biswas}, R., \& {Hlo{\v{z}}ek}, R. 2019, \pasp, 131, 118002

\bibitem[{{Myers} {et~al.}(2023){Myers}, {Moustakas}, {Bailey}, {Weaver}, {Cooper}, {Forero-Romero}, {Abolfathi}, {Alexander}, {Brooks}, {Chaussidon}, {Chuang}, {Dawson}, {Dey}, {Dey}, {Dhungana}, {Doel}, {Fanning}, {Gazta{\~n}aga}, {Gontcho A Gontcho}, {Gonzalez-Morales}, {Hahn}, {Herrera-Alcantar}, {Honscheid}, {Ishak}, {Karim}, {Kirkby}, {Kisner}, {Koposov}, {Kremin}, {Lan}, {Landriau}, {Lang}, {Levi}, {Magneville}, {Napolitano}, {Martini}, {Meisner}, {Newman}, {Palanque-Delabrouille}, {Percival}, {Poppett}, {Prada}, {Raichoor}, {Ross}, {Schlafly}, {Schlegel}, {Schubnell}, {Tan}, {Tarle}, {Wilson}, {Y{\`e}che}, {Zhou}, {Zhou}, \& {Zou}}]{Myers2023}
{Myers}, A.~D., {Moustakas}, J., {Bailey}, S., {et~al.} 2023, \aj, 165, 50

\bibitem[{{Nicolas} {et~al.}(2021){Nicolas}, {Rigault}, {Copin}, {Graziani}, {Aldering}, {Briday}, {Kim}, {Nordin}, {Perlmutter}, \& {Smith}}]{2021A&A...649A..74N}
{Nicolas}, N., {Rigault}, M., {Copin}, Y., {et~al.} 2021, \aap, 649, A74

\bibitem[{{Nugent} {et~al.}(2023){Nugent}, {Polin}, \& {Nugent}}]{Nugent2023}
{Nugent}, A.~E., {Polin}, A.~E., \& {Nugent}, P.~E. 2023, arXiv e-prints, arXiv:2304.10601

\bibitem[{Nugent \& Hamuy(2017)}]{Nugent2017}
Nugent, P., \& Hamuy, M. 2017, in Handbook of Supernovae, ed. A.~W. Alsabti \& P.~Murdin (Cham: Springer International Publishing), 2671--2688

\bibitem[{{Palmese} {et~al.}(2022){Palmese}, {Wang}, {Chen}, {Hu}, {Yang}, {BenZvi}, {Cooke}, {Knop}, {Raichoor}, {Schlafly}, {Aldoroty}, {Baade}, {Brown}, {Chornock}, {Davis}, {Dawson}, {Graham}, {Huang}, {Jiang}, {Kim}, {Koekemoer}, {Margutti}, {Mould}, {Nugent}, {Patat}, {Schlegel}, {Uddin}, \& {Wang}}]{desirt}
{Palmese}, A., {Wang}, L., {Chen}, X., {et~al.} 2022, Transient Name Server AstroNote, 107, 1

\bibitem[{{Pan} {et~al.}(2015){Pan}, {Sullivan}, {Maguire}, {Gal-Yam}, {Hook}, {Howell}, {Nugent}, \& {Mazzali}}]{Pan2015}
{Pan}, Y.~C., {Sullivan}, M., {Maguire}, K., {et~al.} 2015, \mnras, 446, 354

\bibitem[{{Pan} {et~al.}(2014){Pan}, {Sullivan}, {Maguire}, {Hook}, {Nugent}, {Howell}, {Arcavi}, {Botyanszki}, {Cenko}, {DeRose}, {Fakhouri}, {Gal-Yam}, {Hsiao}, {Kulkarni}, {Laher}, {Lidman}, {Nordin}, {Walker}, \& {Xu}}]{PTF2014MNRAS.438.1391P}
---. 2014, \mnras, 438, 1391

\bibitem[{{Patterson} {et~al.}(2019){Patterson}, {Bellm}, {Rusholme}, {Masci}, {Juric}, {Krughoff}, {Golkhou}, {Graham}, {Kulkarni}, {Helou}, \& {Zwicky Transient Facility Collaboration}}]{Patterson2019}
{Patterson}, M.~T., {Bellm}, E.~C., {Rusholme}, B., {et~al.} 2019, \pasp, 131, 018001

\bibitem[{{Perley} {et~al.}(2016){Perley}, {Quimby}, {Yan}, {Vreeswijk}, {De Cia}, {Lunnan}, {Gal-Yam}, {Yaron}, {Filippenko}, {Graham}, {Laher}, \& {Nugent}}]{Perley2016}
{Perley}, D.~A., {Quimby}, R.~M., {Yan}, L., {et~al.} 2016, \apj, 830, 13

\bibitem[{{Perley} {et~al.}(2020){Perley}, {Fremling}, {Sollerman}, {Miller}, {Dahiwale}, {Sharma}, {Bellm}, {Biswas}, {Brink}, {Bruch}, {De}, {Dekany}, {Drake}, {Duev}, {Filippenko}, {Gal-Yam}, {Goobar}, {Graham}, {Graham}, {Ho}, {Irani}, {Kasliwal}, {Kim}, {Kulkarni}, {Mahabal}, {Masci}, {Modak}, {Neill}, {Nordin}, {Riddle}, {Soumagnac}, {Strotjohann}, {Schulze}, {Taggart}, {Tzanidakis}, {Walters}, \& {Yan}}]{Perley2020}
{Perley}, D.~A., {Fremling}, C., {Sollerman}, J., {et~al.} 2020, \apj, 904, 35

\bibitem[{{Pezzotta} {et~al.}(2017){Pezzotta}, {de la Torre}, {Bel}, {Granett}, {Guzzo}, {Peacock}, {Garilli}, {Scodeggio}, {Bolzonella}, {Abbas}, {Adami}, {Bottini}, {Cappi}, {Cucciati}, {Davidzon}, {Franzetti}, {Fritz}, {Iovino}, {Krywult}, {Le Brun}, {Le F{\`e}vre}, {Maccagni}, {Ma{\l}ek}, {Marulli}, {Polletta}, {Pollo}, {Tasca}, {Tojeiro}, {Vergani}, {Zanichelli}, {Arnouts}, {Branchini}, {Coupon}, {De Lucia}, {Koda}, {Ilbert}, {Mohammad}, {Moutard}, \& {Moscardini}}]{Pezzotta2017}
{Pezzotta}, A., {de la Torre}, S., {Bel}, J., {et~al.} 2017, \aap, 604, A33

\bibitem[{{Planck Collaboration} {et~al.}(2016){Planck Collaboration}, {Ade}, {Aghanim}, {Arnaud}, {Ashdown}, {Aumont}, {Baccigalupi}, {Banday}, {Barreiro}, {Bartlett}, {Bartolo}, {Battaner}, {Battye}, {Benabed}, {Beno{\^\i}t}, {Benoit-L{\'e}vy}, {Bernard}, {Bersanelli}, {Bielewicz}, {Bock}, {Bonaldi}, {Bonavera}, {Bond}, {Borrill}, {Bouchet}, {Boulanger}, {Bucher}, {Burigana}, {Butler}, {Calabrese}, {Cardoso}, {Catalano}, {Challinor}, {Chamballu}, {Chary}, {Chiang}, {Chluba}, {Christensen}, {Church}, {Clements}, {Colombi}, {Colombo}, {Combet}, {Coulais}, {Crill}, {Curto}, {Cuttaia}, {Danese}, {Davies}, {Davis}, {de Bernardis}, {de Rosa}, {de Zotti}, {Delabrouille}, {D{\'e}sert}, {Di Valentino}, {Dickinson}, {Diego}, {Dolag}, {Dole}, {Donzelli}, {Dor{\'e}}, {Douspis}, {Ducout}, {Dunkley}, {Dupac}, {Efstathiou}, {Elsner}, {En{\ss}lin}, {Eriksen}, {Farhang}, {Fergusson}, {Finelli}, {Forni}, {Frailis}, {Fraisse}, {Franceschi}, {Frejsel}, {Galeotta}, {Galli}, {Ganga}, {Gauthier}, {Gerbino}, {Ghosh}, {Giard},
  {Giraud-H{\'e}raud}, {Giusarma}, {Gjerl{\o}w}, {Gonz{\'a}lez-Nuevo}, {G{\'o}rski}, {Gratton}, {Gregorio}, {Gruppuso}, {Gudmundsson}, {Hamann}, {Hansen}, {Hanson}, {Harrison}, {Helou}, {Henrot-Versill{\'e}}, {Hern{\'a}ndez-Monteagudo}, {Herranz}, {Hildebrandt}, {Hivon}, {Hobson}, {Holmes}, {Hornstrup}, {Hovest}, {Huang}, {Huffenberger}, {Hurier}, {Jaffe}, {Jaffe}, {Jones}, {Juvela}, {Keih{\"a}nen}, {Keskitalo}, {Kisner}, {Kneissl}, {Knoche}, {Knox}, {Kunz}, {Kurki-Suonio}, {Lagache}, {L{\"a}hteenm{\"a}ki}, {Lamarre}, {Lasenby}, {Lattanzi}, {Lawrence}, {Leahy}, {Leonardi}, {Lesgourgues}, {Levrier}, {Lewis}, {Liguori}, {Lilje}, {Linden-V{\o}rnle}, {L{\'o}pez-Caniego}, {Lubin}, {Mac{\'\i}as-P{\'e}rez}, {Maggio}, {Maino}, {Mandolesi}, {Mangilli}, {Marchini}, {Maris}, {Martin}, {Martinelli}, {Mart{\'\i}nez-Gonz{\'a}lez}, {Masi}, {Matarrese}, {McGehee}, {Meinhold}, {Melchiorri}, {Melin}, {Mendes}, {Mennella}, {Migliaccio}, {Millea}, {Mitra}, {Miville-Desch{\^e}nes}, {Moneti}, {Montier}, {Morgante}, {Mortlock},
  {Moss}, {Munshi}, {Murphy}, {Naselsky}, {Nati}, {Natoli}, {Netterfield}, {N{\o}rgaard-Nielsen}, {Noviello}, {Novikov}, {Novikov}, {Oxborrow}, {Paci}, {Pagano}, {Pajot}, {Paladini}, {Paoletti}, {Partridge}, {Pasian}, {Patanchon}, {Pearson}, {Perdereau}, {Perotto}, {Perrotta}, {Pettorino}, {Piacentini}, {Piat}, {Pierpaoli}, {Pietrobon}, {Plaszczynski}, {Pointecouteau}, {Polenta}, {Popa}, {Pratt}, {Pr{\'e}zeau}, {Prunet}, {Puget}, {Rachen}, {Reach}, {Rebolo}, {Reinecke}, {Remazeilles}, {Renault}, {Renzi}, {Ristorcelli}, {Rocha}, {Rosset}, {Rossetti}, {Roudier}, {Rouill{\'e} d'Orfeuil}, {Rowan-Robinson}, {Rubi{\~n}o-Mart{\'\i}n}, {Rusholme}, {Said}, {Salvatelli}, {Salvati}, {Sandri}, {Santos}, {Savelainen}, {Savini}, {Scott}, {Seiffert}, {Serra}, {Shellard}, {Spencer}, {Spinelli}, {Stolyarov}, {Stompor}, {Sudiwala}, {Sunyaev}, {Sutton}, {Suur-Uski}, {Sygnet}, {Tauber}, {Terenzi}, {Toffolatti}, {Tomasi}, {Tristram}, {Trombetti}, {Tucci}, {Tuovinen}, {T{\"u}rler}, {Umana}, {Valenziano}, {Valiviita}, {Van Tent},
  {Vielva}, {Villa}, {Wade}, {Wandelt}, {Wehus}, {White}, {White}, {Wilkinson}, {Yvon}, {Zacchei}, \& {Zonca}}]{Planck2016}
{Planck Collaboration}, {Ade}, P.~A.~R., {Aghanim}, N., {et~al.} 2016, \aap, 594, A13

\bibitem[{{Popovic} {et~al.}(2021){Popovic}, {Brout}, {Kessler}, {Scolnic}, \& {Lu}}]{pbb+2021}
{Popovic}, B., {Brout}, D., {Kessler}, R., {Scolnic}, D., \& {Lu}, L. 2021, \apj, 913, 49

\bibitem[{{Quimby} {et~al.}(2018){Quimby}, {De Cia}, {Gal-Yam}, {Leloudas}, {Lunnan}, {Perley}, {Vreeswijk}, {Yan}, {Bloom}, {Cenko}, {Cooke}, {Ellis}, {Filippenko}, {Kasliwal}, {Kleiser}, {Kulkarni}, {Matheson}, {Nugent}, {Pan}, {Silverman}, {Sternberg}, {Sullivan}, \& {Yaron}}]{PTF2018ApJ...855....2Q}
{Quimby}, R.~M., {De Cia}, A., {Gal-Yam}, A., {et~al.} 2018, \apj, 855, 2

\bibitem[{{Riess} {et~al.}(1997){Riess}, {Davis}, {Baker}, \& {Kirshner}}]{Riess1997}
{Riess}, A.~G., {Davis}, M., {Baker}, J., \& {Kirshner}, R.~P. 1997, \apjl, 488, L1

\bibitem[{Rigault(2018)}]{Rigault2018}
Rigault, M. 2018, ztfquery, a python tool to access ZTF data, doi:10.5281/zenodo.1345222

\bibitem[{{Rigault} {et~al.}(2013){Rigault}, {Copin}, {Aldering}, {Antilogus}, {Aragon}, {Bailey}, {Baltay}, {Bongard}, {Buton}, {Canto}, {Cellier-Holzem}, {Childress}, {Chotard}, {Fakhouri}, {Feindt}, {Fleury}, {Gangler}, {Greskovic}, {Guy}, {Kim}, {Kowalski}, {Lombardo}, {Nordin}, {Nugent}, {Pain}, {P{\'e}contal}, {Pereira}, {Perlmutter}, {Rabinowitz}, {Runge}, {Saunders}, {Scalzo}, {Smadja}, {Tao}, {Thomas}, \& {Weaver}}]{Rigault2013}
{Rigault}, M., {Copin}, Y., {Aldering}, G., {et~al.} 2013, \aap, 560, A66

\bibitem[{{Rigault} {et~al.}(2015){Rigault}, {Aldering}, {Kowalski}, {Copin}, {Antilogus}, {Aragon}, {Bailey}, {Baltay}, {Baugh}, {Bongard}, {Boone}, {Buton}, {Chen}, {Chotard}, {Fakhouri}, {Feindt}, {Fagrelius}, {Fleury}, {Fouchez}, {Gangler}, {Hayden}, {Kim}, {Leget}, {Lombardo}, {Nordin}, {Pain}, {Pecontal}, {Pereira}, {Perlmutter}, {Rabinowitz}, {Runge}, {Rubin}, {Saunders}, {Smadja}, {Sofiatti}, {Suzuki}, {Tao}, \& {Weaver}}]{2015ApJ...802...20R}
{Rigault}, M., {Aldering}, G., {Kowalski}, M., {et~al.} 2015, \apj, 802, 20

\bibitem[{{Rigault} {et~al.}(2020){Rigault}, {Brinnel}, {Aldering}, {Antilogus}, {Aragon}, {Bailey}, {Baltay}, {Barbary}, {Bongard}, {Boone}, {Buton}, {Childress}, {Chotard}, {Copin}, {Dixon}, {Fagrelius}, {Feindt}, {Fouchez}, {Gangler}, {Hayden}, {Hillebrandt}, {Howell}, {Kim}, {Kowalski}, {Kuesters}, {Leget}, {Lombardo}, {Lin}, {Nordin}, {Pain}, {Pecontal}, {Pereira}, {Perlmutter}, {Rabinowitz}, {Runge}, {Rubin}, {Saunders}, {Smadja}, {Sofiatti}, {Suzuki}, {Taubenberger}, {Tao}, \& {Thomas}}]{Rigault2020}
{Rigault}, M., {Brinnel}, V., {Aldering}, G., {et~al.} 2020, \aap, 644, A176

\bibitem[{{Rigault} {et~al.}(2024){Rigault}, {Smith}, {Goobar}, {Maguire}, {Dimitriadis}, {Johansson}, {Nordin}, {Burgaz}, {Dhawan}, {Sollerman}, {Regnault}, {Kowalski}, {Nugent}, {Andreoni}, {Amenouche}, {Aubert}, {Barjou-Delayre}, {Bautista}, {Bellm}, {Betoule}, {Bloom}, {Carreres}, X., {Copin}, {Deckers}, {de Jaeger}, {Feinstein}, {Fouchez}, {Fremling}, {Galbany}, {Ginolin}, {Graham}, {Groom}, {Harvey}, {Kasliwal}, {Kenworthy}, {Kim}, {Kuhn}, {Kulkarni}, {Lacroix}, {Laher}, {Masci}, {Muller-Bravo}, {Miller}, M., {Perley}, {Popovic}, {Purdum}, {Qin}, {Racine}, {Reusch}, {Riddle}, {Rosnet}, {Rosselli}, {Ruppin}, {Senzel}, {Rusholme}, {Schweyer}, {Terwel}, {Townsend}, {Tzanidakis}, {Wold}, \& {Yan}}]{Rigault2024}
{Rigault}, M., {Smith}, M., {Goobar}, A., {et~al.} 2024, \aap (Submitted)

\bibitem[{{Roman} {et~al.}(2018){Roman}, {Hardin}, {Betoule}, {Astier}, {Balland}, {Ellis}, {Fabbro}, {Guy}, {Hook}, {Howell}, {Lidman}, {Mitra}, {M{\"o}ller}, {Mour{\~a}o}, {Neveu}, {Palanque-Delabrouille}, {Pritchet}, {Regnault}, {Ruhlmann-Kleider}, {Saunders}, \& {Sullivan}}]{2018A&A...615A..68R}
{Roman}, M., {Hardin}, D., {Betoule}, M., {et~al.} 2018, \aap, 615, A68

\bibitem[{{Rubin} {et~al.}(2023){Rubin}, {Aldering}, {Betoule}, {Fruchter}, {Huang}, {Kim}, {Lidman}, {Linder}, {Perlmutter}, {Ruiz-Lapuente}, \& {Suzuki}}]{Rubin2023}
{Rubin}, D., {Aldering}, G., {Betoule}, M., {et~al.} 2023, arXiv e-prints, arXiv:2311.12098

\bibitem[{{Sako} {et~al.}(2018){Sako}, {Bassett}, {Becker}, {Brown}, {Campbell}, {Wolf}, {Cinabro}, {D'Andrea}, {Dawson}, {DeJongh}, {Depoy}, {Dilday}, {Doi}, {Filippenko}, {Fischer}, {Foley}, {Frieman}, {Galbany}, {Garnavich}, {Goobar}, {Gupta}, {Hill}, {Hayden}, {Hlozek}, {Holtzman}, {Hopp}, {Jha}, {Kessler}, {Kollatschny}, {Leloudas}, {Marriner}, {Marshall}, {Miquel}, {Morokuma}, {Mosher}, {Nichol}, {Nordin}, {Olmstead}, {{\"O}stman}, {Prieto}, {Richmond}, {Romani}, {Sollerman}, {Stritzinger}, {Schneider}, {Smith}, {Wheeler}, {Yasuda}, \& {Zheng}}]{Sako2018}
{Sako}, M., {Bassett}, B., {Becker}, A.~C., {et~al.} 2018, \pasp, 130, 064002

\bibitem[{{S{\'a}nchez-S{\'a}ez} {et~al.}(2021){S{\'a}nchez-S{\'a}ez}, {Reyes}, {Valenzuela}, {F{\"o}rster}, {Eyheramendy}, {Elorrieta}, {Bauer}, {Cabrera-Vives}, {Est{\'e}vez}, {Catelan}, {Pignata}, {Huijse}, {De Cicco}, {Ar{\'e}valo}, {Carrasco-Davis}, {Abril}, {Kurtev}, {Borissova}, {Arredondo}, {Castillo-Navarrete}, {Rodriguez}, {Ruz-Mieres}, {Moya}, {Sabatini-Gacit{\'u}a}, {Sep{\'u}lveda-Cobo}, \& {Camacho-I{\~n}iguez}}]{Sanchez-Saez2021}
{S{\'a}nchez-S{\'a}ez}, P., {Reyes}, I., {Valenzuela}, C., {et~al.} 2021, \aj, 161, 141

\bibitem[{{Saulder} {et~al.}(2023){Saulder}, {Howlett}, {Douglass}, {Said}, {BenZvi}, {Ahlen}, {Aldering}, {Bailey}, {Brooks}, {Davis}, {de la Macorra}, {Dey}, {Font-Ribera}, {Forero-Romero}, {Gontcho}, {Honscheid}, {Kim}, {Kisner}, {Kremin}, {Landriau}, {Levi}, {Lucey}, {Meisner}, {Miquel}, {Moustakas}, {Myers}, {Palanque-Delabrouille}, {Percival}, {Poppett}, {Prada}, {Qin}, {Schubnell}, {Tarl{\'e}}, {Vargas Maga{\~n}a}, {Weaver}, {Zhou}, {Zhou}, \& {Zou}}]{Saulder2023}
{Saulder}, C., {Howlett}, C., {Douglass}, K.~A., {et~al.} 2023, MNRAS (submitted), arXiv:2302.13760

\bibitem[{{Schlafly} {et~al.}(2023){Schlafly}, {Kirkby}, {Schlegel}, {Myers}, {Raichoor}, {Dawson}, {Aguilar}, {Allende Prieto}, {Bailey}, {BenZvi}, {Bermejo-Climent}, {Brooks}, {de la Macorra}, {Dey}, {Doel}, {Fanning}, {Font-Ribera}, {Forero-Romero}, {Garc{\'\i}a-Bellido}, {Gontcho A Gontcho}, {Guy}, {Hahn}, {Honscheid}, {Ishak}, {Juneau}, {Kehoe}, {Kisner}, {Kremin}, {Landriau}, {Lang}, {Lasker}, {Levi}, {Magneville}, {Manser}, {Martini}, {Meisner}, {Miquel}, {Moustakas}, {Newman}, {Nie}, {Palanque-Delabrouille}, {Percival}, {Poppett}, {Rockosi}, {Ross}, {Rossi}, {Tarl{\'e}}, {Weaver}, {Y{\`e}che}, {Zhou}, \& {DESI Collaboration}}]{Schlafly2023}
{Schlafly}, E.~F., {Kirkby}, D., {Schlegel}, D.~J., {et~al.} 2023, \aj, 166, 259

\bibitem[{{Schulze} {et~al.}(2018){Schulze}, {Kr{\"u}hler}, {Leloudas}, {Gorosabel}, {Mehner}, {Buchner}, {Kim}, {Ibar}, {Amor{\'\i}n}, {Herrero-Illana}, {Anderson}, {Bauer}, {Christensen}, {de Pasquale}, {de Ugarte Postigo}, {Gallazzi}, {Hjorth}, {Morrell}, {Malesani}, {Sparre}, {Stalder}, {Stark}, {Th{\"o}ne}, \& {Wheeler}}]{Schulze2018}
{Schulze}, S., {Kr{\"u}hler}, T., {Leloudas}, G., {et~al.} 2018, \mnras, 473, 1258

\bibitem[{{Schulze} {et~al.}(2021){Schulze}, {Yaron}, {Sollerman}, {Leloudas}, {Gal}, {Wright}, {Lunnan}, {Gal-Yam}, {Ofek}, {Perley}, {Filippenko}, {Kasliwal}, {Kulkarni}, {Neill}, {Nugent}, {Quimby}, {Sullivan}, {Strotjohann}, {Arcavi}, {Ben-Ami}, {Bianco}, {Bloom}, {De}, {Fraser}, {Fremling}, {Horesh}, {Johansson}, {Kelly}, {Kne{\v{z}}evi{\'c}}, {Kne{\v{z}}evi{\'c}}, {Maguire}, {Nyholm}, {Papadogiannakis}, {Petrushevska}, {Rubin}, {Yan}, {Yang}, {Adams}, {Bufano}, {Clubb}, {Foley}, {Green}, {Harmanen}, {Ho}, {Hook}, {Hosseinzadeh}, {Howell}, {Kong}, {Kotak}, {Matheson}, {McCully}, {Milisavljevic}, {Pan}, {Poznanski}, {Shivvers}, {van Velzen}, \& {Verbeek}}]{PTF2021ApJS..255...29S}
{Schulze}, S., {Yaron}, O., {Sollerman}, J., {et~al.} 2021, \apjs, 255, 29

\bibitem[{{Silber} {et~al.}(2023){Silber}, {Fagrelius}, {Fanning}, {Schubnell}, {Aguilar}, {Ahlen}, {Ameel}, {Ballester}, {Baltay}, {Bebek}, {Benton Beard}, {Besuner}, {Cardiel-Sas}, {Casas}, {Castander}, {Claybaugh}, {Dobson}, {Duan}, {Dunlop}, {Edelstein}, {Emmet}, {Elliott}, {Evatt}, {Gershkovich}, {Guy}, {Harris}, {Heetderks}, {Heetderks}, {Honscheid}, {Illa}, {Jelinsky}, {Jelinsky}, {Jimenez}, {Karcher}, {Kent}, {Kirkby}, {Kneib}, {Lambert}, {Lampton}, {Leitner}, {Levi}, {McCauley}, {Meisner}, {Miller}, {Miquel}, {Mundet}, {Poppett}, {Rabinowitz}, {Reil}, {Roman}, {Schlegel}, {Serrano}, {Van Shourt}, {Sprayberry}, {Tarl{\'e}}, {Tie}, {Weaverdyck}, {Zhang}, {Azzaro}, {Bailey}, {Becerril}, {Blackwell}, {Bouri}, {Brooks}, {Buckley-Geer}, {Castro}, {Derwent}, {Dey}, {Dhungana}, {Doel}, {Eisenstein}, {Fahim}, {Garcia-Bellido}, {Gazta{\~n}aga}, {A Gontcho}, {Gutierrez}, {H{\"o}rler}, {Kehoe}, {Kisner}, {Kremin}, {Kronig}, {Landriau}, {Le Guillou}, {Martini}, {Moustakas}, {Palanque-Delabrouille}, {Peng},
  {Percival}, {Prada}, {Allende Prieto}, {de Rivera}, {Sanchez}, {Sanchez}, {Sharples}, {Soares-Santos}, {Schlafly}, {Weaver}, {Zhou}, {Zhu}, {Zou}, \& {DESI Collaboration}}]{Silber2023}
{Silber}, J.~H., {Fagrelius}, P., {Fanning}, K., {et~al.} 2023, \aj, 165, 9

\bibitem[{{Smartt}(2015)}]{Smartt2015}
{Smartt}, S.~J. 2015, \pasa, 32, e016

\bibitem[{{Smartt} {et~al.}(2009){Smartt}, {Eldridge}, {Crockett}, \& {Maund}}]{Smartt2009}
{Smartt}, S.~J., {Eldridge}, J.~J., {Crockett}, R.~M., \& {Maund}, J.~R. 2009, \mnras, 395, 1409

\bibitem[{{Smith} {et~al.}(2020){Smith}, {Sullivan}, {Wiseman}, {Kessler}, {Scolnic}, {Brout}, {D'Andrea}, {Davis}, {Foley}, {Frohmaier}, {Galbany}, {Gupta}, {Guti{\'e}rrez}, {Hinton}, {Kelsey}, {Lidman}, {Macaulay}, {M{\"o}ller}, {Nichol}, {Nugent}, {Palmese}, {Pursiainen}, {Sako}, {Swann}, {Thomas}, {Tucker}, {Vincenzi}, {Carollo}, {Lewis}, {Sommer}, {Abbott}, {Aguena}, {Allam}, {Avila}, {Bertin}, {Bhargava}, {Brooks}, {Buckley-Geer}, {Burke}, {Carnero Rosell}, {Carrasco Kind}, {Costanzi}, {da Costa}, {De Vicente}, {Desai}, {Diehl}, {Doel}, {Eifler}, {Everett}, {Flaugher}, {Fosalba}, {Frieman}, {Garc{\'\i}a-Bellido}, {Gaztanaga}, {Glazebrook}, {Gruen}, {Gruendl}, {Gschwend}, {Gutierrez}, {Hartley}, {Hollowood}, {Honscheid}, {James}, {Krause}, {Kuehn}, {Kuropatkin}, {Lima}, {MacCrann}, {Maia}, {Marshall}, {Martini}, {Melchior}, {Menanteau}, {Miquel}, {Paz-Chinch{\'o}n}, {Plazas}, {Romer}, {Roodman}, {Rykoff}, {Sanchez}, {Scarpine}, {Schubnell}, {Serrano}, {Sevilla-Noarbe}, {Suchyta}, {Swanson}, {Tarle},
  {Thomas}, {Tucker}, {Varga}, {Walker}, \& {DES Collaboration}}]{2020MNRAS.494.4426S}
{Smith}, M., {Sullivan}, M., {Wiseman}, P., {et~al.} 2020, \mnras, 494, 4426

\bibitem[{{Stahl} {et~al.}(2021){Stahl}, {de Jaeger}, {Boruah}, {Zheng}, {Filippenko}, \& {Hudson}}]{Stahl2021}
{Stahl}, B.~E., {de Jaeger}, T., {Boruah}, S.~S., {et~al.} 2021, \mnras, 505, 2349

\bibitem[{{Sullivan} {et~al.}(2006){Sullivan}, {Le Borgne}, {Pritchet}, {Hodsman}, {Neill}, {Howell}, {Carlberg}, {Astier}, {Aubourg}, {Balam}, {Basa}, {Conley}, {Fabbro}, {Fouchez}, {Guy}, {Hook}, {Pain}, {Palanque-Delabrouille}, {Perrett}, {Regnault}, {Rich}, {Taillet}, {Baumont}, {Bronder}, {Ellis}, {Filiol}, {Lusset}, {Perlmutter}, {Ripoche}, \& {Tao}}]{Sullivan2006rates}
{Sullivan}, M., {Le Borgne}, D., {Pritchet}, C.~J., {et~al.} 2006, \apj, 648, 868

\bibitem[{{Sullivan} {et~al.}(2010{\natexlab{a}}){Sullivan}, {Conley}, {Howell}, {Neill}, {Astier}, {Balland}, {Basa}, {Carlberg}, {Fouchez}, {Guy}, {Hardin}, {Hook}, {Pain}, {Palanque-Delabrouille}, {Perrett}, {Pritchet}, {Regnault}, {Rich}, {Ruhlmann-Kleider}, {Baumont}, {Hsiao}, {Kronborg}, {Lidman}, {Perlmutter}, \& {Walker}}]{Sullivan2010}
{Sullivan}, M., {Conley}, A., {Howell}, D.~A., {et~al.} 2010{\natexlab{a}}, \mnras, 406, 782

\bibitem[{{Sullivan} {et~al.}(2010{\natexlab{b}}){Sullivan}, {Conley}, {Howell}, {Neill}, {Astier}, {Balland}, {Basa}, {Carlberg}, {Fouchez}, {Guy}, {Hardin}, {Hook}, {Pain}, {Palanque-Delabrouille}, {Perrett}, {Pritchet}, {Regnault}, {Rich}, {Ruhlmann-Kleider}, {Baumont}, {Hsiao}, {Kronborg}, {Lidman}, {Perlmutter}, \& {Walker}}]{2010MNRAS.406..782S}
---. 2010{\natexlab{b}}, \mnras, 406, 782

\bibitem[{{Tachibana} \& {Miller}(2018)}]{Tachibana2018}
{Tachibana}, Y., \& {Miller}, A.~A. 2018, \pasp, 130, 128001

\bibitem[{{Taddia} {et~al.}(2019){Taddia}, {Sollerman}, {Fremling}, {Barbarino}, {Karamehmetoglu}, {Arcavi}, {Cenko}, {Filippenko}, {Gal-Yam}, {Hiramatsu}, {Hosseinzadeh}, {Howell}, {Kulkarni}, {Laher}, {Lunnan}, {Masci}, {Nugent}, {Nyholm}, {Perley}, {Quimby}, \& {Silverman}}]{PTF2019A&A...621A..71T}
{Taddia}, F., {Sollerman}, J., {Fremling}, C., {et~al.} 2019, \aap, 621, A71

\bibitem[{{Tripp}(1998)}]{Tripp1998}
{Tripp}, R. 1998, \aap, 331, 815

\bibitem[{{Tully} \& {Fisher}(1977)}]{Tully1977}
{Tully}, R.~B., \& {Fisher}, J.~R. 1977, \aap, 54, 661

\bibitem[{{Tully} {et~al.}(2023){Tully}, {Kourkchi}, {Courtois}, {Anand}, {Blakeslee}, {Brout}, {Jaeger}, {Dupuy}, {Guinet}, {Howlett}, {Jensen}, {Pomar{\`e}de}, {Rizzi}, {Rubin}, {Said}, {Scolnic}, \& {Stahl}}]{Tully2023}
{Tully}, R.~B., {Kourkchi}, E., {Courtois}, H.~M., {et~al.} 2023, \apj, 944, 94

\bibitem[{{Turnbull} {et~al.}(2012){Turnbull}, {Hudson}, {Feldman}, {Hicken}, {Kirshner}, \& {Watkins}}]{Turnbull2012}
{Turnbull}, S.~J., {Hudson}, M.~J., {Feldman}, H.~A., {et~al.} 2012, \mnras, 420, 447

\bibitem[{{Tyson}(2002)}]{Tyson2002}
{Tyson}, J.~A. 2002, in Society of Photo-Optical Instrumentation Engineers (SPIE) Conference Series, Vol. 4836, Survey and Other Telescope Technologies and Discoveries, ed. J.~A. {Tyson} \& S.~{Wolff}, 10--20

\bibitem[{{Uddin} {et~al.}(2017){Uddin}, {Mould}, {Lidman}, {Ruhlmann-Kleider}, \& {Zhang}}]{2017ApJ...848...56U}
{Uddin}, S.~A., {Mould}, J., {Lidman}, C., {Ruhlmann-Kleider}, V., \& {Zhang}, B.~R. 2017, \apj, 848, 56

\bibitem[{{van der Walt} {et~al.}(2019){van der Walt}, {Crellin-Quick}, \& {Bloom}}]{Skyportal}
{van der Walt}, S., {Crellin-Quick}, A., \& {Bloom}, J. 2019, The Journal of Open Source Software, 4, 1247

\bibitem[{{Villar} {et~al.}(2019){Villar}, {Berger}, {Miller}, {Chornock}, {Rest}, {Jones}, {Drout}, {Foley}, {Kirshner}, {Lunnan}, {Magnier}, {Milisavljevic}, {Sanders}, \& {Scolnic}}]{Villar2019}
{Villar}, V.~A., {Berger}, E., {Miller}, G., {et~al.} 2019, \apj, 884, 83

\bibitem[{{Villar} {et~al.}(2020){Villar}, {Hosseinzadeh}, {Berger}, {Ntampaka}, {Jones}, {Challis}, {Chornock}, {Drout}, {Foley}, {Kirshner}, {Lunnan}, {Margutti}, {Milisavljevic}, {Sanders}, {Pan}, {Rest}, {Scolnic}, {Magnier}, {Metcalfe}, {Wainscoat}, \& {Waters}}]{Villar2020}
{Villar}, V.~A., {Hosseinzadeh}, G., {Berger}, E., {et~al.} 2020, \apj, 905, 94

\end{thebibliography}


%

\end{document}